\documentclass[]{jfm}

\usepackage{graphicx}
\usepackage{newtxtext}
\usepackage{newtxmath}
\usepackage{natbib}
\usepackage{hyperref}
\usepackage{ulem}
\usepackage{cancel}
\usepackage{overpic}
\usepackage{comment}
\hypersetup{
    colorlinks = true,
    urlcolor   = blue,
    citecolor  = black,
}

\newcommand{\RomanNumeralCaps}[1]
\linenumbers

\newcommand{\St}{\mathrm{St}}
\newcommand{\upd}{\mathrm{d}}

\newcommand{\blue}[1]{{\color{blue} #1}}

\title{Uncovering flow and deformation regimes in the coupled fluid-solid vestibular system}

\author{Javier Chico-V\'{a}zquez,
 Derek E. Moulton \and Dominic Vella \corresp{\email{dominic.vella@maths.ox.ac.uk}}}

\affiliation{Mathematical Institute, University of Oxford, Woodstock Road, Oxford, OX2 6GG, UK
}

\begin{document}
\maketitle

\begin{abstract}
%Abstract here. 
In this paper, we showcase how flow obstruction by a deformable object can lead to symmetry breaking in curved domains subject to angular acceleration. Our analysis is  motivated  by the deflection of the cupula, a soft tissue located in the inner ear that is used to perceive rotational motion as part of  the vestibular system. The cupula is understood to block the rotation-induced  flow in a toroidal region with the flow-induced deformation of the cupula used by the brain to infer motion. By asymptotically solving the governing equations for this flow, we characterise regimes for which the sensory system is sensitive to either angular velocity or angular acceleration. Moreover, we show the fluid flow is not symmetric in the latter case. Finally, we extend our  analysis of symmetry breaking to understand the formation of vortical flow in cavernous regions within channels. We discuss the implications of our results for the sensing of rotation by mammals.
\end{abstract}

\section{Introduction}
\label{sec:introduction}
\begin{figure}
    \centering
    \includegraphics[width=0.99\linewidth]{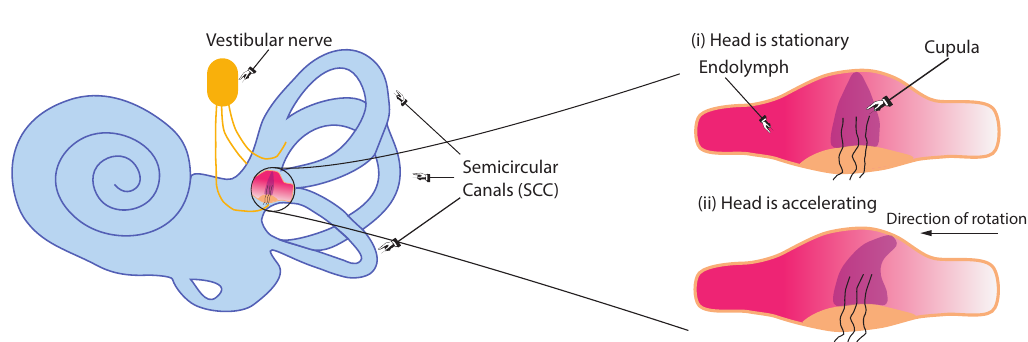}
    \caption{Schematic of the vestibular system. (a) The inner ear and vestibular apparatus: Three mutually orthogonal  semicircular canals (SCCs), each containing a cupula, send information to the nervous system about the rotational motion of the head. (b) Zoom in of the obstruction within each SCC caused by the cupula. Information about the rotation of each SCC is inferred from the deflection of its cupula --- the inertia of the fluid that fills the SCC (endolymph) causes the cupula to deform. (Cupula deformation is sensed via innervated cilia that are embedded within the cupula.)   
    }
    \label{fig:anatomy}
\end{figure}

Rotational motion of the head in humans is perceived through the vestibular system, which is located in the inner ear \citep{paulin_models_2019}. For mathematical modeling purposes, this system can be described as a set of three mutually orthogonal, nearly circular canals, known as the \textit{semicircular canals (SCCs)} in the anatomical literature \citep{oghalai_anatomy_2020, curthoys_dimensions_1987}. These canals resemble deformed tori, where the slender regions are filled with a Newtonian fluid called \textit{endolymph}. The larger region is comprised of two cavities, the \emph{utricle} and the \emph{ampulla}. The latter houses a gelatinous protein-polysaccharide elastic membrane known as the \textit{cupula} \citep{casale_physiology_2024}, which is innervated by hair cells (\textit{cilia}); the innervated cilia  transmit mechanical deflections of the cupula to the nervous system via the vestibular nerve \citep{waxman_vestibular_2024}. 
In mechanical terms, a change in angular velocity about a given axis drives a fluid flow in that canal, which generates a pressure gradient, deforming the cupula (and hence the innervated hairs) and thus allowing the brain to sense the motion. Specifically, as the walls of the canal rigidly rotate with the head, the fluid in the center is left behind, causing the cupula to deform in the opposite direction to the imposed rotation. The mutually orthogonal structure of the semicircular canals allows the detection of any three dimensional rotation of the head. A schematic diagram of the vestibular system is provided in figure~\ref{fig:anatomy}. 

The vestibular system can be affected by several pathologies that disrupt its normal function. One of the most common disorders is benign paroxysmal positional vertigo (BPPV), in which brief episodes of vertigo are triggered by specific head movements. BPPV occurs when calcium carbonate crystals (otoconia) dislodge from the utricle and move into the semicircular canals, causing abnormal stimulation of the vestibular nerve \citep{hornibrook_benign_2011}. Another significant condition is vestibular neuritis, an inflammation of the vestibular nerve, usually caused by viral infections, which leads to acute vertigo, imbalance and nausea \citep{royal_chapter_2014}. Meniere’s disease also affects the vestibular system, causing episodic vertigo due to abnormal fluid buildup in the inner ear, leading to disturbances in balance \citep{harcourt_menieres_2014}. Early diagnosis and appropriate treatment of these vestibular pathologies are essential to improving quality of life and preventing chronic balance issues. Here, mathematical modelling has great potential in enabling for quantitative predictions of balance response and in elucidating the sensitivity of the vestibular system to material changes, for instance as may occur with ageing \citep{konrad_balance_1999}.

A number of mathematical models exist for the vestibular dynamics, both numerical and analytical. On the analytical side, beyond the early phenomenological oscillator models from the 1930's \citep{steinhausen_ueber_1933}, recent models can be classified into two broad categories. The first approach is that of \citeauthor{obrist_fluidmechanics_2008} and co-authors \citep{obrist_fluidmechanics_2008,vega_mathematical_2008,buskirk_fluid_1976} in which the geometry is idealized to allow a solution to be found under arbitrary forcing of the system, i.e.~arbitrary rotational motion. The second approach is that of  \citeauthor{rabbitt_hydroelastic_1992}, who maintain a more realistic geometry but require strong assumptions on the form of the forcing to make analytical progress. In particular, in their series of papers \citep{rabbitt_hydroelastic_1992, damiano_singular_1996, damiano_poroelastic_1999},  \citeauthor{rabbitt_hydroelastic_1992} assume that the forcing is sinusoidal, as might be expected when tilting the head up or down (for instance when nodding). 
Fully numerical investigations of the vestibular system also exist  \citep{boselli_numerical_2009,grieser_validation_2012,boselli_vortical_2013} --- these generally implement a realistic channel geometry but do not model the cupular deformation as a fluid-structure interaction. Instead, they incorporate the effect of cupular deformation via a periodic boundary condition for the flow coupled to a time dependent pressure jump. Recent studies have also modelled the fluid–structure interaction \citep{chung_numerical_2010,goyens_asymmetric_2019,kassemi_fluidstructural_2005,wu_dynamic_2011}; however, the high computational cost of such simulations often restricts them to a single manoeuvre and a single set of parameters, rather than enabling a broader exploration of parameter space to uncover the flow and deformation regimes of the vestibular system.

In this paper, we present both numerical simulations and an analytical approach for cupular dynamics. Our analytical model is derived from first principles, including explicitly both toroidal fluid flow and the mechanics of the cupula. By exploiting the slenderness of the semicircular canals, and applying a detailed asymptotic analysis, we obtain a reduced model that allows us to incorporate both arbitrary geometry and arbitrary forcing, combining the best of previous approaches. Moreover, unlike previous work \citep{rabbitt_hydroelastic_1992,obrist_fluidmechanics_2008} we model the cupula as a full three-dimensional elastic solid, without making a small thickness assumption. We complement this with numerical computations, specifically including fluid-solid couplings. Our combined numerical and analytical approach enables us to validate the reduced analytical model, uncover a number of novel features, including characterizing flow regimes and identifying regions of parameter space with distinct system response.

One of the key motivating issues underlying our study concerns the mechanical properties of the cupula. Although the anatomy of the vestibular system is well understood, the architecture itself is incredibly delicate and fragile, which prohibits the possibility of direct mechanical testing. For this reason, the stiffness of the cupula has only ever been obtained through indirect measurement, a procedure that has produced both some uncertainty and surprisingly low values; for example, a Young's modulus of around 5 Pa has been reported \citep{selva_mechanical_2009}, which is well below values typically associated with soft biological tissues \cite[see][where they estimate the Young's modulus of brain matter to be 1 kPa]{budday_mechanical_2015}. Nevertheless, the stiffness of the cupula is a key mechanical parameter, as it dictates the degree of deformation under a given flow and therefore the potential for and degree of stimulus. In fact, as we will show, this parameter plays an even stronger role, impacting not just the degree of deformation but the qualitative nature of the flow induced by motion as well. By examining the behaviour of our model as the relative stiffness of the cupula varies, we will demonstrate the presence of two distinct regimes: for ``soft" cupulas the deformation follows the angular velocity of the imposed motion, while for ``stiff" cupulas the deformation instead tracks the angular acceleration. 

Moreover, we will explain how the second of these regimes is connected to a symmetry breaking of the flow in the endolymph. As we shall demonstrate in our numerical simulations, presented in Section~\ref{sec:numerical_simulations_surprise}, the flow in the endolymph is only axisymmetric (relative to the duct's center-line or axis) under particular conditions. Despite this observation, many existing models have implicitly assumed a radially symmetric flow. Examination of our analytical solution enables us to explain exactly when and how symmetry breaking occurs. This feature is interesting in the context of the broader literature on flow through curved pipes. Although the effect of curvature on pipe flow was first discussed by \cite{dean_lxxii_1928}, the plethora of subsequent studies have focused mostly on steady flows \citep{chupin_full_2008,pedley_fluid_1980, siggers_steady_2005} --- there have been many fewer investigations into unsteady fluid phenomena \cite[though see ][for an example]{siggers_unsteady_2008}. We shall show that the essential coupling between the Euler force and the (\emph{a priori}) unknown pressure gradient can lead to the annihilation of the symmetric leading order velocity --- a situation that distinguishes this problem from classical studies of flow in curved pipes. We conclude our study with an investigation of the emergence of vortical flow in the utricle. This feature has been  reported previously, but only in numerical simulations \citep{grieser_validation_2012,boselli_vortical_2013}; our model provides both an analytical understanding and an explicit characterization for when vortical flow will emerge.

 \section{Governing equations}\label{sec:governing_equations}
 
 \begin{figure}
     \centering
     {\tiny
    \begin{overpic}[width=0.99\textwidth]{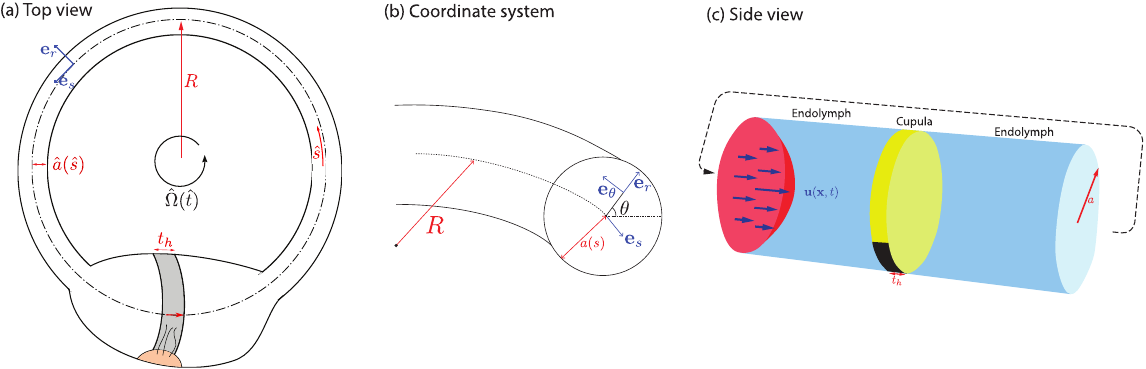}
    \put(8.5,8){Utricle +}
    \put(9,6.7){ampulla}
    \put(13.9,9.2){\blue{$\rho_s$}}
    \put(14.2,7.7){\blue{$E$}}
    \put(14.4,6.7){\blue{$\nu_s$}}
    \put(14.6,5.2){\textcolor{red}{$\hat{z}$}}

    \put(16.5,1.5){\blue{$\mu$, $\rho$}}
    \end{overpic}
    }

     \caption{Problem setup. (a) Plan view of a semicircular canal showing the spatially-varying canal radius, $\hat{a}(\hat{s})$, and the cupula (shaded in grey), which is situated in the enlarged portion, or utricle. (b) Schematic of the chosen coordinate system. (c) Close up of the region around the cupula, highlighting the cupula's thickness, $t_h$, and its attachment to the canal walls via the `crista' (black region). (The toroidal flow is shown schematically here to allow the zoom in on the cupula.)  }
     \label{fig:diagrams}
 \end{figure}

We consider a single semicircular canal, as portrayed schematically in Figure~\ref{fig:diagrams}(a): the endolymph fills a toroidal structure whose centreline forms a circle of radius $R$ and whose radius is small and spatially-varying, denoted $\hat{a}(\hat{s})\ll R$, where $\hat{s}$ is an arc length parameter along the centreline. The canal is subjected to a rotation defined by angular velocity $\hat{\Omega}(\hat{t})$ around the centre of the toroid with rotation axis normal to the plane of the centreline. We remark that this angular velocity will be the same as say the angular velocity of motion not centred around the toroid, like, say, the flow driven by riding a merry-go-round or a cornering car, as explained in Appendix~\ref{sec:angular_velocity_is_the_same}. The endolymph is assumed to be an incompressible Newtonian fluid of dynamic viscosity $\mu$ and density $\rho$. The elastic, gel-like cupula occupies a thin region (shown in grey in figure~\ref{fig:diagrams}) and  has density $\rho_s$, Young's modulus $E$, thickness $t_h$ and Poisson ratio $\nu_s$.  In the absence of detailed observations, the  cupula is assumed to occupy the entire cross-section of the canal, so that it is attached to the wall all the way around its circumference, as can be seen in Figure~\ref{fig:diagrams}(c), where the solid cupula is shaded in yellow and the liquid endolymph is shaded in blue. (The region shaded in black  represents a structure called the crista which attaches the cupula to the canal wall.)

\subsection{Equations for the bulk}
The Navier--Stokes equations for the dimensional fluid velocity $\hat{\mathbf{u}}$ and modified pressure $\hat{p}$ in the co-rotating frame are given by \citep{landau_fluid_1987}
\begin{subequations}\label{eq:navier_stokes_dimensional}
\begin{align}
\boldsymbol{\nabla}\cdot \hat{\mathbf{u}}&=0,\\
    \rho\left(\frac{D \mathbf{\hat{u}}}{D \hat{t}}+\frac{\partial \mathbf{\hat{\Omega}}}{\partial \hat{t}}\times \mathbf{\hat{x}}+2 \mathbf{\hat{\Omega}}\times \mathbf{\hat{u}}+\mathbf{\hat{\Omega}}\times(\mathbf{\hat{\Omega}}\times \mathbf{\hat{x}})\right)&=-\boldsymbol{\nabla} \hat{p}+\mu\boldsymbol{\nabla}^2\mathbf{\hat{u}}.
    \end{align}
\end{subequations}
Here, the first of the extra terms on the left hand side is the Euler force, the second term is the Coriolis force and the final additional term is the centrifugal force, each due to the imposed rotation. The pressure $\hat{p}$ is a modified pressure in the sense that it incorporates the linear fictitious force associated with the linear acceleration of the SCC \citep{buskirk_fluid_1976}. The fluid is assumed to satisfy the no-slip condition at the edges of the walls, so that $\mathbf{\hat{u}}(\hat{r}=\hat{a}(\hat{s}))=\mathbf{0}$ for $\hat{s}\in\,(0,2\pi R)$. Motivated by the small strains in the cupula \citep{selva_mechanical_2009}, it is modeled as a linearly elastic material, satisfying the steady Navier equations:
\begin{subequations}\label{eq:navier_dimensional}
\begin{align}
\rho_s\left(\frac{\partial^2 \boldsymbol{\hat{u}}_s}{\partial t^2}+\frac{\partial \mathbf{\hat{\Omega}}}{\partial \hat{t}}\times \mathbf{\hat{x}}+2\mathbf{\hat{\Omega}}\times \frac{\partial\boldsymbol{\hat{u}}_s}{\partial \hat{t}}+\mathbf{\hat{\Omega}}\times (\mathbf{\hat{\Omega}}\times \mathbf{\hat{x}})\right)
=\boldsymbol{\nabla}\cdot\boldsymbol{\hat{\tau}},
\\
\boldsymbol{\hat{\tau}}=2\mu_s\boldsymbol{\hat{E}}+\lambda_s \mathrm{tr}(\boldsymbol{\hat{E}})\mathbf{1},\quad 2\boldsymbol{\hat{E}}=\boldsymbol{\nabla}\boldsymbol{\hat{u}}_s+(\boldsymbol{\nabla}\boldsymbol{\hat{u}}_s)^\intercal,
\end{align}
\end{subequations}
alongside a linear Hookean constitutive law relating stress $\boldsymbol{\hat{\tau}}$ and the cupular displacement field $\boldsymbol{\hat{u}}_s(\mathbf{x},t)$. The Lame constants are $\mu_s=E/(2(1+\nu_s))$ and $\lambda_s=\nu_s E/((1+\nu_s)(1-2\nu_s))$ \citep{bower_applied_2009}. Finally, we model the fluid structure interaction at the cupula-endolymph boundary in the usual way, imposing continuity of velocity and stress \citep{gkanis_instability_2006}. 

\subsection{Numerical simulations} \label{sec:numerical_simulations_surprise} 

The system of equations \eqref{eq:navier_stokes_dimensional}-\eqref{eq:navier_dimensional} was simulated in COMSOL for different values of the Young's modulus of the cupula and with a Poisson ratio $\nu_s=0.48$. We impose a simple sinusoidal forcing, given by $\hat{
\Omega}(\hat{t}) = \Omega_0\sin (2\pi \hat{t}/\mathcal{T})$, with $\Omega_0=1\,\mathrm{rad/s}$ and $\mathcal{T}=1\mathrm{~s}$. The geometrical parameters are $a = 1.6 \times 10^{-4}$ m, $R=3.2\times 10^{-3}$ m and $t_h = 0.8\times 10^{-4}$ m \citep{daocai_size_2014}. 

The equations were solved for a solid cupula of finite size, i.e.~with no thin cupula assumption. Further details on the numerical simulations are available in Appendix~\ref{sec:COMSOL_details}. This includes deformation profiles of the solid cupula (Figure~\ref{fig:solid_fluid_cupular_profiles}). Moreover,  Figure~\ref{fig:solid_fluid_cupular_profiles} shows that the magnitude of cupula deformation is inversely proportional to Young's modulus.

\begin{figure}
    \centering
    \begin{overpic}[width=0.99\textwidth]{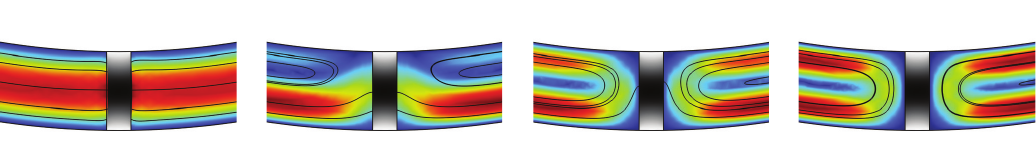}
    \put(0.1,13){\scriptsize(a) $E=10^2$ Pa}

    \put(26,13){\scriptsize(b) $E=10^3$ Pa}

    \put(51.7,13){\scriptsize(c) $E=10^4$ Pa}

    \put(77.5,13){\scriptsize(d) $E=10^5$ Pa}
        
    \end{overpic}
    \caption{Cross-section of the velocity fields in the cupula as computed using COMSOL simulations. Results are shown for a range of cupula stiffnesses. Colour represents the relative magnitude of the fluid speed, with red denoting  regions in which the flow is fast and blue representing stagnant regions; streamlines are represented by solid black curves. As the  stiffness of the cupula increases, a symmetry-breaking of the flow occurs. In particular, for values of the Young's modulus $E>10^3$ Pa the flow is usually not axially symmetric. Here $\hat{
\Omega}(\hat{t}) =\Omega_0 \sin \left(2\pi \hat{t}/\mathcal{T}\right)$ and the snapshots are taken at $\hat{t}=0.25$ s, with $\Omega_0=1$ rad$\cdot$s$^{-1}$ and $\mathcal{T}=1$ sec. The geometrical parameters are $a = 1.6 \times 10^{-4}$ m, $R=3.2\times 10^{-3}$ m and $t_h = 0.8\times 10^{-4}$ m}.

    \label{fig:symmetry_breaking}
\end{figure}

Flow profiles on either side of the cupula produced by these numerical simulations are shown in figure~\ref{fig:symmetry_breaking}. Here a top view of the mid-plane of the flow around the canal is plotted, with colour  indicating the speed of the flow, normalized by the maximum speed throughout the flow and with fast regions coloured red, stagnant regions coloured blue. The cupula appears as the central region and is shown in its deformed configuration. This deflection is imperceptible for all except the $E=10^2\mathrm{~Pa}$ case and so the relative magnitude of the cupula deformation is indicated by the grey scale colouring, which shows that it is maximum at the centre. Figure~\ref{fig:symmetry_breaking} shows the velocity field at time $\hat{t}=0.25$ s, a stage at which transients from the initial condition remain significant. However, it is not necessary to wait for these transients to decay completely to observe the emergence of a distinctly asymmetric velocity profile.

In the panels of figure~\ref{fig:symmetry_breaking}, the Young's modulus, $E$, of the cupula increases from left to right. As should be expected, the magnitude of the deflection decreases as the material becomes stiffer. This is confirmed quantitatively in Figure~\ref{fig:solid_fluid_cupular_profiles} of Appendix~\ref{sec:COMSOL_details}.
Surprisingly, however, we also observe a significant change in flow behaviour: for small values of $E$, the flow is axially symmetric about the centerline of the tube but as $E$ increases the flow transitions, losing its axisymmetry and exhibiting vortical structures.
It is worth noting that these plots use values of the Young's modulus that cover a typical physiological range for soft biological tissues \citep{goriely_mathematics_2017}. Notwithstanding our remark in the introduction that even lower values have been suggested via indirect methods, a flow transition with physiologically relevant values of $E$ suggests that asymmetric flow may be present in a physiological vestibular system. If so, this would contradict the assumption of axial symmetry that is typical in previous models \citep{obrist_fluidmechanics_2008, rabbitt_hydroelastic_1992}, and raises interesting questions about what impact such asymmetry might have on the mechanics of balance and rotational sensing.  To study this behaviour further, we thus turn now to a theoretical analysis of the governing system.

\subsection{Theoretical model}
To investigate the symmetry breaking observed in the numerical simulations, and to obtain a qualitative understanding of flow and deformation characteristics, in this section we use asymptotic analysis to derive a reduced order equation for the deformation of the cupula.

To capture the geometry of the semicircular canal, we introduce toroidal coordinates $(\hat{r},\theta, \hat{s})$, in which $\hat{s}\,\in \,(0,2\pi R)$ is the arc-length along the centreline of the torus. A sketch of this coordinate system is provided in Figure~\ref{fig:diagrams}(b). Cartesian coordinates are related to toroidal coordinates via \citep{pedley_fluid_1980}:
\begin{equation}
    \begin{cases}
        \hat{x} = &(R + \hat{r }\cos{\theta})\cos(\frac{\hat{s}}{R}), \\ 
        \hat{y} = &(R+\hat{r}\cos\theta)\sin(\frac{\hat{s}}{R}),
        \\
        \hat{z} = & -\hat{r} \sin\theta.
    \end{cases}\label{eq:toroidal_coordinates}
\end{equation}
The negative sign in the last equation ensures that the orthonormal basis vectors $\{\mathbf{e}_r, \mathbf{e}_\theta, \mathbf{e}_s\}$ follow the right hand rule. We can now rewrite the Navier-Stokes equations \eqref{eq:navier_stokes_dimensional} in component form. Writing the velocity vector as $\mathbf{\hat{u}}=\hat{u} \mathbf{e}_r +\hat{v}\mathbf{e}_\theta+\hat{w}\mathbf{e}_s$, the continuity equation becomes \cite[see][for example]{pedley_fluid_1980}
 \begin{equation}
     \frac{\partial \hat{u}}{\partial \hat{r}}+\frac{\hat{u}}{\hat{r}}+\frac{1}{\hat{r}}\frac{\partial \hat{v}}{\partial \theta}+\frac{1}{h}\frac{\partial \hat{w}}{\partial \hat{s}}-\frac{\hat{v}\sin\theta }{R h}+\frac{\hat{u}\cos\theta }{R h}=0\label{eq:dimensional_continuity_inertial},
 \end{equation}
where $h=1+\hat{r}\cos(\theta)/R$ is a scale factor. The momentum equations are \citep[see][for the equations expressed in an inertial frame]{pedley_fluid_1980}: 
\begin{subequations}
    \begin{align}
        &\begin{split}
        \rho&\left(\frac{\partial \hat{u}}{\partial \hat{t}} +\hat{u}\frac{\partial \hat{u}}{\partial \hat{r}}+\frac{\hat{v}}{\hat{r}}\frac{\partial \hat{u}}{\partial \theta}+\frac{\hat{w}}{h}\frac{\partial \hat{u}}{\partial \hat{s}}-\frac{\hat{v}^2}{\hat{r}}-\frac{\hat{w}^2}{h}\frac{\cos\theta}{R} -2\hat{\Omega}\hat{w}\cos\theta-\hat{\Omega}^2Rh\cos\theta\right) \\ &= -\frac{\partial \hat{p}}{\partial \hat{r}}+\frac{\mu}{\hat{r}h}\left[\frac{\hat{r}}{h}\frac{\partial }{\partial \hat{s}}\left(\frac{\partial \hat{u}}{\partial \hat{s}}-\frac{\partial }{\partial \hat{r}}(h\hat{w})\right)-\frac{\partial }{\partial \theta}\left(\frac{h}{\hat{r}}\frac{\partial}{\partial \hat{r}}(\hat{r} \hat{v})-\frac{h}{\hat{r}}\frac{\partial \hat{u}}{\partial \theta}\right)\right],
    \end{split}\\
     &\begin{split}
        \rho&\left(\frac{\partial \hat{v}}{\partial \hat{t}}+\hat{u}\frac{\partial \hat{v}}{\partial \hat{r}}+\frac{\hat{v}}{\hat{r}}\frac{\partial \hat{v}}{\partial \theta}+\frac{\hat{w}}{h}\frac{\partial \hat{v}}{\partial \hat{s}}+\frac{\hat{u}\hat{v}}{\hat{r}}+\frac{\hat{w}^2}{Rh}\sin\theta+2\hat{\Omega} \hat{w}\sin\theta+\hat{\Omega}^2 Rh\sin\theta\right) \\&=-\frac{1}{\hat{r}}\frac{\partial \hat{p}}{\partial \theta}+\frac{\mu}{h}\frac{\partial}{\partial \hat{r}}\left[\frac{h}{\hat{r}}\left(\frac{\partial }{\partial \hat{r}}(\hat{r} \hat{v})-\frac{\partial \hat{u}}{\partial \theta}\right)\right]-\frac{\mu}{\hat{r}h^2}\left[\frac{\partial}{\partial \theta}\left(h \frac{\partial \hat{w}}{\partial \hat{s}}\right)-\hat{r}\frac{\partial^2 \hat{v}}{\partial \hat{s}^2}\right],
    \end{split}\\
     &\begin{split}
    \rho&\left(\frac{\partial \hat{w}}{\partial \hat{t}}+\hat{u}\frac{\partial \hat{w}}{\partial \hat{r}}+\frac{\hat{v}}{\hat{r}}\frac{\partial \hat{w}}{\partial \theta}+\frac{\hat{w}}{h}\frac{\partial \hat{w}}{\partial \hat{s}}+\frac{\hat{u}\hat{w}}{Rh}\cos\theta-\frac{\hat{v}\hat{w}}{Rh}\sin\theta\right. \\
    &\left.+\frac{d\hat{\Omega}}{d\hat{t}}(R+\hat{r}\cos\theta)+2\hat{\Omega} \hat{u}\cos\theta-2\hat{\Omega} \hat{v}\sin\theta\right)\\&= -\frac{1}{h}\frac{\partial \hat{p}}{\partial \hat{s}}+\frac{\mu}{\hat{r}^2}\frac{\partial}{\partial 
 \theta}\left[\frac{1}{h}\frac{\partial }{\partial \theta}(h\hat{w})-\frac{\hat{r}}{h}\frac{\partial \hat{v}}{\partial \hat{s}}\right]
    -\frac{\mu}{\hat{r}}\frac{\partial }{\partial \hat{r}}\left[\frac{\hat{r}}{h}\left(\frac{\partial \hat{u}}{\partial \hat{s}}-\frac{\partial }{\partial \hat{r}}(h \hat{w})\right)\right]\label{eq:dimensional_momentum_s}.
    \end{split}
    \end{align}\label{eq:fluid_problem_dimensional}
\end{subequations}
We locate the cupula at arc length position $\hat{s}=0$. 
As indicated above, our numerical simulations have shown that the deformation of the cupula is small compared to the tube radius, suggesting that strains are small and thus that it is sufficient to use a linear equation. This small strain assumption will be confirmed in Section~\ref{sec:scalings} through a scaling argument.
We write the equations for the solid deformation of the cupula in cylindrical coordinates, as the cupula is thin enough that the curvature of the SCC is not important. The cupula is thus modelled as a solid cylinder $0\leq\hat{r}\leq\hat{a}(0)$, $0\leq\theta\leq 2\pi$, $-t_h/2\leq\hat{z}\leq t_h/2$, with the centre of mass of the cupula ($\hat{r}=\theta=\hat{z}=0$) located at position $\hat{r}=\theta=\hat{s}=0$ (or equivalently $\hat{s}=2\pi R$) in terms of the toroidal coordinates of the fluid problem (see figure~\ref{fig:diagrams}(a)). The thickness of the cupula $t_h$ is not assumed to be small when compared to its radius $\hat{a}(0)$. The equations for the solid deformation in component form \eqref{eq:navier_dimensional} are given in \S\ref{sec:solid_mechanics}.

We require boundary conditions for both the fluid problem \eqref{eq:fluid_problem_dimensional} and the elastic problem~\eqref{eq:navier_dimensional}. 
The walls of the endolymph are assumed to have a no-slip condition, so that in the rotating frame, $\mathbf{\hat{u}}(\hat{r}=\hat{a})=0$. We require a second set of boundary conditions where the cupula and the endolymph meet. The kinematic boundary condition requires that 
the cupular velocity and the endolymph velocity at the surface of the cupula must match. As the spatial gradients of the cupula's deformation are small, we can write this as
\begin{subequations}
\begin{align}\label{eq:bdary_cond_velocities_match}
    &\frac{\partial \hat{\boldsymbol{u}}_s}{\partial \hat{t}}(\hat{r},\theta,\hat{z}=t_h/2,\hat t)
    =\mathbf{\hat{u}}(\hat{r}, \hat{s}=t_h/2,\hat{t}),\\
    &\frac{\partial \hat{\boldsymbol{u}}_s}{\partial \hat{t}}(\hat{r},\theta,\hat{z}=2\pi R-t_h/2,\hat t)
    =\mathbf{\hat{u}}(\hat{r},\hat{s}=2\pi R-t_h/2,\hat{t}).
\end{align}
\end{subequations}
We also require boundary conditions for the solid problem ~\eqref{eq:navier_dimensional}. The precise attachment between the cupula and the utricle is an open area of research and the conditions to be satisfied are not immediately clear. We opt to implement a straightforward choice motivated by \cite{rabbitt_hydroelastic_1992} who model the cupula as a clamped plate, but we extend this to include thickness effects. In this regime, the forcing for the cupular displacement is given by the pressure jump across the two sides of the cupula. In particular, at the flat faces of the cylinder we impose the following conditions on the stress tensor

\begin{subequations}
\begin{align}\label{eq:solid_boundary_conditions}
    \hat{\tau}_{zz}=\pm\frac{1}{2}\left[\hat{p}(\hat{r},\theta,\hat{s}=2\pi R-t_h/2,\hat{t})-\hat{p}(\hat{r},\theta,\hat{s}=t_h/2,\hat{t})\right]\equiv\pm\frac{\Delta \hat{p}}{2}, \quad &\text{at }\hat{z}=\pm t_h/2,\\
    \hat{\tau}_{rz}=0,\quad\hat{\tau}_{\theta z}=0, \quad &\text{at }\hat{z}=\pm t_h/2.
\end{align}
\end{subequations}

On the curved face we impose zero displacement, $\boldsymbol{\hat u}_s(\hat{r}=\hat{a}(0))=\boldsymbol 0$.
We also require initial conditions for the velocity, $\mathbf{\hat
{u}}(\mathbf{\hat{x}},\hat{t}=0)$ and the cupular deflection $\boldsymbol{\hat{u}_s}(\hat{t}=0)$. Unless otherwise stated we assume that the system is initially at rest and is undeformed.

\subsection{Scalings and non-dimensionalization}\label{sec:scalings}

The SCCs in mammals are thin and slender: with an aspect ratio $\epsilon=a/R$ between 0.05 and 0.1 \citep{daocai_size_2014}. 
It is therefore natural to exploit $\epsilon\ll1$ and use an asymptotic approach 
to perform a long wavelength asymptotic analysis similar to lubrication theory. We also know from the continuity equation that if $\hat{w}\sim \mathcal{U}$ then $\hat{u},\hat{v}\sim \epsilon \mathcal{U}$. This is required to have a non-trivial balance and is a typical scaling in lubrication theory \citep{craster_viscous_2006,papageorgiou_breakup_1995}.  To make this asymptotic intuition more formal, we non-dimensionalize, scaling the dimensional variables according to:
\begin{align}
\begin{split}
    &\hat{r}=ar,\quad \hat{s} = R s,\quad \hat{z}=az,\quad \hat{t}=\mathcal{T}t,\quad \hat{w} = \frac{R\Omega_0a^2}{\mathcal{T}\nu}w,\\ & \quad\hat{p} =\frac{\mu R \mathcal{U}}{a^2}p,\quad\hat{\boldsymbol{u}}_s= \frac{a^2 R\Omega_0}{\nu}\boldsymbol{u}_s,\quad\hat{\boldsymbol{\tau}}=\frac{EaR\Omega_0}{\nu}\boldsymbol{\tau},\quad \hat{\Omega}=\Omega_0\Omega(t).
    \label{eq:NonDim}
    \end{split}
\end{align} (`Unhatted' variables are therefore dimensionless counterparts of the corresponding hatted variables.)
Here $\mathcal{T}$ is the timescale of variation of the forcing and $\nu=\mu/\rho$ is the kinematic viscosity of the endolymph. 
The velocity scale $\mathcal{U}=R\Omega_0a^2/(\mathcal T\nu)$ is chosen to balance the viscous forces with the Euler force. The deformation scale is chosen to balance the kinematic condition \eqref{eq:bdary_cond_velocities_match}. The pressure scale is chosen viscously to enforce incompressibility. 
Note that $\Omega(t)$ is the dimensionless angular velocity and $\dot{\Omega}(t)$ is the dimensionless angular acceleration.

\subsubsection{Dimensionless equations}\label{sec:nondimensional_3D_equations}

Following the rescaling of equations \eqref{eq:dimensional_continuity_inertial}--\eqref{eq:bdary_cond_velocities_match}, 
the dimensionless continuity equation \eqref{eq:dimensional_continuity_inertial} reads
\begin{equation}
    \frac{\partial u}{\partial r}+\frac{u}{r}+\frac{1}{r}\frac{\partial v }{\partial \theta}+\frac{1}{h}\frac{\partial w}{\partial s}-\epsilon\frac{v\sin\theta}{h}+\epsilon \frac{u\cos\theta}{h}=0\label{eq:toroidal_continuity_non_dimensional},
\end{equation}
where the scale factor $h$ can be expressed as $h=1 + \epsilon r \cos\theta$. The dimensionless Navier--Stokes equations \eqref{eq:fluid_problem_dimensional} along the $\mathbf{e}_r$, $\mathbf{e}_\theta$ and $\mathbf{e}_s$ directions are given, respectively, by:

\begin{subequations}\label{eq:fluid_problem_dimensionless}
\begin{align}
 &\begin{split}
    \epsilon&\St \frac{\partial u}{\partial t} +\epsilon^3  \mathrm{Re}\left( u\frac{\partial u}{\partial r}+\frac{v}{r}\frac{\partial u}{\partial \theta}+\frac{w}{h}\frac{\partial u}{\partial s}- \frac{v^2}{r}-\frac{1}{\epsilon}\frac{w^2}{h}\cos\theta\right)\\&-2\Omega_0 \mathcal{T}\St \Omega(t)w\cos\theta-\Omega_0 \mathcal{T} \Omega(t)^2 h\cos\theta\\ &= -\frac{1}{\epsilon}\frac{\partial p}{\partial r}+\frac{\epsilon}{rh}\left[\frac{r}{h}\frac{\partial }{\partial s}\left(\epsilon^2\frac{\partial u}{\partial s}-\frac{\partial }{\partial r}(hw)\right)-\frac{\partial }{\partial \theta}\left(\frac{h}{r}\frac{\partial}{\partial r}(r v)-\frac{h}{r}\frac{\partial u}{\partial \theta}\right)\right],\label{eq:ns_toroidal_r_non_dimensional}
    \end{split}\\
 &\begin{split}
    \epsilon&\operatorname{St}\frac{\partial v}{\partial t}+\epsilon^2\operatorname{Re}\left(\epsilon u\frac{\partial v}{\partial r}+\epsilon\frac{v}{r}\frac{\partial v}{\partial \theta}+\epsilon\frac{w}{h}\frac{\partial v}{\partial s}+\epsilon\frac{uv}{r}+\frac{w^2}{h}\sin\theta\right)\\&+2\Omega_0\mathcal{T}\St \Omega(t) w\sin\theta+\Omega_0\mathcal{T}\Omega(t)^2 h\sin\theta \\&=-\frac{1}{\epsilon}\frac{1}{r}\frac{\partial p}{\partial \theta}+\frac{\epsilon}{h}\frac{\partial}{\partial r}\left[\frac{h}{r}\left(\frac{\partial }{\partial r}(r v)-\frac{\partial u}{\partial \theta}\right)\right]-\frac{\epsilon}{rh^2}\left[\frac{\partial}{\partial \theta}\left(h \frac{\partial w}{\partial s}\right)-\epsilon^2r\frac{\partial^2 v}{\partial s^2}\right], \label{eq:ns_toroidal_theta}
    \end{split}\\
    &\begin{split}
    \St&\frac{\partial w}{\partial t}+\dot{\Omega}(t)(1+\epsilon r\cos\theta)+\epsilon 2\St\mathcal{T}\Omega_0\Omega(t)(u\cos\theta-v\sin\theta)\\+&\epsilon^2 \operatorname{Re}\left(u\frac{\partial w}{\partial r}+\frac{v}{r}\frac{\partial w}{\partial \theta}+\frac{w}{h}\frac{\partial w}{\partial s}+\epsilon\frac{uw}{h}\cos\theta-\epsilon\frac{vw}{h}\sin\theta\right) \\ &=-\frac{1}{h}\frac{\partial p}{\partial s}+\frac{1}{r}\frac{\partial}{\partial r}\left[\frac{r}{h}\left(\frac{\partial }{\partial r}(h w)-\epsilon^2\frac{\partial u}{\partial s}\right)\right]+\frac{1}{r^2}\frac{\partial}{\partial \theta}\left[\frac{1}{h}\frac{\partial}{\partial \theta}(hw)-\epsilon^2\frac{r}{h}\frac{\partial v}{\partial s}\right].
    \end{split}\label{eq:ns_toroidal_s}
    \end{align}
\end{subequations}

The dimensionless form of \eqref{eq:navier_dimensional} is given by
\begin{align}
\epsilon\frac{\rho_s}{\rho}\frac{1}{\kappa}\left(
\St\frac{\partial^2\boldsymbol{u}_s}{\partial t^2}+\frac{\partial \mathbf{\Omega}}{\partial t}\times \mathbf{x}+2\St\Omega_0\mathcal{T}\mathbf{\Omega}\times\frac{\partial\boldsymbol{u}_s}{\partial t}+\Omega_0\mathcal{T}\mathbf{\Omega}\times(\mathbf{\Omega}\times\mathbf{x})\right)
=\boldsymbol{\nabla}\cdot\boldsymbol{\tau}. 
    \label{eq:nondimensional_navier}
\end{align}
The only inhomogeneous boundary conditions for the solid problem are \begin{align}\tau_{zz}(r,\theta,z=\pm\beta/2,t)=\pm\Delta p(r,\theta,t)/(2\kappa),\end{align}
with $\beta=t_h/a$ the dimensionless thickness of the cupula.
The non-dimensionalization procedure introduces several dimensionless parameters. We shall see that the most important of these are the stiffness \begin{equation}
    \kappa=\frac{E\mathcal{T}a}{R\mu},
    \label{eqn:KappaDefn}
\end{equation} which measures the time scale of forcing to the time scale of the cupula's relaxation, and the Stokes number
\begin{equation}
    \St= \frac{a^2}{\nu\mathcal{T}},
    \label{eqn:StokesDefn}
\end{equation} which measures the time scale of vorticity diffusion across the channel width, $a^2/\nu$, to the time scale of motion, $\mathcal{T}$. (Note that this version of the Stokes number arises in Stokes' second problem and is sometimes replaced by the Womersley number, $\mathrm{Wm}=\St^{1/2}$.)  We also introduce the Reynolds number of the flow, $\mathrm{Re}=\rho \mathcal{U}R/\mu$, as well as the density ratio,  $\rho_s/\rho$ and the timescale ratio $\Omega_0\mathcal{T}$. To understand the relative size (and hence importance) of these parameters, we next discuss characteristic parameter values and typical sizes of dimensionless parameters.

\subsubsection{Parameter values}

First, we consider the geometrical parameters, taken from \cite{daocai_size_2014}: $R\approx 3.2\times 10^{-3}$ m, and $a\approx 1.6\times 10^{-4}$ m, so that the aspect ratio is $\epsilon\sim 0.05$. The cupula's thickness is usually quoted as $t_h\approx 400\,\mu$m.  

The endolymph composition is very close to water, suggesting that the dynamical parameters are similar to water: $\rho = 1000$ kg$\cdot$m$^{-3}$, and the viscosity is $\mu\approx 10^{-3}$ kg$\cdot$m$^{-1}\cdot$s$^{-1}$. Under standard conditions the cupula is neutrally buoyant, so that the solid density is $\rho_s = 1000$ kg$\cdot$m$^{-3}$. 

The most challenging parameter to identify is $E$. We can infer the value of the bending stiffness from the results of \cite{selva_mechanical_2009}, who suggest an extremely low value of the Young's modulus, $E\sim5$ Pa. Other authors quote simply an estimate for the bending stiffness of the cupula when modelled as a thin plate, $B=Et_h^3/(12(1-\nu_s^2))\sim10^{-10}\mathrm{~N\,m}$ \citep{rabbitt_hydroelastic_1992}, from which a similar Pa-like Young's modulus is inferred.

In terms of the motion, usual ranges of operation for humans are $\dot{\Omega}_0\sim 1$ s$^{-2}$ and $\mathcal{T}$ can be anywhere between 0.01 and 10 seconds. We are now in a position to make informed estimates of the sizes of the dimensionless groups appearing in \eqref{eq:fluid_problem_dimensionless}-\eqref{eq:nondimensional_navier}. Both the Stokes number and the relative stiffness $\kappa$ depend on the forcing time scale $\mathcal{T}$. In particular, $\St\ll1$ for $\mathcal{T}>1$ s, but the Stokes number may be large for faster movements.  Similarly,  $\kappa$ can acquire a large range of values. More usefully, the solid's inertia is generally of size $\epsilon \rho_s \St/(\rho\kappa)=\rho_s^2/(E\mathcal{T}) <10^{-2}$ for all physical values of $\mathcal{T}$ and hence may be safely neglected to leading order in $\epsilon$, with similar statements for all the other inertial terms in \eqref{eq:nondimensional_navier}. We will however include this terms in our computation of the first order correction. Similarly, the reduced Reynolds number, $\epsilon^2\mathrm{Re}\approx 10^{-3}$ and nonlinear advection terms in \eqref{eq:fluid_problem_dimensionless} may be neglected. Finally, the dimensionless thickness of the cupula ranges from $\beta=0.1$ to $\beta=2$.

The size of the dimensionless groups discussed above will inform our choices when neglecting terms in the governing equations \eqref{eq:toroidal_continuity_non_dimensional}-\eqref{eq:fluid_problem_dimensionless}. In particular, we will exploit the smallness of $\epsilon$ to neglect terms of order $\epsilon^2\mathrm{Re}$, and the smallness of $\epsilon \rho_s \St/(\rho\kappa)$ to neglect inertial terms in the Navier equation \eqref{eq:nondimensional_navier}. The last simplification is particularly useful as it will allow us to simplify the leading order solid problem significantly to $\boldsymbol{\nabla}\cdot\boldsymbol{\tau}=\boldsymbol 0$.

We now turn to consider in more detail the behaviour of the model for the typical parameter values already discussed. Given the broad range of values that may be taken by the Stokes number, we begin by considering slow movement (i.e.~large $\mathcal{T}$) in\S\ref{sec:asymptotic_solution}, and thus neglect terms of size $\St$. This will allow us to derive a reduced order equation (an ODE) for the deflection of the cupula that can be compared to numerical results in \S\ref{sec:Numerics}. However, in \S\ref{sec:stokes_large} we consider fast movements with finite Stokes numbers, leading to an integro-differential equation for the deflection of the cupula. 

Both the relative stiffness $\kappa$ and the dimensionless cupular thickness $\beta$ are treated as independent parameters. In particular, recalling that $\kappa$ is a key parameter, capturing the relative timescales of the imposed motion and the cupular relaxation, a main objective of our analysis will be to investigate how variations in $\kappa$ give rise to different solution regimes.

\section{Asymptotic solution for negligible Stokes number}\label{sec:asymptotic_solution}
Motivated by the small value of the Stokes number $\St=a^2/(\nu\mathcal{T})\sim 10^{-2}$ for natural movement timescales, $\mathcal{T}\sim 1$ sec, in this section we assume the Stokes number is negligibly small and solve the coupled system \eqref{eq:fluid_problem_dimensionless} asymptotically, expanding the solution in powers of the small aspect ratio $\epsilon$. 
 To this end we introduce the following formal expansions
\begin{equation}
    \begin{split}
        u(r,\theta,s,t)&=u_0(r,\theta,s,t) + \epsilon u_1(r,\theta,s,t)+\epsilon^2 u_2(r,\theta,s,t)+\cdots\\
        v(r,\theta,s,t)&=v_0(r,\theta,s,t) + \epsilon v_1(r,\theta,s,t)+\epsilon^2 v_2(r,\theta,s,t)+\cdots\\
        w(r,\theta,s,t)&=w_0(r,\theta,s,t) + \epsilon w_1(r,\theta,s,t)+\epsilon^2 w_2(r,\theta,s,t)+\cdots\\
        p(r,\theta,s,t)&=p_0(r,\theta,s,t) + \epsilon p_1(r,\theta,s,t)+\epsilon^2 p_2(r,\theta,s,t)+\cdots\\
        \boldsymbol{\tau}(r,\theta,z,t)&=\boldsymbol{\tau}_0(r,\theta,z,t) + \epsilon \boldsymbol{\tau}_1(r,\theta,z,t)+\epsilon^2 \boldsymbol{\tau}_2(r,\theta,z,t)+\cdots \\
        \boldsymbol{u}_s(r,\theta,z,t)&=\boldsymbol{u}_{s0}(r,\theta,z,t) + \epsilon \boldsymbol{u}_{s1}(r,\theta,z,t)+\epsilon^2 \boldsymbol{u}_{s2}(r,\theta,z,t)+\cdots
    \end{split}\label{eq:expansion}
\end{equation}
Substitution of \eqref{eq:expansion} into the dimensionless problem \eqref{eq:fluid_problem_dimensionless} yields a system of linear equations; we shall retain terms up to and including $\mathcal{O}(\epsilon)$ since they are required to explain the symmetry breaking phenomenon that was observed numerically (see figure~\ref{fig:symmetry_breaking}). 
 \subsection{Expanded solution}\label{sec:leading_order_solution}
Substitution of the formal expansion \eqref{eq:expansion} into the governing equations \eqref{eq:fluid_problem_dimensionless} yields the following leading order balance 
\begin{subequations}
\begin{align}
    \dot{\Omega}(t) &=-\frac{\partial p_0}{\partial s}+\frac{1}{r}\frac{\partial}{\partial r}\left(r\frac{\partial w_0}{\partial r}\right)+\frac{1}{r^2}\frac{\partial^2w_0}{\partial \theta^2},\label{eq:leading_order_system}\\
    &\frac{\partial p_0}{\partial r}=\frac{\partial p_0}{\partial \theta}=0,\\
    &\frac{\partial u_0}{\partial r}+\frac{u_0}{r}+\frac{1}{r}\frac{\partial v_0}{\partial \theta}+\frac{\partial w_0}{\partial s}=0,\\
    &\boldsymbol{\nabla}\cdot\boldsymbol{\tau}_0=\boldsymbol 0,\quad
    \tau_{zz0}(r,\theta,z=\pm\beta/2,t)=\pm\frac{1}{2\kappa}\Delta p_0(t),\label{eq:solid_LO}\end{align}
\end{subequations}
We identify the usual lubrication/boundary layer theory result that the leading order pressure is constant along a cross section \citep{craster_viscous_2006,tavakol_extended_2017}. 
This has the direct consequence that the pressure difference $\Delta p_0(t)=\int_{\epsilon\beta/2}^{2\pi-\epsilon \beta/2}\partial p_0/\partial s\,\upd s$ is only a function of time. Hence, we may seek a solution to the solid problem of the form $\boldsymbol{u}_{s0}(r,\theta,z,t)=\Delta p_0(t)\boldsymbol{\bar{u}}_{s0}(r,z)/\kappa$, where $\boldsymbol{\bar{u}}_{s0}=(\bar{u}_{s0}(r,z),0,\bar{w}_{s0}(r,z))^\intercal$ is an axisymmetric solution to \eqref{eq:nondimensional_navier} satisfying the normalized boundary condition $\bar{\tau}_{zz0}=\pm1/2$ at $z=\pm\beta/2$. We can thus solve the leading order solid problem independently from the fluid problem. 
Moreover, motivated by results from the numerical simulations indicating $w_{s0}(r,z)$ varies slowly over $z$, we introduce the depth-averaged cupular deflection $\eta(r,\theta,t)=\eta_0(r,t)+\epsilon\,\eta_1(r,\theta,t)+\cdots$ defined as 
\begin{align}
    \eta(r,\theta,t)=\frac{1}{\beta}\int_{-\beta/2}^{\beta/2}w_s(r,\theta,z,t)\upd z.
\end{align}
In Appendix \ref{sec:solid_mechanics} we show that when $\St\ll1$ we can decompose $\eta$ multiplicatively as 
\begin{align}
    \eta_0(r,\theta,t)=\frac{\Delta p_0(t)}{\kappa}\bar{\eta}(r,\theta),\quad \bar{\eta}(r,\theta)=\frac{1}{\beta}\int_{-\beta/2}^{\beta/2}\bar{w}_s(r,z,t)\upd z,
\end{align}
so that, to leading order, the cupular deflection and the pressure jump are proportional. We may obtain a polynomial approximation for $\bar{\eta}_0(r)$ using techniques from \cite{barber_elasticity_2010}. The final solution (see Appendix \ref{sec:solid_mechanics} for details on the calculation) is 
\begin{align}\label{eq:eta_0_definition}
    \bar{\eta}_0(r;\beta)=\frac{1}{\beta}\int_{-\beta/2}^{\beta/2} w_{s0}(r,z;\beta)\upd z=\frac{3}{16}\frac{1-\nu_s^2}{\beta^3}(1-r^2)^2+\frac{1}{20}\frac{(1+\nu_s)(12-\nu_s)}{\beta}(1-r^2).
\end{align}
Having obtained a form of solution for the solid problem in terms of the pressure jump, we turn our attention to the fluid problem. We remark that the differential operators acting on the continuity and momentum equations are the same as in cylindrical coordinates \citep{batchelor_introduction_1973}, and invoking symmetry we now seek an axisymmetric solution with $v_0=0$ and leading order terms independent of $\theta$, with $w_0=w_0(r,s,t)$. The leading order kinematic boundary condition may be written as 
\begin{align}\label{eq:LO_kinematic}
    \frac{\partial \eta_0}{\partial t}=w_0(r,s=0,t)=w_0(r,s=2\pi,t),\quad u_0(r,s=0,t)=u_0(r,s=2\pi,t)=0.
\end{align}
Our starting point is the continuity equation, which can be integrated over a cross-section to deduce that the flux $Q = \int_0^{2\pi}\int_0^{a(s)}rw(r,\theta, s, t)\,\upd r\,\upd \theta$ is conserved in the $s$ direction, i.e.~$\partial Q/\partial s=0$. This means that the flux is exclusively a function of time, a fact we will exploit to derive a reduced-order equation. Turning our attention to the $\mathcal{O}(\epsilon)$ problem, the equations read
\begin{subequations}
    \begin{align}
        \dot{\Omega}(t) r\cos\theta &=-\frac{\partial p_1}{\partial s}+ r\cos\theta \frac{\partial p_0}{\partial s}+\frac{1}{r}\frac{\partial}{\partial r}\left(r\frac{\partial w_1}{\partial r}\right)+\frac{1}{r^2}\frac{\partial^2w_1}{\partial \theta^2}+\cos\theta \frac{\partial w_0}{\partial r},\label{eq:first_order_system}\\
        &\frac{\partial p_1}{\partial r}=\Omega_0\mathcal{T}\Omega(t)^2\cos\theta,\quad
    \frac{1}{r}\frac{\partial p_1}{\partial \theta}=-\Omega_0\mathcal{T}\Omega(t)^2\sin\theta,\label{eq:3D_pressure_order_one}\\
    &\frac{\partial u_1}{\partial r}+\frac{u_1}{r}+\frac{1}{r}\frac{\partial v_1}{\partial \theta}+\frac{\partial w_1}{\partial s}=\cos\theta \left(r\frac{\partial w_0}{\partial s}-u_0\right),\\
    &\frac{\rho_s}{\rho}\frac{1}{\kappa}\left(\frac{\partial \boldsymbol{\Omega}}{\partial t}\times\mathbf x+\Omega_0\mathcal{T}\boldsymbol{\Omega}\times(\boldsymbol{\Omega}\times \mathbf{x})\right)=\boldsymbol{\nabla}\cdot\boldsymbol{\tau}_1,\label{eq:solid_first_order}
    \\
    &\tau_{zz1}(r,z=\pm\beta/2,t)=\pm\Delta p_1(r,t)/(2\kappa),
    \\
    & \Delta p_1(r,t)=\Delta p_1^{\text{outer}}(t)+\Delta p_1^{\text{BL}}(r,t).
    \end{align}
\end{subequations}
Here, the pressure difference due to the outer flow is $\Delta p_1^{\text{outer}}(t)=\int_{\epsilon\beta/2}^{2\pi-\epsilon\beta/2}\partial p_1/\partial s\,\upd s$ and $\Delta p_1^{\text{BL}}(r,t)$ is a contribution from the boundary layer that forms near the cupula (see Appendix~\ref{sec:boundary_layer} for details). 
We note that $p_1(r,\theta, s,t)$ may be decomposed into a pressure gradient along the duct axis and an $s$-independent pressure variation due to centrifugal effects, so that
\begin{equation}
    p_1(r,\theta,s,t) = \bar{p}_1(s,t)+\Omega_0\mathcal{T}\Omega(t)^2r\cos\theta,
\end{equation}
and as we only require the $s$ component of the pressure gradient in the computation of the axial velocity, we may safely ignore the $s$-independent component of $p_1$ and use $\bar{p}_1(s,t)$ in its place. Moreover, the first order pressure jump across the cupula is $\Delta p_1^\text{outer}=p_1(s=2\pi,t)-p_1(s=0,t)=\Delta \bar{p}_1^\text{outer}$.

The solutions giving the first two orders of the axial velocity in terms of the pressure gradients may be determined directly: $w_0$ is found from \eqref{eq:leading_order_system}, under the assumption of axisymmetry, while $w_1$  may be found by decomposing it into axisymmetric and asymmetric parts and using standard methods. We find that
\begin{subequations}\label{eq:velocity_profiles_instantaneous}
\begin{align}
    &w_0(r,s,t)=-\frac{1}{4}\left(\dot{\Omega}(t)+\frac{\partial p_0}{\partial s}\right)(a(s)^2-r^2),\label{eq:velocity_LO}\\
    &w_1(r,\theta,s,t)= -\frac{1}{4}\frac{\partial \bar{p}_1}{\partial s}(a(s)^2-r^2)-\frac{1}{16}\left(\dot{\Omega}(t)-3\frac{\partial p_0}{\partial s}\right)r(a(s)^2-r^2)\cos\theta.\label{eq:velocity_FO}
    \end{align}
\end{subequations}
\begin{comment}
\begin{figure}
    \centering
    \includegraphics[width=0.99\linewidth]{Figure_4.pdf}
    \caption{Velocity profiles predicted by \eqref{eq:velocity_profiles_instantaneous} as the torus aspect ratio, $\epsilon$, and leading-order term vary. Note how when $\dot{\Omega}(t)$ and $\kappa\Delta p/(2\pi)$ cancel each other, the asymmetric flow dominates. Moreover, the symmetry breaking becomes observable earlier for larger values of $\epsilon$, though we reiterate that our theory is formally valid only for $\epsilon\ll1$.}
    \label{fig:velocity_profiles}
\end{figure}
\end{comment}
\begin{figure}
    \centering
    \begin{overpic}[width=0.95\linewidth]{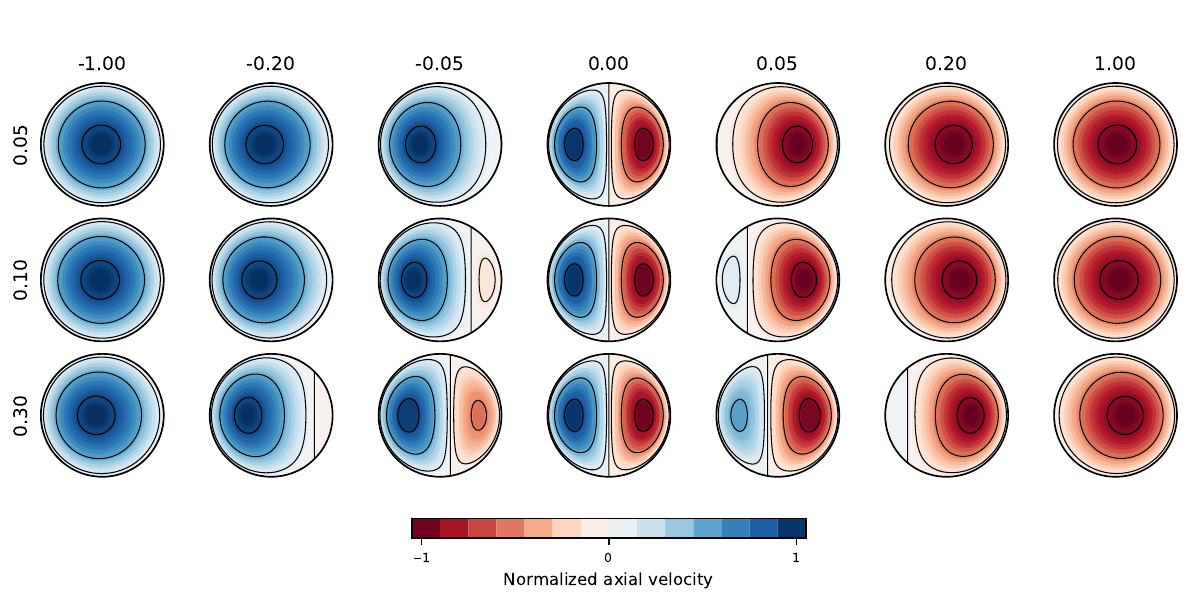}
    %\put(0.5,48.5){\tiny $\epsilon=\frac{a}{R}$}
    \put(44,47){$\dot\Omega(t)+\frac{\Delta p_0}{2\pi}$}
    %\put(41,0.5){\normalem Normalized axial velocity}
    \put(-1,26){\rotatebox{90}{$\epsilon$}}
    \end{overpic}
    \caption{Velocity profiles predicted by \eqref{eq:velocity_profiles_instantaneous} as the torus aspect ratio, $\epsilon$, and leading-order term vary. Note how when $\dot{\Omega}(t)$ and $\kappa\Delta p/(2\pi)$ cancel each other, the asymmetric flow dominates. Moreover, the symmetry breaking becomes observable earlier for larger values of $\epsilon$, though we reiterate that our theory is formally valid only for $\epsilon\ll1$. Note that the horizontal axis may be interpreted as the phase difference between $\dot\Omega$ and $-\Delta p$.}
    \label{fig:velocity_profiles}
    
\end{figure}

We can get a first hint of the symmetry breaking mechanism observed in Figure~\ref{fig:symmetry_breaking} by considering when the asymmetric correction term $\epsilon w_1$ is of a similar size as the symmetric leading order solution $w_0$. Indeed, it is easy to verify that this will be the case when the pressure gradient approximately cancels out the forcing $\dot\Omega(t)$, such that the modulating coefficient $\dot{\Omega}(t)+\frac{\partial p_0}{\partial s}$ in \eqref{eq:velocity_LO} is close to zero. This is visualized in Figure~\ref{fig:velocity_profiles}, where we plot the velocity $w=w_0+\epsilon w_1$ for several values of $\epsilon$ and $\dot{\Omega}(t)+\frac{\partial p_0}{\partial s}$, observing an asymmetric profile when $\dot{\Omega}(t)+\frac{\partial p_0}{\partial s}\ll1$.
We note that the above solutions are only valid far from the cupula, in the slender regions of the SCC ($s\gg\epsilon)$, and thus we refer to \eqref{eq:velocity_profiles_instantaneous} as the outer solution. Closer to the cupula, a boundary layer develops due to the leading order velocity having to adjust from the outer solution $w_0(r,s,t)$ to the cupular deflection profile $\partial \eta_0/\partial t$ as imposed by \eqref{eq:LO_kinematic} (see the curved streamlines close to the cupula in figure~\ref{fig:symmetry_breaking}). The velocity adjustment causes the aforementioned pressure jump $\Delta p_1^\text{BL}(r,t)$, which we compute in Appendix~\ref{sec:boundary_layer} using a matched asymptotic expansion:
\begin{align}\begin{split}
     &\Delta p_1^{\text{BL}}(r,t)=-\frac{\pi }{I_4}\left(\dot{\Omega}(t)+\frac{\Delta p_0(t)}{2\pi}\right)f_3(\beta)\mathrm{Re}\left\{\sum_{n=1}^\infty \bar{A}_nJ_0(\mu_nr)\right\},\\
     &f_3(\beta)=\frac{5(1-\nu_s)}{4\beta^2(12-\nu_s)+10(1-\nu_s)},
     \end{split}
\end{align}
where $\bar{A}_n\in\mathbb{C}$ are constants independent of time and cupular thickness and $\mu_n\in\mathbb{C}$ are the solutions in the first quadrant of $J_1(z)^2=J_0(z)J_2(z)$ \citep[for details see][]{davis_stokes_1990}. From the velocities \eqref{eq:velocity_profiles_instantaneous}, we may compute the flux, noting that the asymmetric components of $w_1(r,\theta,s,t)$ integrate to zero because of the $\cos\theta$ term:
\begin{subequations}
\begin{align}
    Q_0 = -\frac{2\pi}{16}\left(\dot{\Omega}(t)+\frac{\partial p_0}{\partial s}\right)a(s)^4,\label{eq:pressure_grad_flux_LO}\\
    Q_1 = -\frac{2\pi}{16}\frac{\partial \bar{p}_1}{\partial s}a(s)^4.
    \end{align}
\end{subequations}
Since the  flux $Q=Q_0+\epsilon Q_1+...$ is independent of $s$, we can now integrate the above equations along the axis of the duct, obtaining
\begin{subequations}\label{eq:flux_integrated_along_s}
    \begin{align}
        I_4 Q_0 =-\frac{\pi}{8}\left(2\pi \dot{\Omega}(t)+\Delta p_0\right),\\
        I_4 Q_1 = -\frac{\pi}{8}\Delta p_1^\text{outer},
    \end{align}
\end{subequations}
where we have defined $I_4=\int_0^{2\pi}a(s)^{-4}~\upd s$, and used the fact that $\Delta p_1^\text{outer}=\Delta\bar{p}_1^\text{outer}$. We remark that when the thickness of the cupula is appreciable, the integrals should be performed from $s=\epsilon\beta/2$ to $s=2\pi -\epsilon\beta/2$, leading to a slight modification to \eqref{eq:flux_integrated_along_s}, as discussed in Appendix \ref{sec:thickness_correction}.
 
 To connect the flow to the cupula displacement, we evaluate the flux using the velocity $w$ at the cupula, where the fluid velocity must equal the velocity of the cupula: $w=\frac{\partial \eta}{\partial t}$. Therefore, we may write $Q_i=\int_0^{2\pi} \int_0^{a_0}r\frac{\partial \eta_i}{\partial t}\,\upd r\upd \theta$ where $a_0=a(0)$. In Appendix \ref{sec:solid_mechanics} we show that the first order solid problem may be solved as 
 \begin{align}
     \eta_1(r,\theta,t)&=\frac{\Delta p_1^\text{outer}(t)}{\kappa}\bar{\eta}_0(r)+\frac{\rho_s}{\rho\kappa}\dot{\Omega}(t)\bar{\eta}^\text{euler}_1(r)\\
     &+\frac{\rho_s\Omega_0\mathcal{T}}{\rho\kappa}\Omega(t)^2\bar{\eta}^\text{centrif}_1(r)\cos\theta -\frac{\pi}{I_4\kappa}\left(\dot{\Omega}(t)+\frac{\Delta p_0(t)}{2\pi}\right)f_3(\beta)\bar{\eta}^\text{BL}_1(r),
 \end{align}
 where $\bar{\eta}_0$ was defined in \eqref{eq:eta_0_definition} and other $\bar{\eta}_1$ are given in Appendix \ref{sec:solid_mechanics}. Computing the flux we find
 \begin{subequations}
 \begin{align}
     &Q_0=\frac{1}{\kappa}\frac{\upd \Delta p_0}{\upd t}\int_0^{2\pi}\int_0^{a(0)}r\bar{\eta}_0(r;\beta)\,\upd r\,\upd \theta=\frac{2\pi \alpha_0(\beta)}{\kappa}\frac{\upd \Delta p_0}{\upd t},\\
     &Q_1= \frac{2\pi}{\kappa}\left[ 
     \frac{\upd \Delta p_1^\text{outer}}{\upd t}\alpha_0(\beta)+\frac{\rho_s}{\rho}\ddot{\Omega}(t)\alpha_1^\text{euler}(\beta)-\frac{\pi}{I_4}\left(\ddot{\Omega}(t)+\frac{1}{2\pi}\frac{\upd \Delta p_0}{\upd t}\right)f_3(\beta)\alpha_1^\text{BL}(\beta)\right],\end{align}\end{subequations}
where the $\alpha$ factors are
      \begin{align} \alpha_0(\beta)\equiv\int_0^{a(0)}r\bar{\eta}_0(r;\beta)\upd r=\frac{1-\nu_s^2}{32\beta^3}+\frac{(12-\nu_s)(1+\nu_s 
      )}{80\beta},\label{eq:alpha_0_definition}
 \end{align}
and similarly for the other contributions, with explicit forms given in \S\ref{sec:solid_mechanics}. The centrifugal term integrates to zero, and we substitute the flux terms $Q_0$, $Q_1$ into \eqref{eq:flux_integrated_along_s}
from which we obtain a pair of ordinary differential equations for $\Delta p(t)$:
\begin{subequations}
\begin{align}
    &\frac{\alpha_0(\beta)}{\kappa}\frac{\upd \Delta p_0}{\upd t}=-\frac{1}{16 I_4}\left(2\pi \dot{\Omega}(t)+ \Delta p_0\right),\label{eq:3D_equation_for_delta_p}\\
    &\frac{\alpha_0(\beta)}{\kappa}\frac{\upd \Delta p_1^\text{outer}}{\upd t}=-\frac{1}{16 I_4}  \Delta p_1^\text{outer}-\frac{\rho_s}{\rho}\frac{\alpha_1^\text{euler}(\beta)}{\kappa}\ddot{\Omega}(t)+\frac{\pi}{I_4\kappa}\alpha_1^\text{BL}(\beta)f_3(\beta)\frac{\upd }{\upd t}\left(\dot{\Omega}(t)+\frac{\Delta p_0}{2\pi}\right).\label{eq:3D_equation_for_delta_p_correction}
    \end{align}
\end{subequations}
The boundary layer forcing term can be written in terms of $\Delta \ddot{p}_0$ using \eqref{eq:3D_equation_for_delta_p}, but it will be more useful to leave it in terms of $\Delta p_0$. We may solve \eqref{eq:3D_equation_for_delta_p} for any forcing via the integral
\begin{equation}
    \Delta p_0 = -\frac{\pi\kappa}{8\alpha_0(\beta)I_4}\int_0^t\dot{\Omega}(\tau)e^{-\frac{\kappa }{16I_4 \alpha_0(\beta)}(t-\tau)}\upd \tau.\label{eq:3D_pressure_general_solution}
\end{equation}

\subsubsection{Expressions for the velocities}

Once the pressure jump is known from \eqref{eq:3D_equation_for_delta_p}, the pressure gradient may be computed from 
\begin{equation}
    \frac{\partial p_0}{\partial s}=\frac{\Delta p_0}{I_4a(s)^4}+\dot{\Omega}(t)\left[\frac{2\pi}{I_4a(s)^4}-1\right],\quad \frac{\partial \bar{p}_1}{\partial s}=\frac{\Delta p_1^\text{outer}}{I_4 a(s)^4}.
\end{equation}
Substitution into the axial velocity \eqref{eq:velocity_profiles_instantaneous} then yields
\begin{subequations}\label{eq:velocities_from_jump}
\begin{align}
    &w_0 =\frac{\pi}{2I_4a(s)^4}\left[\Omega(t)+\frac{\Delta p_0}{2\pi}\right][r^2-a(s)^2],\\
    &w_1  = \frac{r(r^2-a(s)^2)\cos\theta}{16I_4 a(s)^4}\left[-3\Delta p_0 +(-6\pi+4I_4 a(s)^4)\Omega(t)\right]+\frac{\Delta p_1^\text{outer}}{4I_4a(s)^4}(r^2-a(s)^2).\label{eqn:Jump_w1}
    \end{align}
\end{subequations}
The radial and azimuthal velocities can be recovered from the continuity equation, 
\begin{subequations}
\begin{align}
    u_0 &= \frac{\pi}{2I_4a(s)^5}\left[\Omega(t)+\frac{\Delta p_0}{2\pi}\right]r[r^2-a(s)^2]\frac{\upd a}{\upd s},\\ v_0&=0,\\
    u_1 & = \frac{ 2 \Omega(t) \left[2 I_4 a(s)^4+\pi \right]+  \Delta p_0}{16 I_4 a(s)^5}r^2  \left[r^2-a(s)^2\right] \cos\theta \frac{\upd a}{\upd s}+\frac{\Delta p_1^\text{outer}}{4I_4a(s)^5}r(r^2-a(s)^2)\frac{\upd a}{\upd s},\\
    v_1 & =  -\frac{5  \left[2 \Omega(t) \left(2 I_4 a(s)^4+\pi \right)+  \Delta p_0\right]}{16 I_4 a(s)^5}r^2\left[r^2-a(s)^2\right] \sin\theta\frac{\upd a}{\upd s}.
    \end{align}
\end{subequations}
Note that for a perfect, i.e.~uniform torus, $a'(s)=0$ and there are no radial or azimuthal velocities.

\subsection{Deformation regimes}\label{sec:deformation_regimes}
For a fixed aspect ratio the key remaining dimensionless parameter is the relative stiffness $\kappa$ (which can be varied by changing the bending stiffness or timescale of forcing). We shall see that changing $\kappa$ makes the system sensitive either to the angular acceleration or the angular velocity.
In this section we analyse how these limiting cases arise by solving \eqref{eq:3D_equation_for_delta_p}. Although \eqref{eq:3D_equation_for_delta_p} can be solved analytically for any forcing through \eqref{eq:3D_pressure_general_solution}, qualitative information may be obtained by considering the large and small $\kappa$ limits.

When the cupula is soft ($\kappa\ll1$) the solution to \eqref{eq:3D_equation_for_delta_p} is approximately given by 
\begin{align}
    \Delta p_0 \sim -\frac{\pi\kappa}{8\alpha_0(\beta) I_4}\int_0^t \dot{\Omega}(\tau)\upd \tau = -\frac{\pi\kappa}{8I_4\alpha_0(\beta)}\Omega(t) \quad\text{as}\quad \kappa\to 0.
    \label{eqn:kappaSmall}
    %&\Delta p_1 \sim  \frac{\red{\dot{\Omega}}(t)}{\alpha_0(\beta)}\left(\frac{\pi}{I_4}\alpha_1^\text{BL}(\beta)-\frac{\rho_s}{\rho}\alpha_1^\text{euler}(\beta)\right).
\end{align}
At the other extreme, for a stiff cupula, characterized by $\kappa\gg1$, an approximate solution of \eqref{eq:3D_equation_for_delta_p} is given by 
\begin{align}
    \Delta p_0 \sim -2\pi \dot{\Omega}(t)+\frac{32I_4\pi\alpha_0(\beta)}{\kappa}\ddot{\Omega}\quad\text{as}\quad \kappa\to\infty. 
    \label{eqn:kappaLarge}
\end{align} 

In physical terms, the two results in \eqref{eqn:kappaSmall} and \eqref{eqn:kappaLarge} represent qualitatively distinct regimes for the response of the cupular displacement to the imposed rotation of the canal: for a soft cupula ($\kappa\ll1$), the pressure difference across the cupula, and thus the cupula deformation, is proportional to the angular velocity of the imposed rotation, $\Omega(t)$, while for a stiff cupula ($\kappa\gg1$), the deformation instead follows the angular acceleration $\dot{\Omega(t)}$. Since the cupular deformation is thought to be what is detected by the nerve cells in the cupula, this suggests that the cupula can detect \emph{either} the angular velocity \emph{or} the angular acceleration to which it is subject --- depending on the value of $\kappa$. 

In the latter case, it is easy to verify that the leading order radial and axial velocities suffer a cancellation, as the prefactor of the leading order velocity \eqref{eq:velocity_LO} is $\dot{\Omega}(t)+\frac{\Delta p_0}{2\pi}=\mathcal{O}\left(\frac{1}{\kappa}\right)$, and the asymmetric order $\epsilon$ correction $w_1(r,\theta,s,t)$ dominates for any $\epsilon$ provided $\kappa$ is sufficiently large (the symmetric component of $w_1$ also scales as $\mathcal{O}(\kappa^{-1}))$. This cancellation accounts for  the behaviour observed in Figure~\ref{fig:symmetry_breaking}: as the Young's modulus is increased, making the cupula stiffer to the point that $\kappa\gg1$, the flow ceases to be symmetric. This is intuitive, as a completely rigid cupula does not allow a net flux and hence the axisymmetric leading order flow must vanish.

We emphasize that although symmetry breaking arises from the breakdown of the asymptotic ordering between the first and second terms in the series, this does not imply a loss of asymptotic ordering in the higher-order terms. The symmetry breaking results from a catastrophic cancellation in the leading-order term, while the first correction remains $\mathcal{O}(\epsilon)$, and the subsequent terms are expected to retain their anticipated scaling—indicating that the series remains well-behaved. We confirm this in the next section by numerically solving the full nonlinear problem \eqref{eq:navier_stokes_dimensional}.

To estimate the critical value of $\kappa$ where the transition occurs, we may write $\Delta p_0 =Ae^{2\pi i t}+c.c.$ and $\dot{\Omega}(t)=B e^{2\pi it}+c.c.$, with $A,B\in\mathbb{C}$,and seek the range of $\kappa$ for which $\Delta p$ is in phase with $\Omega$. Direct substitution into \eqref{eq:3D_equation_for_delta_p} yields
\begin{equation}
    A=-\frac{2\pi B}{32 I_4\pi\alpha(\beta)i/\kappa+1}.
\end{equation}
Therefore, the transition change occurs at $\kappa_c = 32 \pi I_4 \alpha(b)$. For $\kappa<\kappa_c$ the phase difference between $A$ and $B$ is more than $\frac{\pi}{4}$, and it will be less than $\frac{\pi}{4}$ for $\kappa>\kappa_c$. When the phase difference is small the response is roughly in phase with the forcing $\Omega(t)$, and vice-versa.

\subsubsection{Sample head rotation}
\begin{figure}
    \centering
    \includegraphics[width=0.99\linewidth]{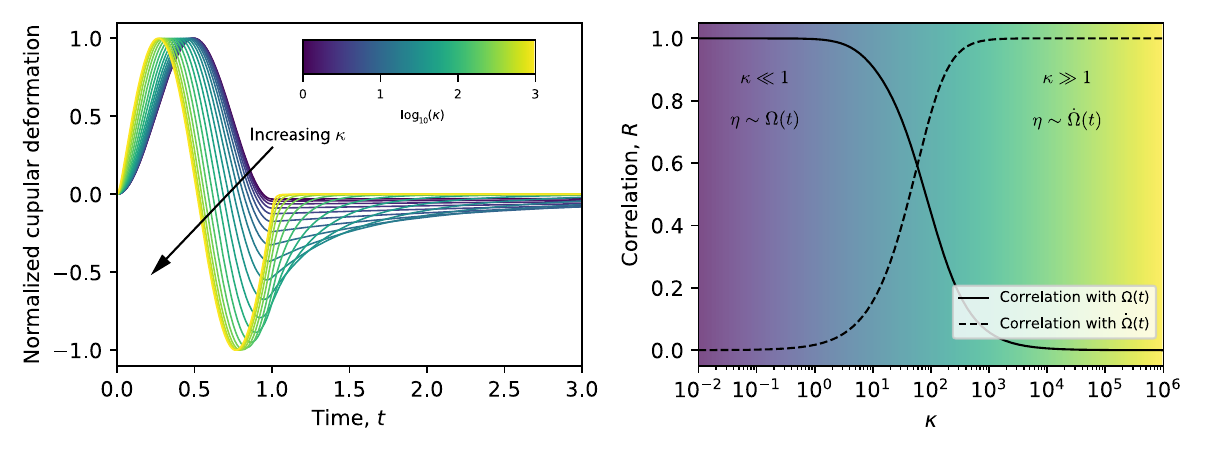}
    \caption{The influence of dimensionless stiffness $\kappa$ on the cupular deformation. (a) As $\kappa$ is increased, the deformation (normalized by the maximum) transitions from following the angular velocity to following the angular acceleration. (b) This transition with $\kappa$ may be shown by plotting the  correlation, $R$, between the deformation and the angular velocity $\Omega(t)$ (solid curve) or the angular acceleration $\dot{\Omega}(t)$ (dashed curve). A transition between the two regimes occurs around $\kappa \approx 100$. In both plots, colour is used to show the value of $\kappa$.}
    \label{fig:two_regimes}
\end{figure}

To visualize the different regimes we have identified above, we solve the equation for the pressure jump \eqref{eq:3D_equation_for_delta_p} with a specific choice for the forcing $\dot\Omega(t)$. While many forms of $\dot\Omega(t)$ could be considered, we are motivated by a clinical head manoeuvre that may be thought of as modelling a slow rotation of the head from right to left. Although other clinical models are described by high-order polynomials \citep{boselli_numerical_2009,boselli_vortical_2013}, we choose a particular form that facilitates an analytical solution of the equations, namely: 
\begin{equation}
    \dot{\Omega}(t)=\begin{cases}
        \sin2\pi t & t\in(0,1)\\
        0 & t>1.
    \end{cases}\label{eq:dimensionless_forcing}
\end{equation}
Solving for the pressure gradient through \eqref{eq:3D_pressure_general_solution} yields
\begin{equation}
    -\Delta p_0 = 2\pi\gamma\kappa\int_0^t\Omega(\tau)e^{-\gamma\kappa(t-\tau)}\upd \tau=\begin{cases}2\pi\gamma\kappa\frac{2 \pi  e^{-\gamma  \kappa t}+\gamma  \kappa  \sin 2 \pi  t-2 \pi  \cos 2 \pi  t}{\gamma ^2 \kappa ^2+4 \pi ^2}& t<1,\\
        \frac{4\pi^2\gamma\kappa}{4\pi^2+\gamma^2\kappa^2}e^{-\gamma\kappa t}(1-e^{-\gamma\kappa})&t>1.
    \end{cases}\label{eq:pressure_solution_example}
\end{equation}
Here, $\gamma = 1/[16\alpha_0(\beta)I_4]$ accounts for domain irregularities and cupular thickness. The cupular deformation, which in this regime is proportional to the pressure jump, is plotted in the left panel of Figure~\ref{fig:two_regimes} for different values of $\kappa$, from which we can clearly see the transition from $\Delta p$ tracking the angular velocity $\Omega(t)$ for small $\kappa$ to tracking the angular acceleration $\dot{\Omega}(t)$ for large $\kappa$, as expected from the preceding analysis. There is an interesting transition region for $\kappa\sim 1$, where we can see an ``overshoot" region at the end of the manoeuvre that has not decayed. This is not the case for either of the limiting regions, where the pressure jump (and cupular displacement) is identically zero after the completion of the head turn.  
In Figure~\ref{fig:two_regimes}(b) we compare how similar the response is to either the angular velocity or the angular acceleration by computing the correlation between the respective functions; for two functions $f(t)$ and $g(t)$, this correlation is defined as
\begin{equation}
    R(f,g)=\frac{\int_0^T f(t)g(t)~\upd t}{\left(\int_0^T f(t)^2~\upd t\cdot\int_0^T g(t)^2~\upd t\right)^{1/2}}.
\end{equation}
As expected from our asymptotic analysis, $\Delta p$ correlates with the angular velocity for small to moderate values of $\kappa$, and the angular acceleration for large values of $\kappa$.  For the parameters used in Figure~\ref{fig:two_regimes}, we compute $\kappa_c=32\pi\approx 100$, in agreement with the transition point observed in the plot.

\section{Numerical simulations\label{sec:Numerics}}

\begin{figure}
    \centering
    \includegraphics[width=0.99\linewidth]{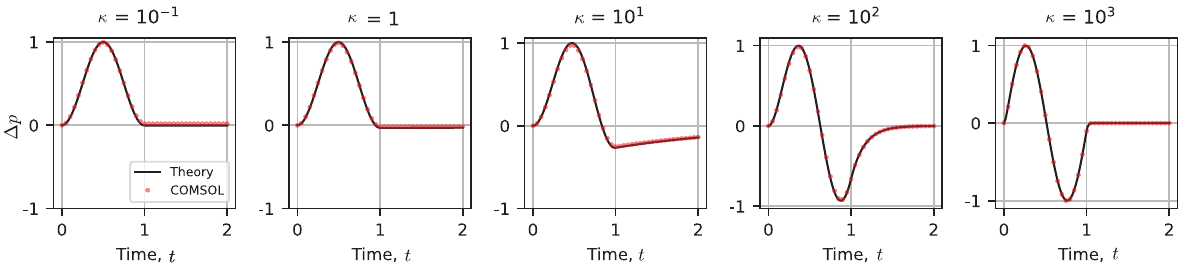}
    \caption{Comparison of {the pressure difference across the cupula as predicted by} COMSOL simulations (markers) and the theoretical prediction from \eqref{eq:pressure_solution_example} (Solid line). As expected from the results in Section~\ref{sec:deformation_regimes}, depending on the value of $\kappa$ the deformation tracks the angular velocity or angular deformation of the forcing (given by \eqref{eq:dimensionless_forcing}). The parameter values used are given in the main text, which correspond to $\St=0.0256$.
    }
    \label{fig:delta_p_comparison}
\end{figure}

\begin{figure}
    \centering
    \includegraphics[width=0.98\linewidth]{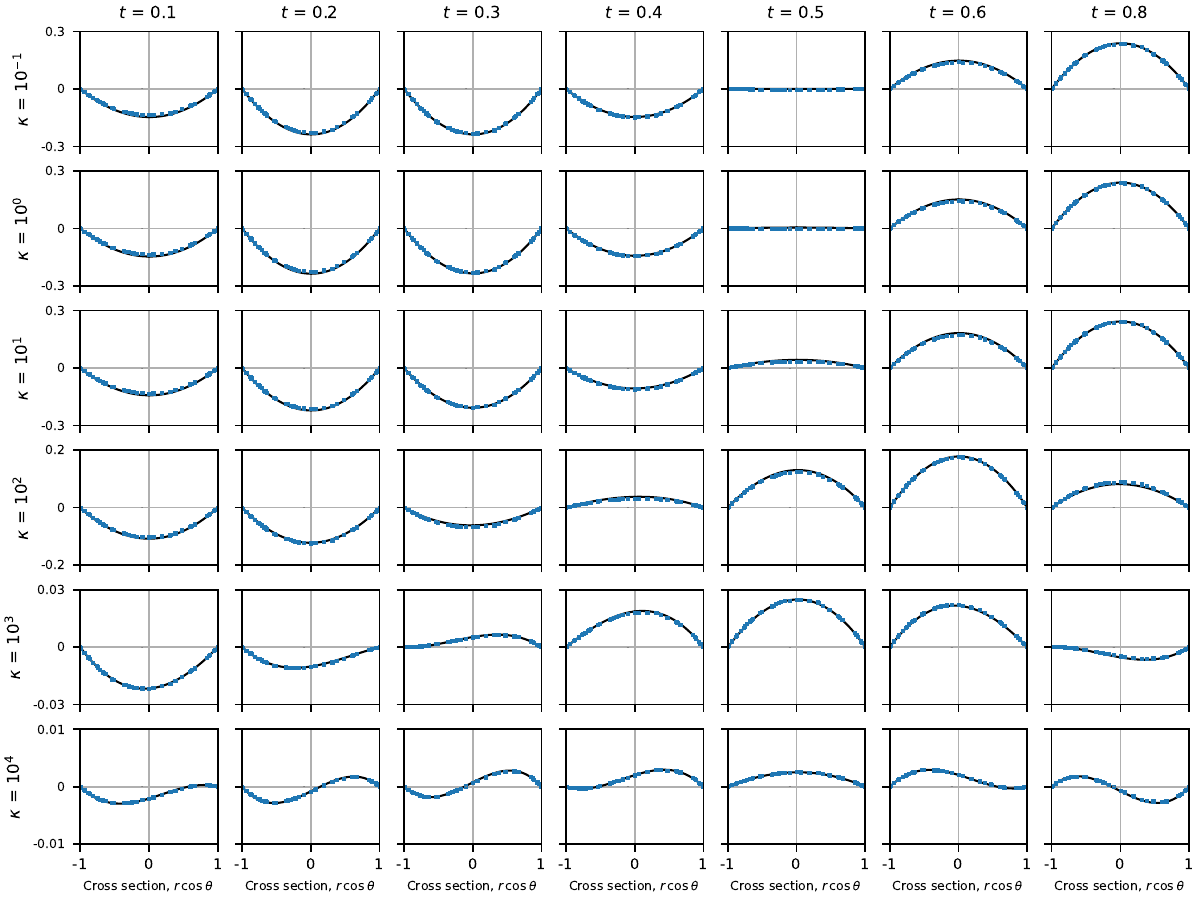}
    \caption{The numerically obtained velocity profiles (blue markers) and theoretical predictions (black solid curves) from \eqref{eq:velocity_LO} and \eqref{eq:velocity_FO}, sampled at $s=\pi$ (the furthest location from the cupula). As $\kappa$ increases, the velocity profile ceases to be symmetric around $\kappa\approx 10^3$. The imposed rotation is given by \eqref{eq:dimensionless_forcing}, with the simulation output sampled at 7 different times. Parameter values are the same as in figure~\ref{fig:delta_p_comparison}.}
    \label{fig:velocity_comparison}
\end{figure}

To test the validity of our asymptotic approach, we return to the numerical simulations in COMSOL as presented in \S\ref{sec:numerical_simulations_surprise} but now imposing within the numerical scheme the forcing given by \eqref{eq:dimensionless_forcing}, and varying the Young's modulus of the solid material  to change $\kappa$. We perform two direct comparisons, appearing in Figures~\ref{fig:delta_p_comparison} and \ref{fig:velocity_comparison}. Figure~\ref{fig:delta_p_comparison} plots the cupular pressure jump $\Delta p$ as a function of time, while Figure~\ref{fig:velocity_comparison} plots the axial velocity profiles across the cross section, sampled at several time points for different values of $\kappa$ in a region of the canal far from the cupula ($s=\pi$). In both figures, we observe excellent agreement between theory and numerics. In particular, the breaking of symmetry in the flow profile for large $\kappa$ is easily observed in Figure~\ref{fig:velocity_comparison}, and the theoretical profile captures the trend and profile shape very well. 

The numerical simulations were performed for $a=1.6\times10^{-4}$ m, $R=3.2\times10^{-3}$ m and $\mathcal{T}=1$ s. As $\kappa$ was varied the Young's modulus of the cupula $E$ was appropriately chosen to match the desired relative stiffness. The fluid is taken to be water ($\mu\approx 10^{-3}$ kg$\cdot$m$^{-1}\cdot$sec$^{-1}$, and $\rho=1000$ kg$\cdot$m$^{-3}$). We consider a uniform semicircular canal with $a(s)=1$.  

We remark that although the symmetry breaking might suggest a breakdown of the asymptotic order, with the second term dominating the first in the series, subsequent terms are well behaved and the series is not divergent. It can be seen from the equations that the $\mathcal{O}(\epsilon^n)$ equations will be of the same form as \eqref{eq:leading_order_system} but with forcing terms depending on the $\mathcal{O}(\epsilon^{m})$ solutions, with $m<n$ and no division by quantities shrinking to zero as $\epsilon\rightarrow0$. This may be inferred from the agreement between the model solution and the numerical solution to the full nonlinear problem, even for large $\kappa$ values where the symmetry is broken. Moreover, as the symmetry breaking occurs because the leading order term shrinks and the correction retains its size (rather than growing), we expect higher order terms to retain their sizes too,  preserving the asymptotic order of the solution.

We have thus far considered slow movements, so that $\mathcal{T}>1\mathrm{~s}$ and $\St\ll1$. However, for faster movements, typically when $\mathcal{T}<1\mathrm{~s}$, the Stokes number is no longer negligible and inertial terms must be retained in the analysis (see \S\ref{sec:scalings}).
However, the Stokes number may also become non-negligible for other reasons --- for example, some authors report slightly thicker semicircular canals ($a$ slightly larger), so that the Stokes number is considerably larger than expected due to its quadratic dependence on radius. To this end, in the next section we consider flows with a finite Stokes number.

\section{Effect of fluid inertia}\label{sec:stokes_large}

While considering the inertialess limit of $\St\ll1$ facilitated analytical progress, there are several circumstances in which inertia may become important, e.g.~faster movements or larger canals. Therefore, we consider the effect of fluid inertia by retaining the $\mathcal{O}(\St)$ terms in the governing equations \eqref{eq:leading_order_system}. Proceeding as in the previous section and seeking a Fourier-Bessel series solution  for the leading order axial velocity of the form
\begin{equation}
    w_0(r,s,t)=\sum_{n=1}^\infty c_n(t,s)\phi_n(r,s), \quad \phi_n(r,s)=J_0\left(\frac{\lambda_n}{a(s)}r\right).\label{eq:w_ansatz}
\end{equation}
Here, $\phi_n$ are the eigenfunctions for the Laplacian in a cylinder of local radius $a(s)$, subject to Dirichlet boundary conditions  \citep{batchelor_introduction_1973}; the $\lambda_n$ are the zeros of the Bessel function of the first kind $J_0(z)$, thereby ensuring that the axial velocity satisfies the no-slip condition at $r=a(s)$ in the rotating frame. 
Thus, substituting \eqref{eq:w_ansatz} into the momentum equation \eqref{eq:ns_toroidal_s}, keeping only the leading order terms, and using the orthogonality properties of Bessel functions we find
\begin{subequations}
\begin{align}
\frac{1}{r}\frac{\partial}{\partial r}&\left(r\frac{\partial \phi_n}{\partial r}\right)=-\frac{\lambda_n^2}{a(s)^2}\phi_n\\
    \int_0^{a(s)}r\phi_m(r,s)\phi_n(r,s)\upd r&=\frac{a(s)^2}{2}\delta_{mn}J_1(\lambda_n)^2,\quad \int_0^{a(s)}r\phi_n(r,s)\upd r=\frac{a(s)^2J_1(\lambda_n)}{\lambda_n},\\
    \St\frac{\partial c_n}{\partial t}&= -\frac{2}{\lambda_n J_1(\lambda_n)}\left(\dot{\Omega}(t)+\frac{\partial p_0}{\partial s}\right) -\frac{\lambda_n^2}{a(s)^2}c_n,\\ c_n(t,s)&=-\frac{2}{\lambda_nJ_1(\lambda_n)}\int_0^t\left(\dot{\Omega}(\tau)+\frac{\partial p_0}{\partial s}\right) \mathcal{K}_n(t-\tau,s;\St)~\upd\tau, \label{eqn:cn_St}
    \end{align}
\end{subequations}
where $\mathcal{K}_n(x,s;\St)=\St^{-1}e^{-\lambda_n^2x/[a(s)^2\St]}$.  The flux may now be computed as
\begin{equation}
    \begin{split}
        Q_0&=2\pi\int_0^{a(s)}r w_0(r,s,t)~\upd r=2\pi a(s)^2\int_0^1 \rho w_0(a(s)\rho,s,t)~\upd\rho \\
        &=-4\pi a(s)^2 \sum_{n=1}^\infty \left[\lambda_n^{-2}\int_0^t\left(\dot{\Omega}(t)+\frac{\partial p_0}{\partial s}\right)\mathcal{K}_n(t-\tau,s;\St)~\upd\tau\right].\label{eq:flux_series}
    \end{split}
\end{equation}
From the continuity equation, the flux $Q_0$ is independent of $s$. Therefore, we can evaluate it at the location of the cupula, where the fluid velocity is known to be equal to the cupular velocity $\frac{\partial \eta_0}{\partial t}$; this gives $Q_0=2\pi \int_0^{a(0)} r\frac{\partial \eta_0}{\partial t}~\upd r$ and allows us to write a reduced system of equations for $\frac{\partial p_0}{\partial s}(s,t)$ and $\eta_0(r,t)$, namely
\begin{subequations}
    \begin{align}
        & \int_0^{a(0)} r\frac{\partial \eta_0}{\partial t}\upd r=-2 a^2\frac{1}{\operatorname{St}} \sum_{n=1}^\infty \left[\frac{1}{\lambda_n^2}\int_0^t\left(\dot{\Omega}(t)+ \frac{\partial p_0}{\partial s}\right)e^{-\lambda_n^2(\tau-t)/(a^2\operatorname{St})}~\upd \tau\right], \label{eq:flux_kernels}\\
        & \eta_0(r,t)=\Delta p_0\frac{\bar{\eta}_0(r)}{\kappa},\quad \Delta p_0=\int_0^{2\pi}\frac{\partial p_0}{\partial s}~\upd s.\label{eq:eta_0_pressure_inertia}
    \end{align} \label{eq:reduced_full_system}
\end{subequations}
Therefore, we may write this as a single equation for the pressure gradient
\begin{align}\label{eq:delta_p_equation_inertia}
    \frac{\alpha_0(\beta)}{\kappa}\frac{\upd \Delta p_0}{\upd t}=-2 a^2\frac{1}{\operatorname{St}} \sum_{n=1}^\infty \left[\frac{1}{\lambda_n^2}\int_0^t\left(\dot{\Omega}(t)+ \frac{\partial p_0}{\partial s}\right)e^{-\lambda_n^2(\tau-t)/(a^2\operatorname{St})}~\upd \tau\right],
\end{align}
where $\alpha_0(\beta)$ is given in \eqref{eq:alpha_0_definition}.
For an arbitrary inner radius $a(s)$, \eqref{eq:delta_p_equation_inertia} can be solved using the Laplace transform, as shown in Appendix~\ref{sec:laplace_transform}. Before tackling the general case, we focus on the simple case where the tube radius is uniform, $a(s)\equiv1$.

\subsection{A simple example}\label{sec:inertia_simple_example}
For the special case when $a(s)\equiv1$, i.e.~the tube radius is constant, the pressure gradient can be assumed to be independent of $s$, that is $\frac{\partial p_0}{\partial s}=\frac{\Delta p_0}{2\pi}$, and \eqref{eq:flux_kernels} simplifies to
\begin{equation}
  \frac{\alpha_0(\beta)}{\kappa}\frac{\upd \Delta p_0}{\upd t}=-2  \sum_{n=1}^\infty \left[\lambda_n^{-2}\int_0^t\left(\dot{\Omega}(\tau)+ \frac{\Delta p_0}{2\pi}\right)\mathcal{K}_n(t-\tau;\St)~\upd\tau\right]. \label{eq:appendix_flux_series1}
\end{equation} 
To transform this equation into a more manageable form, we define a complete kernel $\mathcal{K}(x;\St)=\St^{-1}\sum_{n=1}^\infty\lambda_n^{-2}e^{-\lambda_n^2x/\St}$. 
\begin{align}\label{eq:3D_integral_equation_Stokes}
    \frac{\alpha_0(\beta)}{\kappa}\frac{\upd \Delta p_0}{\upd t}=-2\bar{\dot{\Omega}}(t)-\frac{1}{\pi}\int_0^t\Delta p_0(\tau)\mathcal{K}(t-\tau;\St)\upd \tau.
\end{align}
where we have introduced $\bar{\dot{\Omega}}(t)=\int_0^t\dot{\Omega}(\tau)\mathcal{K}(t-\tau;\St)~\upd \tau$.
Equation \eqref{eq:3D_integral_equation_Stokes} may be efficiently solved numerically by truncating the infinite series in the kernel and transforming the integral equation into a system of ODEs. This is a standard calculation, with details given in Appendix~\ref{sec:app_integro_ODE}.

\subsection{Fluid inertia can make the cupula underdamped}
 To understand the effect of inertia in the cupular response, we first consider the case of a cupula that is initially stretched by a pressure jump $\Delta p_0(t=0)=1$ in a frame rotating at constant speed, so that $\Omega(t)=0$. 
 
\begin{figure}
    \centering
    {\scriptsize
    \begin{overpic}[width=0.99\textwidth, tics=10]{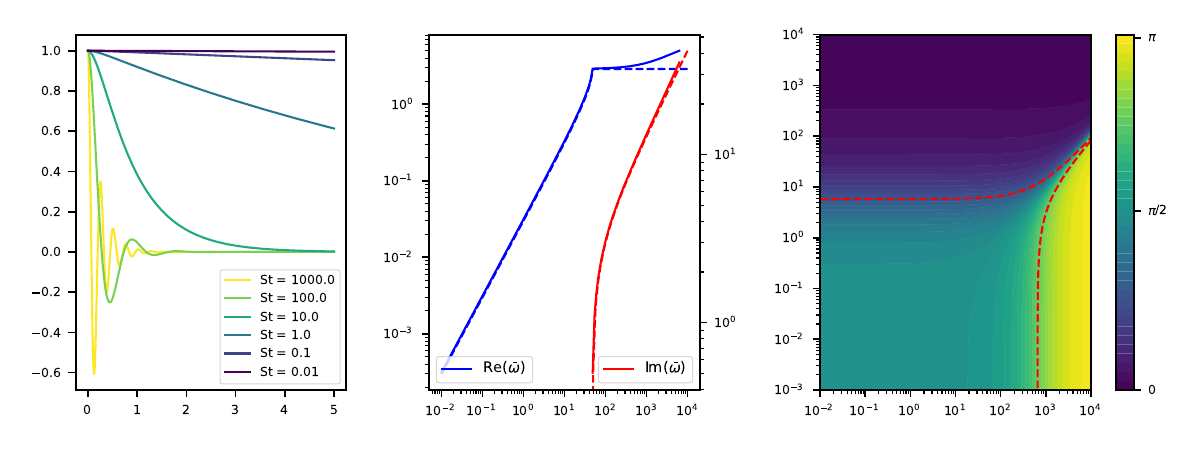}
    \put(16,0.5){$t/\St$}
    \put(0.5,17){\rotatebox{90}{$\Delta p_0(t)$}}
    \put(9,36){Relaxation of the cupula}
    \put(30.25,17){\rotatebox{90}{$\mathrm{Re}(\bar{\omega})$}}
    \put(40,0.5){$\kappa\St/(\pi\alpha)=1/q$}
    \put(60,23){\rotatebox{270}{$\mathrm{Im}(\bar{\omega})$}}
    \put(38,36){Underdamping for high $\St$}
    \put(80,0.5){$\kappa$}
    \put(63.5,19){\rotatebox{90}{$\St$}}
    \put(97.5,17){\rotatebox{90}{$\arg(\chi)$}}
    \put(69,36){Input-output phase difference}
    \end{overpic}
    }
    \caption{Left panel: Solution to \eqref{eq:3D_integral_equation_Stokes}, when $\dot\Omega(t)=0$ and $\Delta p_0 (t=0)=1$ for different values of the Stokes number, showing underdamped dynamics for large enough $\mathrm{St}$. Central panel: Bifurcation diagram, showing the evolution of $\mathrm{Re}(\bar{\omega})$ (blue) and $\mathrm{Im}(\bar{\omega})$ (red). Markers represent the numerically obtained solution from \eqref{eq:Stokes_growth_rate} and dashed lines the analytical approximation \eqref{eq:stokes_growth_approx}. 
    Right panel: bifurcation diagram for $\dot\Omega(t)\sim e^{it}$, showing how the transition between the three regimes depends on both $\kappa$ and $\mathrm{St}$. 
    Colour represents the complex angle of $\chi$, with purple representing $\arg(\chi)=0$, green is $\arg(\chi)=\pi/2$ and yellow is $\arg(\chi)=\pi$.}
    \label{fig:fluid_inertia}
\end{figure}

 The numerical solution to \eqref{eq:3D_integral_equation_Stokes} for a range of Stokes numbers is given in Figure~\ref{fig:fluid_inertia}(a). Notice that for sufficiently large $\mathrm{St}$, $\Delta p(t)$ exhibits decaying oscillatory behaviour, meaning that the cupula is underdamped; this is in contrast to smaller values of $\mathrm{St}$, in which the cupula dynamics show an exponential decay whose rate of decay increases with $\mathrm{St}$.
 \begin{comment}
\begin{figure}
    \centering
    \includegraphics[width=0.99\linewidth]{Figure_8.pdf}
    \caption{Left panel: Solution to \eqref{eq:3D_integral_equation_Stokes}, when $\dot\Omega(t)=0$ and $\Delta p_0 (t=0)=1$ for different values of the Stokes number, showing underdamped dynamics for large enough $\mathrm{St}$. Central panel: Bifurcation diagram, showing the evolution of $\mathrm{Re}(\bar{\omega})$ (blue) and $\mathrm{Im}(\bar{\omega})$ (red). Markers represent the numerically obtained solution from \eqref{eq:Stokes_growth_rate} and dashed lines the analytical approximation \eqref{eq:stokes_growth_approx}. 
    Right panel: bifurcation diagram for $\dot\Omega(t)\sim e^{2\pi it}$, showing how the transition between the two regimes $\Delta p_0\sim \dot\Omega(t)$ and $\Delta p_0\sim \dot\Omega(t)$ depends on both $\kappa$ and $\mathrm{St}$. 
    Colour represents the complex angle of $\chi$, with blue $\arg(\chi)=\pi/2$ and red representing $\arg(\chi)=0$.}
    
    \label{fig:fluid_inertia}
\end{figure}
\end{comment}
To understand this change in behaviour as $\mathrm{St}$ is increased, we seek an exponential ansatz to solve \eqref{eq:3D_integral_equation_Stokes}, with $\Delta p_0 \sim e^{-\omega t}$ in the simple case when $\dot\Omega(t)=0$. Direct substitution yields the following condition: 
\begin{equation}
    \frac{\alpha_0(\beta) \pi }{\kappa \mathrm{St}}\bar{\omega}=\sum_{n=1}^\infty \frac{1}{\lambda_n^2}\frac{1}{\lambda_n^2-\bar{\omega}},\label{eq:Stokes_growth_rate}
\end{equation}
where the rescaled growth rate is $\bar{\omega}=\mathrm{St}\,\omega$. Equation \eqref{eq:Stokes_growth_rate} provides an equation for $\bar{\omega}$ depending on the parameter $q=\alpha_0(\beta)\pi/(\kappa \mathrm{St})=\alpha_0(\beta) \pi\rho\nu^2R/(Ea^3)$, which is independent of the forcing timescale $\mathcal{T}$. Analytical progress can be made by truncating the sum at $n=1$, i.e.~considering only the first term. This leads to 
\begin{equation}
    \bar{\omega}=\frac{\lambda_1^2}{2}\left[1\pm \left(1-\frac{4}{q\lambda_1^6}\right)^{1/2}\right].\label{eq:stokes_growth_approx}
\end{equation}
From this, we can identify the critical value for the parameter $q$ at which the transition to underdamped dynamics occurs: $q_c = 4/(\lambda_1)^6\approx0.0207$. 

A natural question to ask now is if fluid inertia can alter the development of the two flow regimes outlined previously in Section~\ref{sec:deformation_regimes}? Proceeding as before, we assume a forcing $\dot{\Omega}(t)=B e^{ i t}$, and try an ansatz $\Delta p_0 =  Ae^{ it}$. Substitution into the integral equation \eqref{eq:3D_integral_equation_Stokes} and neglecting contributions from the initial conditions yields 
\begin{equation}
    \frac{i\alpha_0(\beta) A}{\kappa} = -\sum_{n=1}^\infty\frac{2B+A/\pi}{\lambda_n^2(i\mathrm{St}+\lambda_n^2)}
\end{equation}
It is convenient to define the function $\mathcal{G}:\mathbb{R}\rightarrow\mathbb{C} $, given by $\mathcal{G}(\mathrm{St})=\sum_{n=1}^\infty\lambda_n^{-2}/( i\mathrm{St}+\lambda_n^2)$. We find that the response (characterized by $A$) is related to the forcing (characterized by $B$) through
\begin{equation}
    A=\frac{-2B \mathcal{G}(\mathrm{St})}{ i\alpha_0(\beta)/\kappa +\mathcal{G}(\mathrm{St})/   \pi}.
\end{equation}
Therefore, we see that the angle of the complex quantity 
\begin{equation}
\chi=\frac{-\mathcal{G}(\mathrm{St})}{\pi i\alpha_0(\beta) +\kappa \mathcal{G}(\mathrm{St})}=\frac{-1}{\kappa+\pi i\alpha_0(\beta)/\mathcal{G}(\St)}\end{equation}
will determine if the deformation follows the angular velocity (if the angle is close to $\pi/2$) or the angular acceleration (when the angle is close to $0$ or multiples of $\pi$). To achieve analytical progress we truncate the sum in $\mathcal{G}$, keeping only the first term, and we find
\begin{eqnarray}
    \frac{\mathrm{Re}(\chi)}{\mathrm{Im}(\chi)}=-\frac{\kappa-\pi\alpha_0(\beta)\lambda_1^2\St}{\pi\lambda_1^4\alpha_0(\beta)}.
\end{eqnarray}
Therefore, the curve in the parameter space $(\kappa, \mathrm{St})$ separating the two regimes is given implicitly by
\begin{eqnarray}
    \left\vert \frac{\kappa}{\pi\lambda_1^4\alpha_0(\beta)}-\frac{\St}{\lambda_1^2}\right\vert=1.\label{eq:inertia_two_regimes_curves}
\end{eqnarray}
In figure~\ref{fig:fluid_inertia}(c), we plot the argument of $\chi$ as a function of $\kappa$ and $\mathrm{St}$, indicating as well the approximate bifurcation curves given by \eqref{eq:inertia_two_regimes_curves}. 
This diagram indicates where in the $\kappa$-$\mathrm{St}$ phase space the cupula deflection follows the angular acceleration, angular velocity or the angular displacement. 
For small values of the Stokes number $\mathrm{St}$ we recover the previous picture, where $\kappa\ll1$ indicates the deformation follows the angular velocity ($\mathrm{arg}(\chi)=\pi/2$) and $\kappa\gg1$ indicates the response is guided by the angular acceleration ($\mathrm{arg}(\chi)=\pi$). 
But we also find a dependence on $\mathrm{St}$ for non-small Stokes numbers. For given $\kappa$ less than about 100, a transition to angular displacement tracking occurs for $\mathrm{St}$ greater than about 1, meaning that the cupula system may be tuned to follow angular displacement even for small cupula stiffness if the Stokes number is high enough (characterized by $\mathrm{arg}(\chi)=0$ in figure~\ref{fig:fluid_inertia}(c)). 
This transition point increases for larger $\kappa$, as indicated by the blue wedge region in figure \ref{fig:fluid_inertia}(c), meaning that an orders of magnitude higher $\kappa$ is possible with cupular deflection still tracking angular velocity, if the Stokes number is accordingly increased in a very particular way. In the new regime for $\St\gg1$ (and moderate $\kappa$) we find the pressure jump is approximately:
\begin{align}
    \Delta p_0(t)=-\frac{\kappa}{2\alpha_0(\beta)\St}\int_0^t\Omega(\tau)\upd \tau,
\end{align}
and indeed proportional to the angular displacement.

We may interpret the effect of high fluid inertia by considering the response of the cupula as the forcing frequency is increased. For small forcing frequencies the deformation will follow the angular acceleration, and as the forcing frequency is increased the cupula will start deforming in phase with the angular velocity, as expected from \S\ref{sec:deformation_regimes}; this is well known in the vestibular literature \citep{benson_sensory_1990}. However, our results from this section suggest that when the forcing frequency is further increased (in humans, to about 100 Hz), the cupular deformation will be in phase with angular displacement. This sort of high frequency motion could be expected for example during impacts or equipment malfunction.

\subsection{Non-uniform channel widths}
As noted in the introduction, an advantage of our theoretical formulation is that it is also compatible with a non-uniform and arbitrary channel width, described by the function $a(s)$, so long as the small aspect ratio between channel width and length is maintained. This case is more delicate than the one we saw in the last subsection, as the pressure gradient is no longer constant but depends on $a(s)$, and must be integrated along the channel to obtain the pressure jump across the cupula. In Appendix~\ref{sec:laplace_transform} a method based on the Laplace transform and the convolution theorem is developed, through which we obtain an approach to solving the problem for both non-uniform channel widths and $\St>0$. The main result is that we obtain the same form of equation \eqref{eq:3D_integral_equation_Stokes} for the deflection $\eta_0(r,t)$, with the kernel $\mathcal{K}(x;\St)$ given by the (temporal) inverse Laplace transform of
\begin{equation}
    \tilde{\mathcal{K}}(\sigma;\St)=2\pi\left(\int_0^{2\pi}\frac{\upd s}{a(s)^4\sum_{n=1}^\infty\lambda_n^{-2}\left[a(s)^2\St\sigma +\lambda_n^2\right]^{-1}}\right)^{-1}.
\end{equation}
For a given canal profile $a(s)$, the Kernel may be numerically obtained by fixing a discretization of $\sigma$ into a finite number of points. For each point, the integral can be populated for $a(s)$ and computed using standard quadrature methods. This will yield $\tilde{\mathcal{K}}(\sigma;\St)$ for a finite number of $\sigma$. The equation for the pressure jump \eqref{eq:3D_integral_equation_Stokes}, can then be either solved in Laplace space, inverting the transformed solution using an efficient algorithm \citep{kuhlman_review_2013}, or in real space using a trapezoidal method.

We have developed a general framework that allows us to solve for the cupular displacement and pressure jump for complicated canal geometries allowing also for the possibility of fluid inertia. As an example of the scenarios in which this approach might be useful, we now reconsider some of the numerical results presented in figure~\ref{fig:velocity_profiles}.

\subsection{Fluid inertia explains discrepancies between numerics and model}

\begin{figure}
    \centering
    \includegraphics[width=0.95\linewidth]{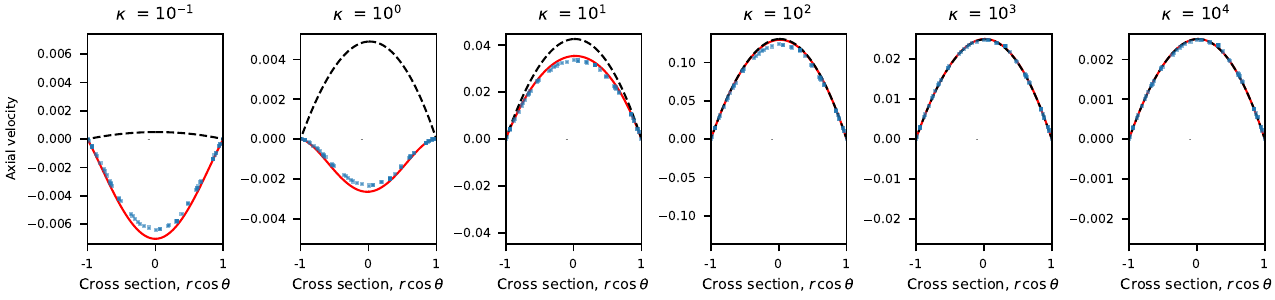}
    \caption{Velocity profiles for $t=0.5$, including the finite fluid inertia correction. Dashed black line: $\St =0$, blue markers: numerics, solid red line: $\St>0$ model. The parameter values are the same as in figure~\ref{fig:velocity_comparison}.}
    \label{fig:Stokes_fixes_numerics}
\end{figure}

Figure~\ref{fig:velocity_profiles} generally shows excellent agreement between the $\St=0$ model presented in \S\ref{sec:asymptotic_solution} and our COMSOL numerics, especially at the scale of the largest velocities, which were used for comparison in figure \ref{fig:velocity_profiles}.  However, if we zoom in to situations where the velocity is small, such as when $t=0.5$, then we might expect to observe differences caused by small errors in the phase of the motion.  Figure~\ref{fig:Stokes_fixes_numerics} shows just such an effect: the agreement between the predictions of the $\St=0$ asymptotics (dashed black curves) and COMSOL simulations (blue markers) is no longer satisfactory for small and moderate values of the stiffness $\kappa\lesssim1$: while the absolute error \emph{is} small, the relative error is very large. This is because even for small Stokes numbers, the exact time at which the velocity vanishes is different from that predicted by the $\St=0$ asymptotics of \S\ref{sec:asymptotic_solution}, which essentially assume that the motion is quasi-steady.

Figure \ref{fig:Stokes_fixes_numerics} also shows the predictions of the leading-order in $\St$ asymptotic results presented in this section. As might be expected, we see much better agreement between the prediction accounting for $\St>0$ via \eqref{eqn:cn_St} and \eqref{eq:3D_integral_equation_Stokes} (shown by solid red curves in figure \ref{fig:Stokes_fixes_numerics}) and the results of COMSOL simulations (points) than with the earlier result, which neglected the effects of fluid inertia entirely. We emphasize that this finite inertia case requires the numerical calculation of the integrals in \eqref{eqn:cn_St} and \eqref{eq:3D_integral_equation_Stokes}. 

We now consider if domain irregularities can give rise to interesting flow phenomena, in particular unexpected symmetry breaking and vortical flows when $\kappa\ll1$, which have been reported previously \citep{boselli_vortical_2013}.

%\newpage
\section{Analytical description of vortical flow}
As a final point of analysis, we turn our attention to the possibility of vortical flow:
several authors have reported the existence of vortical structures in computational studies of flow in the utricle, the enlarged portion of the semicircular canal \cite[see][for example]{boselli_numerical_2009,goyens_asymmetric_2019}. Vortices appear to occur even in the analogue of our limit $\kappa\ll1$, when the flow in the slender portion of the canals is \emph{largely} symmetric. In this section we use our asymptotic analysis to give an analytical description of such vortical flow structures and determine the geometrical conditions required for their emergence. 

Mathematically, the reason why we might expect enlarged regions of the canal to experience symmetry breaking may be seen from the form of the leading order and $O(\epsilon)$ axial velocity terms in \eqref{eq:velocities_from_jump}. In particular, in regions where $a(s)>1$, the magnitude of the leading order velocity $w_0$ is proportional to $1/[I_4\,a(s)^4]$, while the correction is of order $\epsilon$ the $\dot{\Omega}(t)$ term in \eqref{eqn:Jump_w1} remains $O(\epsilon)$ even as $a(s)^4$ becomes large. (We retain the $I_4$ term here, since it is also dependent on $a(s)$.) Hence, we might expect noticeable asymmetry in the flow to develop in regions where $1/[I_4 a(s)^4]$ is comparable to $\epsilon$. Note that this can be achieved in this way without requiring that we are in the stiff cupula regime $\kappa\gg1$ that led to the symmetry-breaking discussed in Section~\ref{sec:deformation_regimes}.

\begin{figure}
    \centering
    \includegraphics[width=0.99\linewidth]{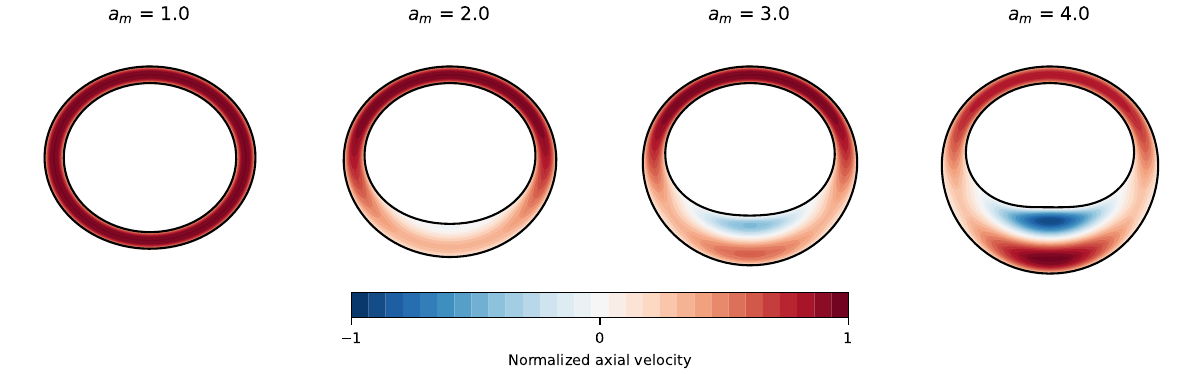}
    \caption{Analytical reconstruction of vortical flow in the utricle as the maximum channel radius, $a_m$, increases. The channel has a largely uniform radius but is wider in the vicinity of the utricle --- see \eqref{eqn:UtricleProfile} for the detailed profile of the tube. Here, we observe how as the size of the utricle is augmented the vortex develops. The forcing is given by \eqref{eq:dimensionless_forcing} and we show the solution at time $t=0.25$. The parameters used are $\epsilon=0.05$, and $\kappa=0.1$. Furthermore we use the solution from \S\ref{sec:asymptotic_solution} that assumes the fluid inertia is vanishingly small, $\St=0$.} 
    \label{fig:vortical_flow}
\end{figure}

To demonstrate this possibility, we consider the predictions of our asymptotic theory for a canal with a localized bulge by taking the radius $a(s)$ to be the sum of a constant and a Gaussian: 
\begin{equation}
    a(s)=1+(a_m-1)e^{-\gamma s^2}. 
    \label{eqn:UtricleProfile}
\end{equation}
Here, the cupula and utricle are located at $s=0$, $\gamma$ is a parameter controlling the width of the enlargement and $a_m$ is the maximum inner radius of the tube. This choice is motivated by the qualitative agreement with the imaging from \cite{daocai_size_2014}. In Figure~\ref{fig:vortical_flow} we plot the velocity distribution for several channel geometries, in particular a top view of the mid plane of the flow around the canal, with colour indicating the magnitude of the axial velocity. 
The size of the enlarged region is increased from left to right (with $\gamma=1$ and $a_m =1,2,3,4$), and the appearance of the vortical flow is clear. The forcing is given by \eqref{eq:dimensionless_forcing} and we visualize the solution at time $t=0.25$. Here we have used a small value of $\kappa=0.1$ so that the flow in the slender regions is predicted to remain largely symmetric, a feature that we will verify below. 

\subsection{Conditions required for the formation of the utricular vortex}
As noted above, symmetry breaking in the utricle occurs when the first order correction $\epsilon w_1$ is comparable with the leading order velocity $w_0$.  Within a cross section, the maximum value of $w_0$ is attained at $r=0$, and is given by
\begin{equation}
    \vert w_0^*(s,t)\vert = \frac{\pi}{2I_4a(s)^2}\left\vert \dot{\Omega}(t)+\frac{\Delta p_0}{2\pi}\right\vert\sim \frac{\pi}{2I_4a(s)^2}\vert \dot{\Omega}(t)\vert,
\end{equation}
where the last approximation follows when $\kappa\ll1$. The maximum value of $w_1$ is attained at $r=\frac{a(s)}{\sqrt{3}}$ and $\theta =0,\pi$, and is given by 
\begin{equation}
    \vert w_1^*\vert = \frac{1}{24\sqrt{3}I_4 a(s)}\left\vert-3\Delta p_0+\left(-6\pi +4 I_4a(s)^4 \right)\dot{\Omega}(t)\right\vert.\label{eq:size_of_w0}
\end{equation}
The symmetric component of $w_1$ scales as $\epsilon \Delta p_1^{\text{outer}} / a(s)^2$, and since it is much smaller than both the asymmetric component of $w_1$ and the leading-order flow $w_0$, it is neglected in the present analysis (we set $\Delta p_1^\text{outer}=0$).
Focusing on the utricle, where $a(s)$ is largest, $\vert w_1^*\vert$ is dominated by 
\begin{equation}
  \vert w_1^*\vert\sim  \frac{a_m^3}{6\sqrt{3}}\left\vert  \dot{\Omega}(t)\right\vert.\label{eq:size_of_w1}
\end{equation}
Comparing $w_0$ and the next term in the expansion, $\epsilon w_1$, we conclude that \emph{noticeably} asymmetrical flow in the utricle first emerges when
\begin{equation}
\frac{\pi}{2I_4 a_m^2}\sim \epsilon\frac{a_m^3}{6\sqrt{3}},\Longleftrightarrow \frac{3\sqrt{3}\pi}{\epsilon }\sim a_m^5 \int_0^{2\pi}\frac{\upd s}{a(s)^4}.\label{eq:utricular_vortex_condition}
\end{equation} 

\begin{figure}
    \centering
    \begin{overpic}
    [width=0.99\linewidth]{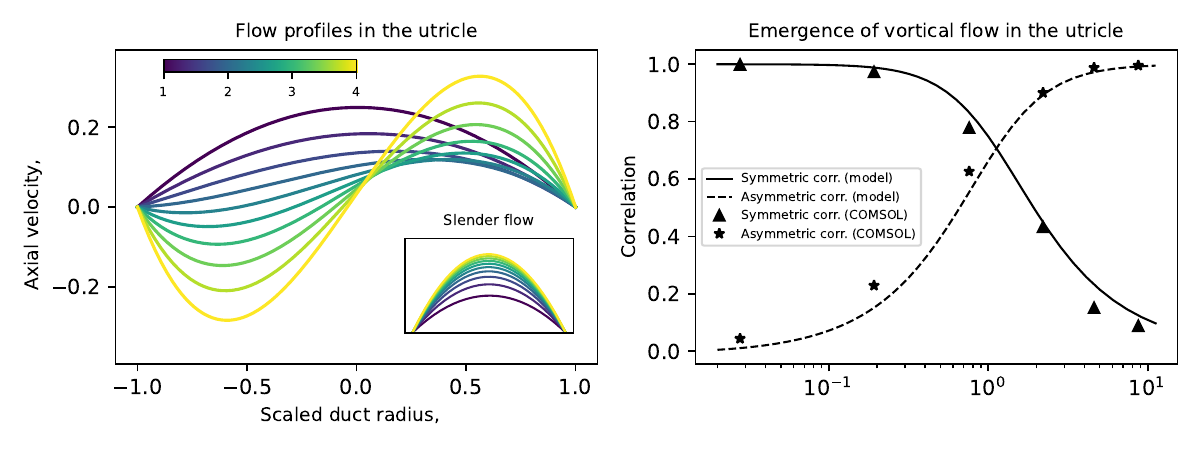}
    \put(21,29){\tiny $a_m$}
    \put(37.7,2.3){\scriptsize $r/a_m$}
    \put(73,2.3){\scriptsize $\xi=\frac{\epsilon}{3\sqrt{3}\pi}I_4 a_m^5$}
    \put(2.15,24.5){\rotatebox{90}{\scriptsize $w$}}
    \end{overpic}
    \caption{Left panel: analytical reconstruction of flow profiles in the wide region of the channel (representing the utricle), $w(r,\theta,s=\pi,t)$ for different values of the maximum enlargement $a_m$. The inset shows the flow profiles in the thin region of the flow $w(r,\theta,s=0,t)$; these remain symmetric, confirming the symmetry breaking mechanism is not the same as the global symmetry-breaking mechanism discussed in \S\ref{sec:deformation_regimes}. Right panel: correlation (as defined in \eqref{eqn:CorrelationDefns}) between the axial velocity in the utricle $w(r)=w_0(r)+\epsilon w_1(r,\theta)$ and the symmetric (solid) and asymmetric flow profile (dashed). Curves show the results of the analytical computation and triangles and stars show the correlations computed from the COMSOL solution.  We find that the transition occurs when $\xi= a_m^5 I_4\epsilon /(3\sqrt{3}\pi) \sim 1$, as predicted by our analysis.}
    \label{fig:utricular_flow_profiles}
\end{figure}

This motivates the introduction of a transition parameter $\xi= a_m^5 I_4\epsilon /(3\sqrt{3}\pi)$. The transition can be seen qualitatively in the left panel of Figure~\ref{fig:vortical_flow}, where the flow in the utricle (bottom half of the torus) transitions from symmetric to asymmetric as the bulge is increased. In Figure~\ref{fig:utricular_flow_profiles} this transition is analysed quantitatively. In the left panel, the flow profile in the utricle is plotted for varying $a_m$. Observe that for large $a_m$, the utricle flow is asymmetric while the flow in the slender part of the canal remains symmetric. To quantify the transition to noticeably asymmetrical flow, we compute the correlation between the velocity profile $w= w_0 +\epsilon w_1$ and the symmetric ($w_0$) and asymmetric ($w_1$) solutions.

The correlations may be written as 
\begin{subequations}
    \begin{align}\label{eqn:CorrelationDefns}
        &C(w,w_0)=\frac{R(w,w_0)}{\sqrt{R(w,w)R(w_0,w_0)}},\quad C(w,w_1)=\frac{R(w,w_1)}{\sqrt{R(w,w)R(w_1,w_1)}},\\
        & R(u(r,\theta),v(r,\theta))=\int_0^{2\pi}\int_0^{a_m}u(r,\theta)v(r,\theta)r~\upd r\upd \theta,
        \end{align}\end{subequations}

and we find 
\begin{subequations}
    \begin{align}
 &C(w,w_0)=\sqrt{\frac{R_0}{R_0+\epsilon^2 R_1}},\quad C(w,w_1)=\sqrt{\frac{\epsilon^2 R_1}{R_0+\epsilon^2 R_1}},\\
 &R_0 = R(w_0,w_0)=2\pi\int_0^{a_m}w_0(r)^2r~\upd r=\frac{\pi^3}{12I_4^2a_m^2}\left(\dot{\Omega}(t)+\frac{\Delta p_0}{2\pi}\right)^2,\\
 \begin{split}&R_1=R(w_1,w_1)=\int_0^{2\pi}\cos^2\theta\upd\theta\int_0^{a_m}\left[\frac{r(r^2-a_m^2)(-3\Delta p_0+(4a_m^4I_4-6\pi)}{16I_4 a_m^4}\right]^2r~\upd r\\
 &\quad \,=\frac{\pi}{24\cdot16^2I_4^2}\left(-3\Delta p_0+(4a_m^4I_4-6\pi)\dot{\Omega}(t)\right)^2\end{split}
\end{align}
\end{subequations}

Here, we have assumed that $\Delta p_1^\text{outer}=0$ and used the fact that $R(w_0,w_1)=0$, which is trivial since $w_0w_1\sim\cos\theta$. The correlation $C(w,w_0)$ will be close to 1 when the flow is \emph{largely} symmetrical, and closer to zero when the flow is \emph{noticeably} asymmetrical. The opposite is true of the correlation $C(w,w_1)$. In the the right panel of figure~\ref{fig:utricular_flow_profiles} we plot the symmetrical correlation $C(w,w_0)$ (solid line) and the asymmetrical correlation $C(w,w_1)$ (dashed line) as predicted from our analytical model.

Based on the definition of $\xi$ we expect a transition when $\xi \sim 1$. figure~\ref{fig:utricular_flow_profiles} confirms this expectation, showing that a transition indeed occurs at this point when the geometry $a(s)$ satisfies \eqref{eqn:UtricleProfile}.  We note that for times when $\dot{\Omega}(t) = 0$, our analysis does not apply, and the flow remains symmetric, even for large utricles. This follows from \eqref{eq:size_of_w1}, as it is clear that $w_1 = 0$ when $\dot{\Omega}(t) = 0$.

Our analysis of the onset of vortical flow is further supported by numerical simulations, represented by triangles and stars in the righthand panel of figure~\ref{fig:utricular_flow_profiles}. These simulations were conducted using COMSOL, modelling the utricle as an ellipsoidal expansion of the toroidal geometry. We find good agreement between the analytical predictions and COMSOL simulations regarding the emergence of vortical flow. (We attribute the small discrepancy between the COMSOL and analytical model  to the fact that the geometry is not identical in both cases, and furthermore because the analytical result does not have a solid obstacle disrupting the flow.) Furthermore, the consistency between the numerical simulations and analytical solutions suggests that, although the first-order correction may be larger than the leading-order term, the subsequent terms in the series remain well behaved, ensuring that the asymptotic ordering is preserved.

\section{Discussion and conclusion}
In this study, we developed a mathematical framework to model fluid flow in the semicircular canals of the vestibular system, focussing in particular on the interaction between the fluid motion and cupular deformation. Through a systematic analytical and numerical investigation, we identified distinct physical regimes and key mechanisms that govern the fluid-structure response to an imposed rotation. Our results not only advance the understanding of flow dynamics in these biologically relevant systems but also provide a simple framework with the potential for analysing vestibular function and dysfunction in response to head movements. 

Our analytical approach consisted in solving the Navier-Stokes equations via an asymptotic series in the small aspect ratio of the semicircular canals. Through asymptotic analysis, and by connecting the fluid flow at the cupula to the cupular deformation, we reduced the vestibular dynamics to an ODE system for the cupular pressure jump, whose behaviour could  easily be characterized. In this way, we established three primary regimes of flow-cupula interaction, depending on the value of the relative stiffness parameter $\kappa$: 

\begin{itemize}
    \item \textbf{Soft cupula regime}, $\kappa\ll1$: When the cupular stiffness is relatively low, the deformation of the cupula closely follows the angular velocity of the head. In this regime, the flow in the canal exhibits symmetry about the centreline, and in dimensional terms the magntiude of the cupular deformation scales as $\hat{\eta}\sim a^2R\Omega_0/\nu$.
    \item \textbf{Stiff cupula regime}, $\kappa\gg1$: As the stiffness of the cupula increases, the deformation transitions to follow the angular acceleration of the head. In this regime, the symmetry of the fluid flow about the centreline is broken, creating distinct zones of differential flow. This transition highlights the importance of structural properties of the cupula in shaping the dynamic response of the vestibular system. The magnitude of the cupular deformation scales as $\hat{\eta}\sim a^2R\Omega_0/(\nu\kappa)=a R^2\Omega_0\rho/(E \mathcal{T})$ in this regime.
    \item \textbf{High-frequency regime}, $\St\gg\max\{\kappa,1\}$: if the forcing frequency increases substantially inertial effects in the fluid must be considered and the deformation follows the angular displacement of the head. This type of motion could be expected when e.g. driving over an uneven surface at speed, or in a collision. In this regime the magnitude of the deformation is $\hat{\eta}\sim R^2\Omega_0\mu/(E a)$.

\end{itemize}

In this work the cupula was modelled as an elastic solid of finite thickness, $t_h > 0$, in contrast to previous analytical models that either treated it as a plate of vanishing thickness \citep{rabbitt_hydroelastic_1992} or that did not include the cupula as a physical solid structure  \citep{obrist_fluidmechanics_2008}. Experimental observations indicate that the cupula is not thin \citep{selva_mechanical_2009}, motivating the inclusion of thickness as a parameter to more accurately capture its deformation in an analytically tractable framework. Our theoretical model suggests that:
\begin{itemize}
    \item For small thickness ratios $\beta=t_h/a$, the deformation resembles that of a clamped plate (a quartic profile). In this case, the cupular deformation scales as $\beta^{-3}$ for $\beta\ll1$.
    \item Increasing $\beta$ leads to a transition towards a quadratic deformation that matches the radially symmetric leading-order velocity profile $w_0$. In this case the deformation is smaller, scaling with $\beta^{-1}$.
\end{itemize}

To verify the analytical findings, we conducted numerical simulations of the reduced equations using COMSOL Multiphysics. The numerical results showed excellent agreement with the asymptotic predictions, confirming the validity of the analytical approximations across a wide parameter space. Importantly, the numerical approach enabled us to also account for the influence of complex fluid--solid interaction boundary conditions and the nonlinear, advection, term in the Navier-Stokes equations that are otherwise intractable analytically, and comparison with the analytical solution points to their influence being small (and $\mathrm{Re}$ not being a relevant parameter in the problem).

When the inertial terms were incorporated into the governing equations via inclusion of a finite Stokes number, $\St$, we observed important modifications to the system's behaviour. For small Stokes numbers, and in studying the relaxation of the cupula to an initial deformation, the system exhibited overdamped dynamics. This behaviour is also predicted when inertia is neglected, and is consistent with the low-Reynolds-number assumption inherent to the vestibular fluid dynamics. However, for sufficiently large Stokes numbers, the system exhibited underdamped oscillations, with the cupular deformation following angular displacement even in the soft cupula regime for sufficiently high $\St$. This transition highlights the interplay between inertial and viscous forces in shaping the dynamic response of the system. Physiologically, this finding suggests that under certain conditions, such as during rapid head movements (e.g.~during an impact), the vestibular system may exhibit enhanced sensitivity to displacement due to inertial effects. 

The assumption of an idealized toroidal geometry allowed for significant analytical simplifications but is also a significant simplification of the true anatomy of the semicircular canals, which exhibit variations in cross-sectional shape. To address this, we extended our analysis to more realistic geometries, focusing on domains with a single enlarged region that deviates from the perfect torus. In these regions, we found that significant deviation in radius gives another mechanism for the breaking of flow symmetry (in addition to the rigidity--induced effect already discussed). These results provide new insights into the functional implications of anatomical variability in the semicircular canals. For example, variations in canal geometry across species or due to developmental differences may influence the sensitivity and response characteristics of the vestibular system.

In each of the scenarios considered, our analytical approach enabled us to derive explicit expressions for the transition point between physical regimes, that is we obtained formulas for the critical values of the relevant system parameters at which the transition between different regimes occurs. These formulas lend insight into the fine balance between different components of the system, and enable us to speculate on how the vestibular system may have been fine tuned by evolution in different organisms, and/or key considerations in engineering an artificial vestibular system. We turn to such considerations next.

\subsection{Implications and applications}

The findings of this study have several implications for both biology and engineering. In the context of vestibular physiology, our results contribute to a deeper understanding of how the semicircular canals transduce head motion into neural signals. The distinction between velocity-sensitive and acceleration-sensitive regimes suggests that while the mechanical properties of the cupula, combined with canal geometry, enable the system to function under a wide range of motion frequencies \citep{bronstein_vertigo_2013} it emphasizes that what is sensed differs markedly across this parameter space. 
This flexibility of sensing may be useful for maintaining balance and spatial orientation across diverse locomotor activities \citep{golding_motion_2005,golding_pathophysiology_2015}.

Here, it is worth considering the distinction between the soft cupula ($\kappa\ll1$) and stiff cupula ($\kappa\gg1$) regimes in terms of dimensional quantities. Recall that $\kappa$ is defined as $\kappa =E \mathcal{T}a/(R\mu)$, where $\mathcal{T}$ is the timescale for the head motion, $a$ and $R$ are respectively the small and large radii defining the canals, and $E$ is the cupula's Young's modulus. We see that the soft cupula regime may be attained for fast movements (small $\mathcal{T}$), highly viscous media ($\mu$ large) and of course soft cupulas in absolute terms (small $E$). The converse holds for the stiff cupula regime. Inserting parameter values for $a$, $R$, $E$ and $\mu$ into the transition value predicted by our model,  $\kappa_c = 32\pi I_4\alpha_0(\beta)$, we obtain a critical value
\begin{equation}\label{eq:Tscale}
   \mathcal{T}=\frac{\kappa_c R\mu}{E a}%\frac{a^2 R\mu \kappa_c}{B}=\frac{12(1-\nu^2)\, a^2R\mu\,\kappa_c}{E\,t_h^3}\sim \frac{a^2 R\mu }{E t_h^3},
\end{equation}
which may be interpreted as a critical timescale of head motion below which the system responds to angular velocity, and above which the system responds to angular acceleration. Inserting typical values for a human adult, we compute a transition frequency of 0.27 Hz. Interestingly, human experiments with controlled oscillation frequencies have reported a maximum motion sickness when the frequency is around 0.2 Hz \citep{golding_motion_2001}. Our analysis suggests an intriguing possible explanation for this maximal sickness at intermediate frequencies: intermediate frequencies correspond to motion for which the response of the cupular system follows neither the angular acceleration nor velocity. The ``neural mismatch" hypothesis predicts that motion sickness is induced in situations where there is a disagreement between visual or vestibular cues and the information anticipated by the nervous system \citep{benson_sensory_1990}. Since the vestibular system may be expected to provide information that matches neither the true acceleration or the true velocity around the transition point we suggest that it may be the transition point between small and large $\kappa$ that causes motion sickness. 

Evidence suggests that susceptibility to motion sickness peaks at a higher frequency for animals smaller than humans \cite{golding_biodynamic_2016, javid_variables_1999}. Given the preceding discussion, this is a little surprising: if we assume that the size scales of the vestibular system scale in proportion then  \eqref{eq:Tscale} shows that the critical frequency should remain the same. \emph{If} the preceding hypothesis is correct, it would suggest that either the material parameters of the cupula change or that some non-trivial allometric scaling of the dimensions of the canal must occur. (We are unaware of any data on the allometric scaling.) 

In \S\ref{sec:deformation_regimes} we also mentioned that in the transition region (when $\kappa$ is nether large nor small) the response develops an overshoot at the end of the manoeuvre: the deformation, and hence a sensing signal persists 
after the motion has concluded. This feature seems undesirable (the ``neural mismatch'' hypothesis would predict a high likelihood of experiencing motion sickness, as the vestibular input will disagree with the visual input) but also in line with everyday experience of dizziness.

A natural question to ask is whether a similar system could be conceived to detect linear acceleration, rather than rotational motion. This might be expected, for instance, in the limiting case where $\epsilon = 0$, in which the velocity profile must remain exactly symmetric. In this case, the dimensional cupular deflection can be expressed as
\begin{align}
\hat{\eta} \sim a^2 R \Omega_0 / \nu = \epsilon^2 R^3 \Omega_0 / \nu \rightarrow 0, \quad \text{as} \quad \epsilon \rightarrow 0
\end{align}
for fixed $R$ (which must remain fixed to maintain physiological realism). Thus, in the limit $\epsilon =0$, the semicircular canals become ineffective at detecting angular motion. Consistent with this prediction of our model, linear motion of the head is indeed detected by a different component of the vestibular system—the otolith organs—which rely on solid inertia \citep{rabbitt_biomechanics_2004}.

Ageing significantly impairs balance and vestibular function, as evidenced by widespread declines in vestibulo‑ocular reflexes, hair cell counts, and neural processing with advancing age 
\citep{anson_perspectives_2016}. While the hypothesis that cupular stiffening contributes to vestibular decline is plausible, no direct physiological or medical evidence currently links ageing to such mechanical changes. Instead, clinical data indicate that balance deficits in older adults are more often driven by hormonal or neurochemical modulation and sensory cell loss (\cite{iwasaki_dizziness_2014} report 25\% fewer hair cells in nonagenarians versus individuals in their 50s). Nonetheless, the idea of altered cupular mechanics remains an interesting direction for future research, and the model we have presented forms an attractive framework for investigating hypotheses and linking with clinical observations.  

From an engineering perspective, the insights gained from this study could inform the design of biomimetic sensors, prosthetics and systems, for example the MEMS prototype from \cite{raoufi_development_2019}. For instance, understanding the interplay between fluid dynamics and flexible structures in the semicircular canals could inspire the development of flow sensors or inertial measurement devices that mimic the sensitivity and robustness of the vestibular system, for instance, biologically-inspired inertial navigation systems. Additionally, the analytical framework developed here could be extended to other biological systems involving thin fluid-filled structures, such as the cochlea or cardiovascular vessels.

\subsection{Limitations and future directions}

Our model has been based on a number of approximations that have facilitated the analysis that we have presented here. Of these, perhaps the most important is our use of a disk shaped model of the cupula --- this is mathematically convenient but is likely to be an over-simplification of the true geometry of the cupula \citep{selva_mechanical_2009}. Related to the previous point, the semicircular canals (SCCs) are modelled with circular cross-sections, although anatomical observations indicate they are elliptical \citep{curthoys_dimensions_1987}. While the leading-order equations can be readily adapted to more general cross-sectional geometries, this is not straightforward at $\mathcal{O}(\epsilon)$. At the same time, we note that our existing analysis assumes that  the cupula's thickness is constant, but photographic evidence suggests this is likely not the case \citep{rabbitt_biomechanics_2004}. In particular, the cupula seems to be thinner in the centre and thicker towards the edge. Whilst this will influence the shape of the deflection profile of the cupula, it is unlikely to give rise to new phenomena.

Another simplification in our model is that that the cupula is uniformly clamped to the crista and canal walls. However, previous results based on this assumption \cite{selva_mechanical_2009} required a very small Young's modulus, close to $5\mathrm{~Pa}$, to match experimentally observed deformations \cite[][]{selva_mechanical_2009}. This is an extremely low value, perhaps indicating the softest material in the human body, and is unrealistic when compared to other ``soft" biological tissues \citep{goriely_mathematics_2017}. We suggest that this anomalous stiffness of the cupula may be a result of the clamped boundary conditions on all sides of the cupula, as used here and as usual in the vestibular literature \citep{rabbitt_hydroelastic_1992}, may be incorrect; typical anatomical drawings suggest that the cupula is only clamped on a part of its boundary, and is free to move on other regions of the boundary. This would increase the apparent flexibility of the cupula, creating similar system behaviour with without requiring an unusually small Young's modulus. Moreover, materials with similarly low Young’s moduli are often viscoelastic \citep[see, for example, hydrogels in][]{ahearne_characterizing_2005}. Incorporating a viscoelastic term into our linear model of cupular deformation is in principle straightforward. However, the absence of specific data on the cupula and its material properties would significantly complicate parameter tuning. Additionally, since the system already includes a source of damping through fluid viscosity, disentangling the respective contributions of fluid and solid damping would be non-trivial.

Our model may also allow for other phenomena within the vestibular system to be investigated. An interesting avenue using the techniques developed here is the light cupula phenomenon~\cite[see][for a review of the concept]{lee_light_2024}, as well as related concepts such as the buoyancy hypothesis to explain balance loss after alcohol intake~\citep{nieschalk_effects_1999}. These essentially state that when alcohol is consumed, ethanol diffuses faster into the cupula than the surrounding endolymph, changing their density ratio (which under normal functioning is very close to one, so that the cupula is neutrally buoyant). As ethanol is less dense than water, the cupula would then become negatively buoyant, deforming differently than ordinarily and sending incorrect signals to the nervous system. This can be accounted for in our model by including a buoyancy term so that the cupula can float or sink through the endolymph. An analysis along these lines is presented in Appendix~\ref{sec:light_cupula}, where we show that even a small density change (on the order of $1\%$) might lead to a cupular deformation comparable to that induced by a rotation of $\Omega_0\approx 0.3$ rad$\cdot$s$^{-1}$ --- this estimate suggests that alcohol intake may indeed lead to additional cupular deflections (and hence sensory mismatch) consistent with the observed effects of alcohol on balance \citep{hegeman_even_2010}.

\backsection[Supplementary data]{\label{SupMat}Supplementary material and movies are available at \\https://doi.org/10.1017/jfm.2019...}

\backsection[Acknowledgements]{We acknowledge fruitful conversations with Dr Miguel Vaca, Dr Eduardo Martin Sanz, Prof.~Michael Gresty and Prof.~Adolfo Bronstein. We are further grateful to Prof.~Sarah L. Waters and Prof.~Ian M. Griffiths for their invaluable feedback on an early version of the project.  For the purpose of Open
Access, the authors will apply a CC BY public copyright license to any Author
Accepted Manuscript version arising from this submission.} 

\backsection[Funding]{JCV is funded by a St. John's College scholarship.}

\backsection[Declaration of interests]{The authors report no conflict of interest.}

\backsection[Data availability statement]{The data that support the findings of this study will be made openly available in the Oxford Research Archive.}

\backsection[Author ORCIDs]{Javier Chico-Vazquez, https://orcid.org/0009-0008-7536-6978. \\ 
Dominic Vella, https://orcid.org/0000-0003-1341-8863\\
Derek Moulton, https://orcid.org/0000-0003-3597-7973}

%\backsection[Author contributions]{Authors may include details of the contributions made by each author to the manuscript'}

\appendix

\section{Angular velocity of the SCC}\label{sec:angular_velocity_is_the_same}

We consider the motion of a (rigid-body) human standing vertically on a rotating object, such that the person (and hence the SCC) are not positioned on the axis of rotation. For instance, we could have in mind being on a merry-go-round or going around a corner in a vehicle (which are common situations that lead to motion sickness). In this appendix we show that the angular velocity of the SCC is the same as that of the rotating object.

\begin{figure}
    \centering
    \begin{overpic}[width=0.7\textwidth]{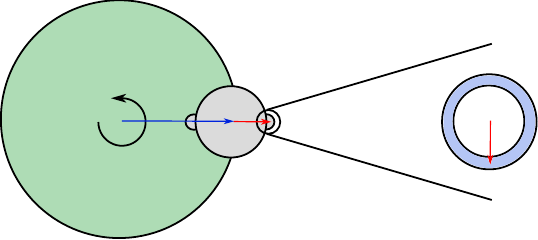}
    \put(21,14){$\Omega$}
    \put(31,24){\textcolor{blue}{$\mathbf{R}$}}
    \put(45.5,24){\textcolor{red}{$\mathbf{r}$}}
    \put(88,18){\textcolor{red}{$\mathbf{x}$}}
    
    \put(11,35){Merry-go-round}
    \put(39.5,17){Head}
    \put(88,32){SCC}
        
    \end{overpic}
    \caption{Diagram for rotational motion not centered on the SCCs. }
    \label{fig:angular_velocity}
\end{figure}

The centre of the human head has position $\mathbf{R}(t)=R\mathbf{e}_r$ measured from the centre of the rotating object (see figure~\ref{fig:angular_velocity}). The human is rotating with angular velocity $\boldsymbol{\Omega}(t)=\Omega(t)\mathbf{e}_z$ about this point, so that $\dot{\boldsymbol{R}}=\boldsymbol{\Omega} \times \mathbf{R}$. Then, $\dot{\boldsymbol{R}}=R\Omega\, \mathbf{e}_\theta$. Now, we consider a horizontal semicircular canal, whose central location is at position $\mathbf{R}+\mathbf{r}$, where $\mathbf{r}$ is a position vector from the centre of the head to the canal centre (see figure~\ref{fig:angular_velocity}). We can write $\mathbf{r}=r(\cos\varphi \mathbf{e}_r+\sin\varphi \mathbf{e}_\theta)$ for some $\varphi$. If the body is rotating with the moving object, then it is effectively rigid body rotation, so that $\dot{r}=\dot{\varphi}=0$. Since the frame $\{\mathbf{e}_r,\mathbf{e}_\theta\}$ rotates with angular velocity $\Omega(t)\mathbf{e}_z$, it follows that
\begin{align}
    \dot{\mathbf{r}}=\Omega \mathbf{e}_z\times \mathbf{r},
\end{align}
meaning that the vector $\mathbf{r}$ undergoes rotation with the same angular velocity as the rotating object in which it sits.
%Here, we have used $\dot{\mathbf{e}}_r=\Omega\mathbf{e}_\theta$ and $\dot{\mathbf{e}}_\theta=-\Omega\mathbf{e}_r$. 
Similarly, if we define $\mathbf{x}$ as a vector pointing from the centre of the semicircular canal to a point inside the canal (see figure~\ref{fig:angular_velocity}), we have $\dot{\mathbf{x}}=\boldsymbol{\Omega}\times\mathbf{x}$. Writing the position of a part of the semicircular canal as measured from a lab frame as $\mathbf{p}=\mathbf{R}+\mathbf{r}+\mathbf{x}$, we find
\begin{align}
    \dot{\mathbf{p}}=\dot{\mathbf{R}}+\dot{\mathbf{r}}+\dot{\mathbf{x}}=\boldsymbol{\Omega}\times\mathbf{p},
\end{align}
and hence there is no additional linear velocity. 

%%%%%%%%%%%%%%%%%%%%%%%%%%%%%%%%%%

\section{On the deformation of the cupula}

In this section we discuss the details for the numerical simulations from figure~\ref{fig:symmetry_breaking}, and present a method to obtain a polynomial approximation for the deformation of the cupula. 

\subsection{Details for numerical solutions}\label{sec:COMSOL_details}
The governing equations from Section~\ref{sec:governing_equations} were solved in COMSOL for different values of the Young's modulus. The equations were solved on a moving grid, without neglecting the geometric nonlinearity and including the nonlinear terms in the Navier--Stokes equations. As mentioned previously, the cupula is modelled as a full three-dimensional solid, without assuming it is a thin structure and with a finite thickness, $t_h$.

\begin{figure}
    \centering
    \includegraphics[width=0.99\linewidth]{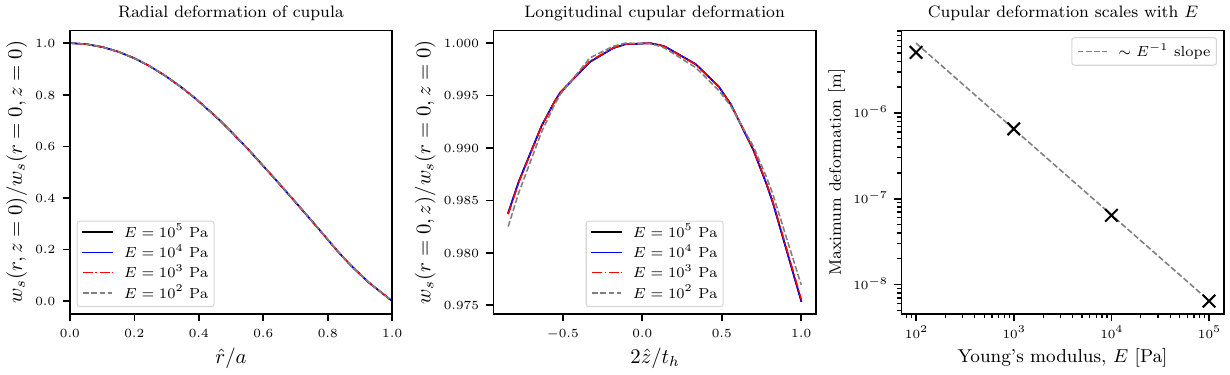}
    \caption{Numerically obtained deformation of the solid material at $\hat{t}=0.25$ s, normalized by the maximum deformation. Left panel: deformation along the direction perpendicular to the canal centre-line (in the model's coordinate system this is along $r$). Middle panel: deformation in the direction parallel to the centreline (i.e.~along $z$). Right panel: deformation at the centre of the cupula plotted as a function of the Young's modulus. All deformations shown have been averaged over the azimuthal direction $\theta$.}
    \label{fig:solid_fluid_cupular_profiles}
\end{figure}

In Figure~\ref{fig:solid_fluid_cupular_profiles} we plot the deformation of the cupula, by showing (left panel) the deformation of the cupula in the direction normal to the flow and (middle panel) the deformation of the cupula in the direction along the flow. In the first case, we observe that the deformation is large in the centre of the structure and zero on the edges, as expected, moreover we observe all curves collapse. This indicates the deformation regime is linear. In the second plot, we see that the deformation only varies by around 2\% in the direction parallel to  the centreline. Again, this is indicates the use of a depth averaged measure for the cupula's deformation, i.e. $\eta(r,t)=\beta^{-1}\int_{-\beta/2}^{\beta/2}w_s(r,z,t)\upd z$ is appropriate. (The forcing used to generate  Figure~\ref{fig:solid_fluid_cupular_profiles} is that given by \eqref{eq:dimensionless_forcing}.) Finally, we note that the magnitude of the deformation is inversely proportional to the stiffness $E$, as can be seen in Figure~\ref{fig:solid_fluid_cupular_profiles} (right panel).

\subsection{Solid deformation of non-slender cupulas}\label{sec:solid_mechanics}
The cupula is a moderately thick elastic solid, and hence we require a full linearly elastic model to compute its deformation. Here, we compute the deformation of the cupula for $\mathcal{O}(1)$ and $\mathcal{O}(\epsilon)$, and compare our analytical approximation to a numerically obtained deformation using the finite element method. The dimensionless equations for the solid deformation in component form are:
\begin{subequations}
\begin{align}
&\begin{split}\frac{\epsilon}{\kappa}\frac{\rho_s}{\rho}\left[\St\frac{\partial^2u_s}{\partial t^2}-\Omega_0\mathcal{T}\Omega(t)^2\cos\theta-2\St\Omega_0\mathcal{T}\Omega(t)\frac{\partial w_s}{\partial t}\cos\theta\right]
\\=\frac{1}{r}\frac{\partial }{\partial r}(r\tau_{rr})+\frac{1}{r}\frac{\partial \tau_{r\theta }}{\partial \theta}+\frac{\partial \tau_{rz}}{\partial z}-\frac{\tau_{\theta\theta}}{r}\end{split}\\
&\begin{split}\frac{\epsilon}{\kappa}\frac{\rho_s}{\rho}\left[\St\frac{\partial^2u_s}{\partial t^2}+\Omega_0\mathcal{T}\Omega(t)^2\sin\theta+2\St\Omega_0\mathcal{T}\Omega(t)\frac{\partial w_s}{\partial t}\sin\theta\right]
\\=\frac{1}{r}\frac{\partial }{\partial r}(r\tau_{r\theta})+\frac{1}{r}\frac{\partial \tau_{\theta\theta }}{\partial \theta}+\frac{\partial \tau_{\theta z}}{\partial z}+\frac{\tau_{r\theta}}{r}\end{split}\\
    &\begin{split}\frac{\epsilon}{\kappa}\frac{\rho_s}{\rho}\left[\St\frac{\partial^2 w_s}{\partial t^2}+\frac{\upd \Omega}{\upd t}+2\St\Omega_0\mathcal{T}\Omega(t)\left(\frac{\partial u_s}{\partial t}\cos\theta-\frac{\partial v_s}{\partial t}\sin\theta\right)\right]
    \\=\frac{1}{r}\frac{\partial }{\partial r}(r\tau_{rz})+\frac{1}{r}\frac{\partial \tau_{\theta z}}{\partial \theta}+\frac{\partial \tau_{zz}}{\partial z}\end{split}
\end{align}
\end{subequations}
with the constitutive law,
\begin{subequations}
    \begin{align}
        &\tau_{rr}=\frac{\lambda_s}{E} \boldsymbol{\nabla}\cdot\boldsymbol{u}_s+2\frac{\mu_s}{E} \frac{\partial u_s}{\partial r},&\tau_{r\theta}=\frac{\mu_s}{E} \left(\frac{1}{r}\frac{\partial u_s}{\partial\theta}+\frac{\partial v_s}{\partial r}-\frac{v_s}{r}\right),\\
        &\tau_{\theta\theta}=\frac{\lambda_s}{E} \boldsymbol{\nabla}\cdot\boldsymbol{u}_s+2\frac{\mu_s}{E}  \frac{1}{r}\left(\frac{\partial v_s}{\partial \theta}+u_s\right),
        &\tau_{rz}=\frac{\mu_s}{E} \left(\frac{\partial u_s}{\partial z}+\frac{\partial w_s}{\partial r}\right),\\
        &\tau_{zz}=\frac{\lambda_s}{E} \boldsymbol{\nabla}\cdot\boldsymbol{u}_s+2\frac{\mu_s}{E}  \frac{\partial w_s}{\partial z}, &\tau_{\theta z}=\frac{\mu_s}{E} \left(\frac{1}{r}\frac{\partial v_s}{\partial \theta}+\frac{\partial w_s}{\partial z}\right),
    \end{align}
\end{subequations}
and boundary conditions
\begin{subequations}
    \begin{align}
    \tau_{zz}=\frac{\Delta p}{2\kappa},\quad\tau_{rz}=\tau_{\theta z}=0 ,\quad z=\beta/2\\
    \tau_{zz}=-\frac{\Delta p}{2\kappa},\quad \tau_{rz}=\tau_{\theta z}=0 ,\quad z=-\beta/2\\
    u_s=v_s=w_s=0,\quad r=a(0). 
\end{align}
\end{subequations}
For the purpose of computing the pressure difference across the cupula, it suffices to use the equations for the solid deformation in cylindrical coordinates. The additional terms due to the toroidal curvature can be shown to be $\propto\sin\theta$ or $\cos\theta$ and therefore do not contribute to the flux when integrated over a cross section.
In this appendix we will solve the solid equations up to and including $\mathcal{O}(\epsilon)$ using the polynomial method presented in \cite{barber_elasticity_2010}. To keep the algebra more palatable we will take $a(0)=1$ without loss generality, but results for important volume displacement coefficients $\alpha$ can be obtained for general $a(0)$ by multiplying by $a(0)^3$.

\subsubsection{Leading order solution}
From lubrication theory, we know that $\Delta p_0$ is a function of time only. Accordingly, we seek an axisymmetric solution to the Navier equations, independent of $\theta$. This is an assumption of our model: although a spatially uniform pressure could, in principle, generate circumferential instabilities such as wrinkling, we assume such effects do not arise in this context. Moreover, since the precise attachment of the cupula to the canal walls (and hence the associated boundary conditions) remains poorly understood, our assumption yields the simplest plausible solution that retains physiological realism. The axisymmetric Navier equations to leading order are:
\begin{subequations}
\begin{align}
&0
=\frac{1}{r}\frac{\partial }{\partial r}(r\tau_{rr0})+\frac{1}{r}\frac{\partial \tau_{r\theta 0}}{\partial \theta}+\frac{\partial \tau_{rz0}}{\partial z}-\frac{\tau_{\theta\theta0}}{r},\\
    &0
    =\frac{1}{r}\frac{\partial }{\partial r}(r\tau_{rz0})+\frac{1}{r}\frac{\partial \tau_{\theta z0}}{\partial \theta}+\frac{\partial \tau_{zz0}}{\partial z},
\end{align}
\end{subequations}
with boundary conditions
\begin{subequations}
\begin{align}\label{eq:thickness_boundary_conditions}
    \tau_{zz0}&=\frac{\Delta p_0}{2\kappa},\quad \tau_{rz0}=0, \quad z=\beta/2,\\
    \tau_{zz0}&=-\frac{\Delta p_0}{2\kappa},\quad \tau_{rz0}=0, \quad z=-\beta/2,\\
    u_{s0}&=v_{s0}=w_{s0}=0,\quad r=1.
\end{align}
\end{subequations}
$\beta=t_h/a$ is the dimensionless thickness of the cupula and $\kappa=E\mathcal{T}a/(R\mu)$ is the relative stiffness for a cupula of arbitrary thickness. As this is an axisymmetric problem, we take $v_s=0$. As the problem is linear suffices to solve it for $\Delta p/\tilde{\kappa}=1$. We proceed by using the technique from \cite{barber_elasticity_2010}, writing potentials $\phi$ and $\omega$ as
\begin{align}
    \phi_0 = \sum_{j=1}^5A_jP_j(r,z),\quad \omega_0 = \sum_{j=1}^4B_jP_j(r,z)
\end{align}
where the harmonic polynomials are \citep{barber_elasticity_2010}
\begin{align}
\begin{split}
    P_1(r,z)=z,\quad P_2(r,z)=(2z^2-r^2)/2,\quad P_3(r,z)=(2z^3-3zr^2)/2,\\
    P_4(r,z)=(8z^4-24z^2r^2+3r^4)/8,\quad P_5(r,z)=(8z^5-40z^3r^2+15zr^4)/8.
    \end{split}
\end{align}
Following example 25.2.1 and Table 21.1 from \cite{barber_elasticity_2010}, we compute the stresses $\tau_{rr0},\tau_{rz0},\tau_{zz0}$ and $\tau_{\theta\theta0}$, and formulate the (strong) boundary conditions at $z=\pm\beta/2$, as in the example. Instead of using simply supported boundary conditions at $r=1$, we use the following (weak) boundary conditions to fix the displacement at $r=1$.
\begin{align}
    \int_{-\beta/2}^{\beta/2}u_{s0}(r=1,z)\upd z=\int_{-\beta/2}^{\beta/2} zu_{s0}(r=1,z)\upd z= \int_{-\beta/2}^{\beta/2} w_{s0}(r=1,z)\upd z=0.
\end{align}
Following \cite{barber_elasticity_2010}, we formulate a linear system of eleven equations for the nine unknowns $\{A_1,A_2,A_3,A_4,A_5,B_1,B_2,B_3,B_4\}$, however these conditions are not all independent: the coefficient matrix has rank 9, which allows us to solve for the unknowns uniquely. Once the constants are known, we find the solution for the deformation profile:
\begin{align}\label{eq:deformation_profile_with_thickness}
    \bar{\eta}_0(r;\beta)=\frac{1}{\beta}\int_{-\beta/2}^{\beta/2} w_{s0}(r,z;\beta)\upd z=\frac{3}{16}\frac{1-\nu_s^2}{\beta^3}(1-r^2)^2+\frac{1}{20}\frac{(1+\nu_s)(12-\nu_s)}{\beta}(1-r^2).
\end{align}
The first term dominates for $\beta\rightarrow0^+$ when the cupula is very thin and is exactly the deformation profile for a thin clamped plate under uniform loading. The second term dominates when the cupula is thick and $\beta\sim 1$, leading to a quadratic deformation profile for thick cupulas. $\bar{\eta}(r;\beta)$ may be used to compute the coefficient $\alpha_0(\beta)$,
\begin{align}\label{eq:alpha_with_thickness}
    \alpha_0(\beta)=\int_0^1 r\bar{\eta}_0(r;\beta)\upd r=\frac{1-\nu_s^2}{32\beta^3}+\frac{(12-\nu_s)(1+\nu_s)}{80\beta}.
\end{align}

\subsubsection{Comparison with numerics and thick cupula limit}\label{sec:thick_cupula_limit}
\begin{figure}
    \centering
    \begin{overpic}[width=0.99\textwidth]{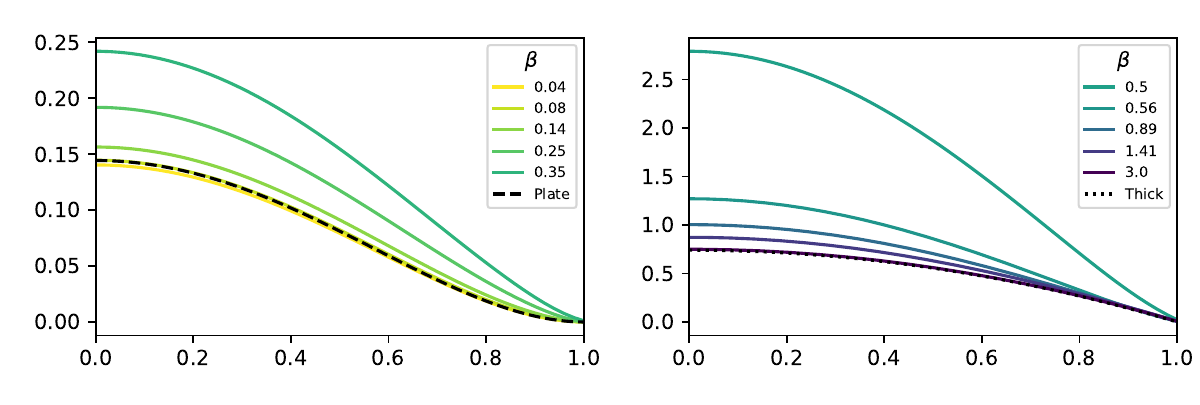}
    \put(28,1){$r$}
    \put(77,1){$r$}

    \put(-0.5, 11){\rotatebox{90}{$\hat\eta(\hat r)\cdot E t_h^3/\Delta p$}}
    \put(50, 11){\rotatebox{90}{$\hat\eta(\hat r)\cdot E t_h/\Delta p$}}
    
    \put(14,31.5){Cupular deformation ($\beta\ll1$)}
    \put(64,31.5){Cupular deformation ($\beta\gg1$)}
        
    \end{overpic}
    \caption{Left: numerically obtained cupular deformation $\eta(r)$ for $\beta<0.5$, scaled by deformation of a plate, and plate deformation (black dashed line) for reference. Right: numerically obtained cupular deformation $\eta(r)$ for $\beta\geq0.5$, scaled by deformation of a plate, and the thick cupula limit \eqref{eq:thick_cupula_limit} (black dotted line) for reference. Colour indicated the dimensionless thickness of the cupula for both panels.}
    \label{fig:deformation_profiles}
\end{figure}

\begin{figure}
    \centering
    \begin{overpic}[width=0.99\textwidth]{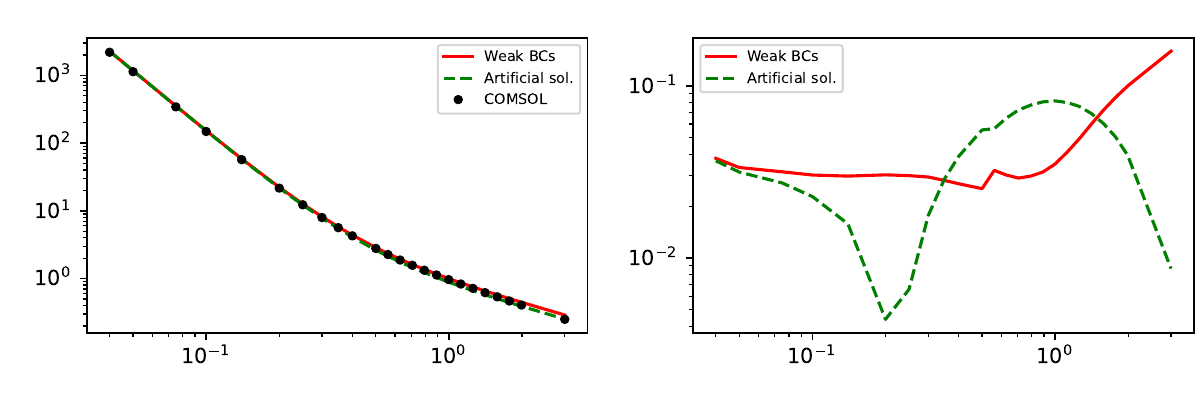}
    \put(28,1){$\beta$}
    \put(78,1){$\beta$}

    \put(-0.5, 12){\rotatebox{90}{$\hat\eta_\mathrm{m}(\beta)\cdot E/\Delta p$}}
    \put(50, 6){\rotatebox{90}{$\vert\hat\eta_\mathrm{m}(\beta) -\hat\eta^\ast_\mathrm{m}(\beta)\vert/\hat\eta^\ast_\mathrm{m}(\beta)$}}
    
    \put(18,31.5){Maximum deflection}
    \put(70,31.5){Relative error}
        
    \end{overpic}
    \caption{Left: evolution of the maximum deformation (at $r=1$) as a function of the thickness for: (a) numerically obtained profiles (black markers), (b) analytical solution obtained using weak boundary conditions (solid red line) and (c) composite solution obtained by summing the thin and thick cupula limits (dashed green line). Right: Relative error between the (a) weak boundary conditions and numerics (solid red line) and (b) composite solution and numerics (dashed green line)}
    \label{fig:solid_numerics_comparison}
\end{figure}

In figure~\ref{fig:deformation_profiles}, we present numerically computed deformation profiles $\eta^\ast(r)$, obtained using COMSOL, for a range of dimensionless thicknesses $\beta \in [0.04, 3]$. The results reveal a clear qualitative transition in the deformation behavior. For small values of $\beta$ (left panel), corresponding to thin a cupula, the deformation resembles that of a clamped elastic plate, exhibiting a characteristic fourth-order polynomial profile with vanishing radial slope at the edge, i.e., $\upd \eta / \upd r = 0$. As the thickness increases, the profile progressively approaches a quadratic form, consistent with the analytical prediction given by equation~\eqref{eq:deformation_profile_with_thickness}. In the left panel, the deformation is non-dimensionalized using the plate scaling $E t_h^3 / \Delta p$, while in the right panel, it is scaled using the membrane-like scaling $E t_h / \Delta p$.
To examine the influence of thickness on the magnitude of deformation, we plot in the left panel of figure~\ref{fig:solid_numerics_comparison} the maximum deformation $\eta_\mathrm{m}(\beta) = \eta(r=1,\beta)$ as a function of the dimensionless thickness $\beta$. The numerical results (black markers) are compared against the analytical solution from \citet{barber_elasticity_2010} (solid red line). For a thin cupula ($\beta \ll 1$), the numerical results exhibit the expected plate-like scaling $\eta_\mathrm{m} \sim \beta^{-3}$, while for a thick cupula ($\beta \gg 1$), the deformation follows the scaling $\eta_\mathrm{m} \sim \beta^{-1}$, consistent with the prediction from equation~\eqref{eq:deformation_profile_with_thickness}. However, the agreement with the analytical solution deteriorates for larger values of $\beta$, which can be attributed to the weak enforcement of boundary conditions at $r=1$. This discrepancy is quantified in the right panel of figure~\ref{fig:solid_numerics_comparison}, where we plot the relative error between the analytical solution $\eta_\mathrm{m}$ and the numerical result $\eta^\ast_\mathrm{m}$ (red line). The relative error remains small for $\beta \ll 1$, but exceeds 10\% for $\beta > 1$.

To improve our analytical solution for larger $\beta\gg1$, we seek an averaged deformation straight from the Navier equations. Starting from the axial stress balance, 
\begin{align}
    \frac{1}{r}\frac{\partial}{\partial r}\left(r\tau_{rz0}\right)+\frac{\partial \tau_{zz0}}{\partial z}=0,
\end{align}
we integrate from $z=-\beta/2$ to $z=\beta/2$, labelling $\langle f(r)\rangle=\beta^{-1}\int_{-\beta/2}^{\beta/2}f(r,z)\upd z$,
\begin{align}
    \frac{1}{r}\frac{\partial}{\partial r}\left(r\langle \tau_{rz0}\rangle\right)+\frac{\Delta p_0}{\beta\kappa}=0,
\end{align}
where we have used the boundary conditions for $\tau_{zz}$ at $z=\pm \beta/2$. The averaged shear stress is thus
\begin{align}
    \langle \tau_{rz0}\rangle=-\frac{r\Delta p_0}{2\beta\kappa}+\cancel{\frac{\text{constant}}{r}},
\end{align}
and from the constitutive law we find that 
\begin{align}
    \tau_{rz}=\frac{\mu_s}{E}\left(\partial_rw_s+\partial_z u_s\right) \Longrightarrow \bar{\tau}_{rz}=\frac{\mu_s}{E}(\partial_r \bar{w}_s+\cancel{[u_s]^{z=\beta/2}_{z=-\beta/2}}),
\end{align}
where we have used symmetry of the displacing mode to cancel the $u_s$ terms. Thus, using the zero displacement condition $w_s(r=a_0)=0$ we obtain
\begin{align}\label{eq:thick_cupula_limit}
    \eta_0(r)=\bar{w}_s(r)=\frac{E\Delta p_0}{4\mu_s \beta \kappa}(a_0^2-r^2)=\frac{\Delta p_0 (1+\nu_s)}{2\beta \kappa}(a_0^2-r^2),
\end{align}
where we observe the required scaling $\beta^{-1}$ from figure~\ref{fig:solid_numerics_comparison}. Comparing this result with the large $\beta$ limit of \eqref{eq:eta_0_definition},
\begin{align}
    \bar{\eta}_0(r)\sim \frac{1}{20}\frac{(12-\nu_s)(1+\nu_s)}{\beta}(a_0^2-r^2),
\end{align}
we see that they differ by a factor of $(12-\nu_s)/20\approx 0.57$ vs $1/2$. This is because we are only enforcing boundary conditions in the weak sense at $r=a_0$, which is a good approximation for $\beta\ll1$ but looses accuracy for thicker cupulas. Hence, an argument could be made that a better solution to the leading order solid problem is actually
\begin{align}
    \eta_0(r)=\frac{3}{16}\frac{(1-\nu_s^2)}{\beta^3}(a_0^2-r^2)^2+\frac{1+\nu_s}{2\beta}(a_0^2-r^2),
\end{align}
as it agrees better with the numerics (see figure~\ref{fig:solid_numerics_comparison}, dashed green line) and has the correct asymptotic dependence (and not just the scaling) for both $\beta\ll1$ and $\beta\gg1$.

\subsubsection{First order solution}
The first order problem for $\boldsymbol{\tau}_1(r,\theta,z,t)$ is
\begin{subequations}
\begin{align}
&\frac{\epsilon}{\kappa}\frac{\rho_s}{\rho}\left[-\Omega_0\mathcal{T}\Omega(t)^2\cos\theta\right]
=\frac{1}{r}\frac{\partial }{\partial r}(r\tau_{rr1})+\frac{1}{r}\frac{\partial \tau_{r\theta1}}{\partial \theta}+\frac{\partial \tau_{rz1}}{\partial z}-\frac{\tau_{\theta\theta1}}{r},\\
&\frac{\epsilon}{\kappa}\frac{\rho_s}{\rho}\left[\Omega_0\mathcal{T}\Omega(t)^2\sin\theta\right]
=\frac{1}{r}\frac{\partial }{\partial r}(r\tau_{r\theta1})+\frac{1}{r}\frac{\partial \tau_{\theta\theta1 }}{\partial \theta}+\frac{\partial \tau_{\theta z1}}{\partial z}+\frac{\tau_{r\theta1}}{r},\\
    &\frac{\epsilon}{\kappa}\frac{\rho_s}{\rho}\left[\frac{\upd \Omega}{\upd t}\right]
    =\frac{1}{r}\frac{\partial }{\partial r}(r\tau_{rz1})+\frac{1}{r}\frac{\partial \tau_{\theta z1}}{\partial \theta}+\frac{\partial \tau_{zz1}}{\partial z},
\end{align}
\end{subequations}
with boundary conditions
\begin{subequations}
\begin{align}\label{eq:thickness_boundary_conditions_F0}
    \tau_{zz1}&=\frac{\Delta p_1^\text{outer}(t)}{2\kappa}+\frac{\Delta p_1^\text{BL}(r,t)}{2\kappa},\quad \tau_{rz1}=0, \quad z=\beta/2,\\
    \tau_{zz1}&=-\frac{\Delta p_1^\text{outer}(t)}{2\kappa}-\frac{\Delta p_1^\text{BL}(r,t)}{2\kappa},\quad \tau_{rz1}=0, \quad z=-\beta/2,\\
    u_{s1}&=v_{s1}=w_{s1}=0,\quad r=1.
\end{align}
\end{subequations}
The Navier equations at this order have a conservative body force that may be written as the gradient of the following potential
\begin{align}
\begin{split}
    &V(r,\theta,z,t)=V_\text{euler}(z,t)+V_\text{centrifugal}(r,\theta,t),\\
    &V_\text{euler}(z,t)=-\frac{\rho_s}{\rho \kappa}\dot{\Omega}(t)z,\quad V_\text{centrifugal}(r,\theta,t)=\frac{\rho_s \Omega_0\mathcal{T}}{\kappa\rho}\Omega(t)^2 r\cos\theta.
    \end{split}
\end{align}
We first consider the solution for $\Delta p_1^\text{BL}=0$. To this end we introduce three pairs of potentials: $\{\phi_1^u,\omega_1^u\}$ to deal with $\Delta p_1^\text{outer}$ (this is the same as the leading order problem), $\{\phi_1^e,\omega_1^e\}$ an axisymmetric potential to deal with $V_\text{euler}$ and $\{\phi_1^c,\omega_1^c\}$, a non-axisymmetric potential to deal with $V_\text{centrifugal}$. The potentials for $V_\text{euler}$ and $V_\text{centrifugal}$ are used with homogeneous boundary conditions, and we use weak boundary conditions on $r=a$ as in the leading order problem:
\begin{align}
\int_{-\beta/2}^{\beta/2}u_{s1}(r=1,z)~\upd z=\int_{-\beta/2}^{\beta/2} zu_{s1}(r=1,z)\upd z=\int_{-\beta/2}^{\beta/2} w_{s1}(r=1,z)\upd z=0.
\end{align}
Therefore, we may recycle the leading order solution for $\{\phi_1^u,\omega_1^u\}$, giving us 
\begin{align}
    \eta_1^u(r,t)=\frac{\Delta p_1(t)}{\kappa}\bar{\eta}_0(r;\beta).
\end{align}
\cite{barber_elasticity_2010} shows that for a conservative body force a potential $\phi$ must satisfy $\nabla^2\phi=(1-2\nu_s)V/(1-\nu_s)$, from where we deduce particular solutions for $\phi_1^e$ and $\phi_1^c$, which allows us to write 
\begin{align}
    \phi_1^e=-\frac{\rho_s}{\rho\kappa}\frac{(1-2\nu_s)}{1-\nu_s}\dot{\Omega}(t)\frac{z^3}{6}+\sum_{j=1}^5A_jP_j(r,z),\quad  \omega_1^e = \sum_{j=1}^4B_jP_j(r,z).
\end{align}
Using the strong and weak boundary conditions to formulate a linear system for the 9 constants we find
\begin{subequations}
    \begin{align}
        \eta_1^e(r,z,t)=\frac{1}{\beta}\int_{-\beta/2}^{\beta/2}w_{s1}^e(r,z,t)\,\upd z = \frac{\rho_s}{\rho \kappa}\dot{\Omega}(t)\bar{\eta}_1^e(r),\\
        \bar{\eta}_1^e(r)=\frac{1}{20}(12-\nu_s)(1+\nu_s)(1-r^2)+\frac{3}{16}\frac{(1-\nu_s^2)}{\beta^2}(1-r^2)^2,\\
        \alpha_1^e(\beta)=\int_0^1 r\bar{\eta}_1^e(r;\beta)\upd r=\frac{(1-\nu_s^2)}{32\beta^3}+\frac{(12-\nu_s)(1+\nu_s)}{80\beta}.
    \end{align}
\end{subequations}
The solution for $V_\text{centrifugal}$ is more convoluted, as it involves a non-axisymmetric deformation. Although it is possible to find the deflection it is sufficient for us to note that the solution is of the form $w_{s1}^c,u_{s1}^c,\tau_{rr1}^a,\tau_{\theta\theta1}^c,\tau_{zz1}^c,\tau_{rz1}^c\sim\cos\theta$ and $v_{s1}^c,\tau_{r\theta1}^c,\tau_{\theta z1}^c\sim \sin\theta$. Therefore, we have
\begin{align}
    \eta_1^c(r,\theta,t)=\frac{1}{\beta}\int_{-\beta/2}^{\beta/2}w_{s1}^c(r,\theta,z,t)\upd z=\frac{\rho_s\Omega_0\mathcal{T}}{\rho \kappa}\Omega(t)^2\bar{\eta}_1^c(r)\cos\theta,
\end{align}
for some function $\bar{\eta}_1^c(r)$. The contribution to the flux is thus
\begin{align}
Q_1^c=\int_0^{2\pi}\int_0^1r\eta_1^c(r,\theta,t)\upd r\upd\theta\sim \int_0^{2\pi}\cos\theta\upd \theta=0.
\end{align}
Therefore the non-axisymmetric solution does not contribute to the fluid solid coupling.

\textbf{Cupular deflection due to the boundary layer pressure jump.}
In Appendix~\ref{sec:boundary_layer} we show the pressure jump due to the presence of a cupular boundary layer is of the form
\begin{align}
    \Delta p_1^{\text{BL}}(r,t)=f_1(t)g_1(\beta)\mathrm{Re}\left\{\sum_{n=1}^\infty \bar{A}_nJ_0(\mu_nr)\right\},
\end{align}
where the complex eigenvalues satisfy $J_2(\mu_n)J_0(\mu_n)-J_1(\mu_n)^2=0$ and the constants $\bar{A}_n$ are independent of $t$ and $\beta$ (for details see Appendix~\ref{sec:boundary_layer}). The precise functional form of $f_1(t)$ and $g_1(\beta)$ is not relevant here, as the solid problem is linear, so it suffices to solve the problem for $\bar{\eta}_{1n}^\text{BL}(r;\beta)$ with boundary condition
\begin{align}
    \tau_{zz}\vert_{z=-\beta/2}=J_0(\mu_n r),
\end{align}
and the full solution can be obtained by invoking superposition, 
\begin{align}
\eta_1^\text{BL}(r,t;\beta)=f_1(t)g_1(\beta)\mathrm{Re}\left\{\sum_{n=1}^\infty\bar{A}_n \bar{\eta}_{1n}^\text{BL}(r;\beta).\right\}
\end{align}
An approach using potentials to account for the cupular thickness is complicated for this radially dependent pressure jump, but motivated by the success of the artificial solution in \S\ref{sec:thick_cupula_limit}, we may solve the problem in the $\beta\ll1$ and $\beta\gg1$ limits and obtain an approximate solution for arbitrary thickness by combining both solutions. The thin-cupula regime is given by a plate equation, which under our dimensionless variables reads
\begin{align}
    \frac{\beta^3}{12(1-\nu_s^2)}\frac{1}{r}\frac{\upd }{\upd r}\left(r\frac{\upd }{\upd r}\left[\frac{1}{r}\frac{\upd }{\upd r}\left(r\frac{\upd \bar{\eta}_1^\text{BL}}{\upd r}\right)\right]\right)=J_0(\mu_n r),
\end{align}
which has a clamped ($\eta = \upd \eta/\upd r=0$ at $r=1$) solution:
\begin{align}
    \bar{\eta}_{1n}^\text{BL}(r;\beta)=\frac{12(1-\nu_s^2)}{\beta^3}\left[\frac{(r^2-1)J_1(\mu_n)}{2\mu_n^3}+\frac{1}{\mu_n^4}(J_0(\mu_n r)-J_0(\mu_n))\right].
\end{align}
In the thick cupula regime, following \S\ref{sec:thick_cupula_limit}, we have 
\begin{align}
    \frac{1}{r}\frac{\partial}{\partial r}\left(r\bar{\tau}_{rz1}^\text{BL}\right)+\frac{J_0(\mu_n r)}{\beta}=0,\quad \Longrightarrow\frac{1}{r}\frac{\partial}{\partial r}\left(r \frac{\partial \bar{\eta}_{1n}^\text{BL}}{\partial r}\right)=-\frac{E}{\mu_s\beta}J_0(\mu_n r),
\end{align}
with solution 
\begin{align}
\bar{\eta}_{1n}^\text{BL}(r;\beta)=\frac{2(1+\nu_s)}{\beta\mu_n^2}\left[J_0(\mu_n r)-J_0(\mu_n)\right].
\end{align}
Therefore, we may write the solution for $\eta_1^\text{BL}(r,t;\beta)$ by combining the thin and thick cupula solutions additively:
\begin{align}
\begin{split}
    &\eta_1^\text{BL}(r,t;\beta)=f_1(t)g_1(\beta)\bar{\eta}_1^\text{BL}(r;\beta),\\
    &\bar{\eta}_1^\text{BL}(r;\beta)=\frac{12(1-\nu_s^2)}{\beta^3}\bar{\eta}_1^\text{BL, thin}(r)+\frac{2(1+\nu_s)}{\beta}\bar{\eta}_1^\text{BL, thick}(r),\\
    &\bar{\eta}_1^\text{BL, thin}(r)=\mathrm{Re}\left\{\sum_{n=1}^\infty\bar{A}_n\left[\frac{(r^2-1)J_1(\mu_n)}{2\mu_n^3}+\frac{1}{\mu_n^4}(J_0(\mu_n r)-J_0(\mu_n))\right]\right\}, \\
    &\bar{\eta}_1^\text{BL, thick}(r)=\mathrm{Re}\left\{\sum_{n=1}^\infty\frac{\bar{A}_n}{\mu_n^2}\left[J_0(\mu_n r)-J_0(\mu_n)\right]\right\}.
    \end{split}
\end{align}
more importantly, the coefficient $\alpha_1^\text{BL}(\beta)=\int_0^{1}r \bar{\eta}_1^\text{BL}(r)\upd r$ is
\begin{align}
\begin{split}
    &\alpha_1^\text{BL}(\beta)=\frac{12(1-\nu_s^2)}{\beta^3}\alpha_1^\text{BL, thin}+\frac{2(1+\nu_s)}{\beta}\alpha_1^\text{BL, thick},\\
    &\alpha_1^\text{BL, thin}=\mathrm{Re}\left\{\sum_{n=1}^\infty\bar{A}_n\left[\frac{J_2(\mu_n)}{2\mu_n^4}-\frac{J_1(\mu_n)}{8\mu_n^3}\right]\right\}\approx 0.00408,\\
    &\alpha_1^\text{BL, thick}=\mathrm{Re}\left\{\sum_{n=1}^\infty\bar{A}_n \frac{J_2(\mu_n)}{2\mu_n^2}\right\}\approx 8.5525\times 10^{-9},
    \end{split}
\end{align}
where we have used the $\bar{A}_n$ computed in Appendix~\ref{sec:boundary_layer} to compute $\alpha_1^\text{BL, thin}$ and $\alpha_1^\text{BL, thick}$.

%%%%%%%%%%%%%%%%%%%%%%%

\section{Numerical procedure for integro-differential equations}\label{sec:app_integro_ODE}

When the Stokes number of the flow is no longer negligible, the deformation of the cupula satisfies \eqref{eq:3D_integral_equation_Stokes}, which is a Volterra integro-integral equation\citep{polianin_handbook_1998}, and so  may be solved numerically using the trapezoidal method. For small values of the Stokes number this solution procedure requires an increasingly fine temporal discretization, making the problem computationally intensive. Therefore, an alternative numerical scheme is required.

Integral equations with exponential kernels can be transformed into systems of ODEs by introducing additional variables \citep{wazwaz_volterra_2011}; although the kernel $\mathcal{K}(x,\St)$ in \eqref{eq:3D_integral_equation_Stokes} is not strictly exponential, it may be seen as a linear combination of exponential Kernels. To this end, we may define a sequence of auxiliary variables
\begin{align}
    z_n(t) = \frac{1}{\St\lambda_n^2}\int_0^t\left(\dot{\Omega}(t)+\frac{1}{2\pi}\Delta p_0\right)e^{-\lambda_n^2(t-\tau)/\St}~\upd \tau,\quad n=0,\dots, N-1.
\end{align}
Upon truncation of the infinite series \eqref{eq:3D_integral_equation_Stokes} reads
\begin{align}
    \alpha_0(\beta)\frac{1}{\kappa}\frac{\upd \Delta p_0}{\upd t}=-2\sum_{n=0}^{N-1} z_n.
\end{align}
Considering $\frac{\upd z_n}{\upd t}$ and differentiating under the integral sign we find that
\begin{align}
    \frac{1}{\kappa}\frac{\upd z_n}{\upd t}=\frac{1}{\St\lambda_n^2}\left(\dot{\Omega}(t)+\frac{1}{2\pi}\Delta p_0\right)-\frac{\lambda_n^2}{\St}z_n.
\end{align}
Therefore we have a system of $N+1$ differential equations for the $N+1$ unknowns, which may be solved efficiently for any value of the Stokes number. 

\section{Laplace transform approach for finite Stokes number}\label{sec:laplace_transform}
The general equation determining the shape of the cupular deflection is \eqref{eq:reduced_full_system}, which we Laplace transform in time to obtain:
 \begin{subequations}
     \begin{align}
         \int_0^{a_0}r\sigma \tilde{\eta}_0\upd r &= -2a(s)^2\frac{1}{\St}\sum_{n=1}^\infty\frac{1}{\lambda_n^2}\left(\tilde{\Omega}(\sigma)\sigma+ \frac{\partial \tilde{p}_0}{\partial s}\right)\frac{1}{\sigma +\lambda_n^2/(a^2\St)},\\
         &\tilde{\eta}_0(r,\sigma)=\Delta \tilde{p}_0(\sigma)/\kappa    \bar{\eta}_0(r)\label{eq:pressure_deflection_laplace_transformed}
         \end{align}
 \end{subequations}
(Here the convolution theorem has been used to compute the transform of the convolution integral.) We may now isolate the pressure gradient in the first equation as it may be factored out of the sum
\begin{equation}
    \tilde{\Omega}(\sigma)\sigma+\frac{\partial \tilde{p}_0}{\partial s}= -\frac{\St \sigma}{2a^2}\int_0^{a_0}r \tilde{\eta}_0~\upd r \frac{1}{\sum_{n=1}^\infty\lambda_n^{-2}\left(\sigma +\lambda_n^2/(a^2\St)\right)^{-1}}. 
\end{equation}
Hence, after integrating along the length of the duct, 
\begin{equation}
    2\pi\tilde{\Omega}(\sigma)\sigma+ \Delta \tilde{p}_0= -\frac{ \sigma}{2}\int_0^{a_0}r \tilde{\eta}_0\upd r \int_0^{2\pi}\frac{\upd s}{a(s)^4\sum_{n=1}^\infty\lambda_n^{-2}\left[a(s)^2\St\sigma +\lambda_n^2\right]^{-1}}. 
\end{equation}
Substituting the pressure jump using \eqref{eq:pressure_deflection_laplace_transformed} leads to a single equation for $\Delta\tilde{ p}_0$, 
\begin{equation}
    2\pi\tilde{\Omega}(\sigma)\sigma+ \Delta \tilde{p}_0= -\frac{ \sigma}{2}\frac{\alpha_0(\beta)\Delta \tilde{p}_0}{\kappa}\frac{2\pi}{\tilde{\mathcal{K}}(\sigma;\St) },
\end{equation}
where the transformed kernel is 
\begin{equation}
    \tilde{\mathcal{K}}(\sigma;\St)=2\pi\left(\int_0^{2\pi}\frac{\upd s}{a(s)^4\sum_{n=1}^\infty\lambda_n^{-2}\left[a(s)^2\St\sigma +\lambda_n^2\right]^{-1}}\right)^{-1},
\end{equation}
and $\alpha_0(\beta)$ is given in \eqref{eq:alpha_0_definition}.
Multiplication by the transformed kernel, followed by the inversion of the transform and the application of the convolution theorem leads to 
\begin{equation}
    \frac{1}{\kappa}\alpha_0(\beta)\frac{\upd \Delta p_0}{\upd t}=-2\int_0^t\left(\dot{\Omega}(\tau)+\frac{1}{2\pi}\Delta p_0(\tau)\right)\mathcal{K}(t-\tau;\St)\upd \tau,
\end{equation}
where $\mathcal{K}(t;\mathrm{St})=\mathcal{L}^{-1}[\tilde{\mathcal{K}}(\sigma;\St)]$ is the kernel.

This calculation is crucial because it systematically reduces the governing equation for the cupular deflection into a solvable integral equation by leveraging the Laplace transform. By transforming the original time-dependent equations, the problem is converted into an algebraic form where the pressure gradient can be explicitly isolated, and integrated in space to obtain the pressure jump. Furthermore, inverting the transform and applying the convolution theorem ultimately yields an explicit time-domain equation governing the evolution of $\eta_0$. This final equation is particularly useful, as it expresses the cupular deflection in terms of a convolution integral. This approach allows for the computation of the solution in arbitrary domains, linking the problem to that solved in a simpler domain via the transformed kernel.

\section{A thick cupula experiences a first order correction}\label{sec:thickness_correction}

We can estimate effect of the cupula's thickness on the leading order pressure jump by removing the cupula from the integrations along the duct. In particular, the pressure jump is now defied as:
\begin{align}
    \Delta p=\int_{\epsilon\beta/2}^{2\pi-\epsilon\beta/2}\frac{\partial p}{\partial s}\upd s =p(s=2\pi-\epsilon \beta/2)-p(s=\epsilon \beta/2),%=p(s=2\pi)-p(s=0)-\epsilon\beta\left(\left.\frac{\partial p}{\partial s}\right\vert_{s=2\pi}+\left.\frac{\partial p}{\partial s}\right\vert_{s=0}\right)+\mathcal{O}(\epsilon^2).
\end{align}
where $\beta=t_h/a$ is the dimensionless cupular thickness. We treat $\epsilon\beta$ as its own asymptotic scale in the expansion, so that $\Delta p_0=\int_{\epsilon\beta/2}^{2\pi-\epsilon\beta/2}\frac{\partial p_0}{\partial s}\upd s$ and similar for higher order, without considering the integration bounds for the asymptotic series. We show here the modifications for the leading order problem for $\Delta p_0(t)$.
\eqref{eq:pressure_grad_flux_LO} is modified to 
\begin{align}
    \tilde{I}_4Q_0&=-\frac{\pi}{8}(\dot{\Omega}(t)(2\pi-\epsilon \beta)+\Delta p_0),
\end{align}
The remainder of the calculation is the same as in the main text, leading to an alternate form for \eqref{eq:3D_equation_for_delta_p}: 
\begin{align}
    \frac{\alpha}{\kappa}\frac{\upd \Delta p_0}{\upd t}&=-\frac{1}{16 \tilde{I}_4}(\dot{\Omega}(t)(2\pi-\epsilon\beta)+\Delta p_0).\label{eq:pressure_jump_leading_order_with_thickness}
\end{align}
Hence, we have obtained (slightly) modified reduced order equations for the pressure jump that account for the cupular thickness $\epsilon\beta$, and we still find the correction vanishes, $\Delta p_1(t)=0$

\section{Boundary layer calculation}\label{sec:boundary_layer}

Close to the cupula the leading order velocity field $w_0$ (a quadratic in $r$) has to adjust to the axially averaged profile of the cupula, $\partial\eta/\partial t$. This gives rise to a boundary layer close to the cupula, first studied by \cite{damiano_singular_1996}. The pressure gradient required to adjust to the cupular flow profile in a distance $\sim\epsilon$ gives rise to an $\mathcal{O}(\epsilon)$ correction to the pressure jump.

Here, we present a simpler method for obtaining the correction due to the boundary layer when $\St\ll1$. We introduce rescaled arc length coordinates $ S^+=\epsilon^{-\alpha} s$ and $S^-=\epsilon^{-\alpha}(s-2\pi) $ (on one side of the cupula $S^+\geq0$, while, on the other, $S^-\leq0$). Following \citep{damiano_singular_1996} we find the distinguished limit for $\alpha=1$. For notational convenience we use the letter $S$ to refer to $S^\pm$ interchangeably; the boundary layer equations will be identical and the context makes it clear which we mean. For example taking the limit as $S\rightarrow\infty$ means $S^+\rightarrow\infty$ and conversely $S\rightarrow-\infty$ means $S^-\rightarrow-\infty$. 
We either focus on the case or a perfect torus or located the cupula at the point of maximum enlargement of the canal. Under those conditions, $a(s)=a(0)+\mathcal{O}(\epsilon^2)$ close to the cupula, so that the canal radius changes slowly. For the purpose of this calculation, we take $a(0)=1$ without loss of generality. Moreovoer, we consider a thin cupula $\beta\rightarrow 0$, so that the matching conditions simplify 
\begin{align}\begin{split}
    &w_0(r,s=\epsilon\beta+\delta \eta,t)=w_0(r,s=2\pi-\epsilon\beta+\delta\eta,t)=\int_{-\beta/2}^{\beta/2}\frac{\partial w_{s0}}{\partial t}(r,z,t)\upd z=\frac{\partial \eta}{\partial t}\\
    &\Longrightarrow w_0(r,s=0,t)=w_0(r,s=2\pi,t)=\frac{\partial \eta}{\partial t}.
    \end{split}
\end{align}
where we have used $\delta=a^2\Omega_0/\nu\ll1$ and $\epsilon\beta\rightarrow0$ (if the latter is not satisfied it is not a problem for the leading order computation that follows, as we only consider the leading order equations). We further note that for consistency with the weak boundary conditions used to obtain the solid solution, the matching involves axially averaged solid deformations. Similarly, for the radial velocity,
\begin{align*}
u_0(r,s=0,t)=u_0(r,s=2\pi,t)=\int_{-\beta/2}^{\beta/2}u_{s0}(r,z,t)\upd z=0
\end{align*}
The scalings for the boundary layer are
\begin{align}\label{eq:BL_scalings}
\begin{split}
     u&=\frac{1}{\epsilon}\left(U_0(r,\theta,S,t)+\epsilon U_1(r,\theta,S,t)+\cdots \right)\\
     v &= \frac{1}{\epsilon}\left(V_0(r,\theta,S,t)+\epsilon V_1(r,\theta,S,t)+\cdots\right)\\
     w&=\,\,\,W_0(r,\theta,S,t)+\epsilon W_1(r,\theta,S,t)+\cdots\\
     p &= P_0(r,\theta,S,t)+\epsilon P_1(r,\theta,S,t)+\epsilon^2 P_2(r,\theta,S,t)\cdots 
     \end{split}
\end{align}
The $\epsilon^{-1}$ scaling for the cross-sectional velocities is informed by the continuity equation \eqref{eq:toroidal_continuity_non_dimensional}. We will focus on the leading order matching, for which we require the following matching between layers for the velocities
\begin{align}
    \lim_{S^\pm \rightarrow \pm \infty}U_0(r,S,t)=0,\quad\lim_{S^\pm \rightarrow \pm \infty}V_0(r,S,t)=0,\\
    \lim_{S^\pm \rightarrow \pm \infty}W_0(r,S,t)=w_0(r,s=0,t)=\frac{\pi}{2I_4a(0)^4}\left[\dot{\Omega}(t)+\frac{\Delta p_0}{2\pi}\right](a(0)^2-r^2).
\end{align}
Substituting \eqref{eq:BL_scalings} into \eqref{eq:fluid_problem_dimensionless}, we find that the $\mathcal{O}(\epsilon^{-1})$ equations, $\partial_rP_0=\partial_{\theta}P_0=\partial_SP_0=0$, imply that the pressure is spatially constant across the boundary layers, and in particular
\begin{subequations}
\begin{align}
&P_0(S^+=0,t)=P_0(S^+=\infty,t)=p_0(s=0, t),\\
    &P_0(S^-=0,t)=P_0(S^-=-\infty,t)=p_0(s=2\pi,t),
\end{align}\end{subequations}
and in particular the leading order pressure jump is the same, i.e. $\Delta P_0(t)=P_0(S^-=0,t)-\Delta P_0(S^+=0,t)=\Delta p_0^\mathrm{outer}(t)$, in agreement with \cite{damiano_singular_1996}.
\subsection{Boundary layer equations}
At the following order, substitution of \eqref{eq:BL_scalings} into the Navier-Stokes equations \eqref{eq:fluid_problem_dimensionless} after incorporating the identically zero term $\boldsymbol{\nabla}(\boldsymbol\nabla\cdot\boldsymbol{u})$, yields \cite[see ][where the equations are given with different scalings]{damiano_singular_1996}
\begin{subequations}
\begin{align}
    0&=\frac{1}{r}\frac{\partial }{\partial r}\left(r U_0\right)+\frac{1}{r}\frac{\partial V_0}{\partial\theta} +\frac{\partial W}{\partial S},\\
    \frac{\partial P_1}{\partial r}&=\frac{1}{r}\frac{\partial }{\partial r}\left(r\frac{\partial U_0}{\partial r}\right)+\frac{1}{r^2}\frac{\partial^2 U_0}{\partial\theta^2}+\frac{\partial^2U_0}{\partial S^2}-\frac{2}{r^2}\frac{\partial V_0}{\partial\theta}-\frac{U_0}{r^2},
    \\
    \frac{1}{r}\frac{\partial P_1}{\partial\theta}&=\frac{1}{r}\frac{\partial }{\partial r}\left(r\frac{\partial V_0}{\partial r}\right)+\frac{1}{r^2}\frac{\partial^2 V_0}{\partial\theta^2}+\frac{\partial^2V_0}{\partial S^2}+\frac{2}{r^2}\frac{\partial U_0}{\partial \theta}-\frac{V_0}{r^2},
    \\
    \dot\Omega(t)+\frac{\partial P_1}{\partial S}&=\frac{1}{r}\frac{\partial }{\partial r}\left(r\frac{\partial W_0}{\partial r}\right)+\frac{1}{r^2}\frac{\partial^2 W_0}{\partial\theta^2}+\frac{\partial^2W_0}{\partial S^2}.\label{eq:axial_momentum_BL}
\end{align}
\end{subequations}
we recognize the Stokes equations in cylindrical coordinates. As the outer solution is axisymmetric, we seek an axisymmetric boundary layer velocity profile, with $V_0=0$ and $\partial_\theta=0$ for all other variables. We remove the inhomogeneous $S\rightarrow\pm\infty$ matching by introducing $\bar{W}_0=W_0-w_0(r,s=0,t)=W_0+\frac{1}{4}(\dot\Omega(t)+\partial p/\partial s\vert_{s=0})(a(0)^2-r^2)$, so that the axial momentum equation \eqref{eq:axial_momentum_BL} reads
\begin{align}\label{eq:axial_momentum_BL_W_bar}
    -\left.\frac{\partial p}{\partial s}\right\vert_{s=0}+\frac{\partial P_1}{\partial S}&=\frac{1}{r}\frac{\partial }{\partial r}\left(r\frac{\partial \bar{W}_0}{\partial r}\right)+\frac{\partial^2\bar{W}_0}{\partial S^2}.
\end{align}
The conditions at the wall of the cupula transform to
\begin{align}
    \bar{W}_0(r,S=0,t)=\frac{\partial \eta_0}{\partial t}-w_0(r,s=0,t),\quad U_0(r,S=0,t)=0,
\end{align}
alongside homogeneous matching conditions, $\Bar{W_0},U_0\rightarrow0$ as $S\rightarrow\pm\infty$. It is convenient to introduce a Stokes streamfunction $\psi(r,S,t)$ such that \citep{davis_stokes_1990}
\begin{subequations}
    \begin{align}
    U_0=-\frac{1}{r}\frac{\partial \psi}{\partial S},\quad \bar{W}_0=\frac{1}{r}\frac{\partial \psi}{\partial r},\\
    L_{-1}(L_{-1}\psi)=0,\label{eq:BL_eigenproblem}\\
    L_1 = \frac{\partial^2}{\partial r^2}-\frac{1}{r}\frac{\partial }{\partial r}+\frac{\partial^2}{\partial S^2},
\end{align}
\end{subequations}
with $L_{-1}$sometimes known in the literature as the Stokes operator \citep{payne_stokes_1960}. 
\subsection{Eigenfunction expansion}
Seeking a separable solution of \eqref{eq:BL_eigenproblem} that decays as $\vert S\vert\rightarrow\infty$ of the form $\psi(r,S)=\psi_n(r)e^{-\pm \mu_n S}$, we find a family of solutions which satisfies the no-slip condition $\psi=\partial_r\psi=0$ at the walls, $r=1$ while $\frac{\psi}{r}$ and $\frac{\partial_r \psi}{r}$ are bounded as $r\rightarrow 0$. \cite{katopodes_piston_2000} computed these functions:
\begin{equation}\label{eq:3D_BL_eigenfunctions}
\psi_n(r)=\frac{J_2(\mu_n)}{2\mu_n}\left[\frac{r J_1(\mu_n)}{J_1(\mu_n)}-\frac{r^2J_2(\mu_n r)}{J_2(\mu_n)}\right],%\quad\psi_n(r)=r [r J_1(\mu_n)J_0(\mu_n r)-J_0(\lambda_n)J_1(\lambda_n r)],
\end{equation}
where the eigenvalues $\mu_n\in\mathbb{C}$ are solutions of the transcendental equation located in the first quadrant \cite[see][for details]{katopodes_piston_2000,davis_stokes_1990}
\begin{align} 
J_1(\mu_n)^2&=J_0(\mu_n)J_2(\mu_n),%\quad\lambda_n  \left[J_0(\lambda_n)^2+J_1(\lambda_n)^2\right]-2J_0(\lambda_n)J_1(\lambda_n)=0,
\end{align}
and we express the streamfunction $\psi$ as a superposition of eigenfunctions,
\begin{equation}
    \psi^\pm(r,S,t) = \mathrm{Re}\left\{\sum_{n=0}^\infty A_n(t;\beta) \psi_n(r)e^{\mp\mu_n S}\right\},
\end{equation}
where $\mathrm{Re}\{\cdot\}$ denotes the real part and $A_n(t;\beta)\in\mathbb{C}$. We can determine the coefficients by enforcing the velocities at the cupula are as required:
\begin{subequations}
\begin{align}
    U_0(r,S=0,t) &= -\frac{1}{r}\frac{\partial \psi}{\partial S} =\mp \mathrm{Re}\left\{\sum_{n=0}^\infty  \mu_n A_n(t) \frac{\psi_n(r)}{r}\right\}=0,\label{eq:BL_U_boundary_condition}\\
    \bar{W}_0(r,S=0,t)&=\frac{1}{r}\frac{\partial \psi}{\partial r}= \mathrm{Re}\left\{\sum_{n=0}^\infty  A_n(t) \frac{\psi_n'(r)}{r}\right\}=\frac{\partial \eta_0}{\partial t}-w_0(r,s=0,t).\label{eq:BL_W_bar_boundary_condition_derivative}
    \end{align}
\end{subequations}
\subsection{Boundary layer forcing}
Before computing the $A_n(t)$ it is informative to write the forcing in \eqref{eq:BL_W_bar_boundary_condition_derivative} as
\begin{align}\label{eq:BL_forcing}
    \frac{\partial \eta_0}{\partial t}-w_0(r,s=0,t)=\frac{\upd \Delta p_0}{\upd t}\frac{\bar{\eta}_0(r)}{\kappa}-w_0(r,s=0,t).
\end{align}
Substituting $\upd \Delta p_0/\upd t$  from \eqref{eq:3D_equation_for_delta_p}, then \eqref{eq:BL_forcing} simplifies to 
\begin{align}
\begin{split}
\frac{\bar{\eta}_0(r)}{\kappa}\frac{\kappa}{16I_4\alpha_0(\beta)}(2\pi \dot{\Omega}(t)+\Delta p_0) -\frac{\pi}{2I_4a(0)^4}\left[\dot{\Omega}(t)+\frac{\Delta p_0}{2\pi}\right](a(0)^2-r^2)\\=\frac{\pi}{2I_4}\left(\dot{\Omega}(t)+\frac{\Delta p_0}{2\pi}\right)\left[\frac{\bar{\eta}_0(r;\beta)}{4\alpha_0(\beta)}+\frac{1}{a(0)^4}\left(r^2-a(0)^2\right)\right]\equiv f_1(t)f_2(r)f_3(\beta,\nu_s).
\end{split}
\end{align}
This means we have decomposed $\partial\eta/\partial t-w_0(r,s=0,t)=f_1(t)f_2(r)f_3(\beta,\nu_s)$ multiplicatively in a fashion that is valid for all values of the relative stiffness $\kappa$, with
\begin{subequations}
\begin{align}
    f_1(t)&=\frac{\pi}{2I_4}\left(\dot{\Omega}(t)+\frac{\Delta p_0}{2\pi}\right),\\
    f_2(r)&=1-4r^2+3r^4,\\ f_3(\beta,\nu_s)&=\frac{5(1-\nu_s)}{4\beta^2(12-\nu_s)+10(1-\nu_s)},
\end{align}\end{subequations}
where we have used the solution for using the solution for $\bar{\eta}_0(r;\beta)$ from \S\ref{sec:solid_mechanics}.

\subsection{Computation of coefficients}
\begin{figure}
    \centering
    \begin{overpic}[width=0.99\textwidth]{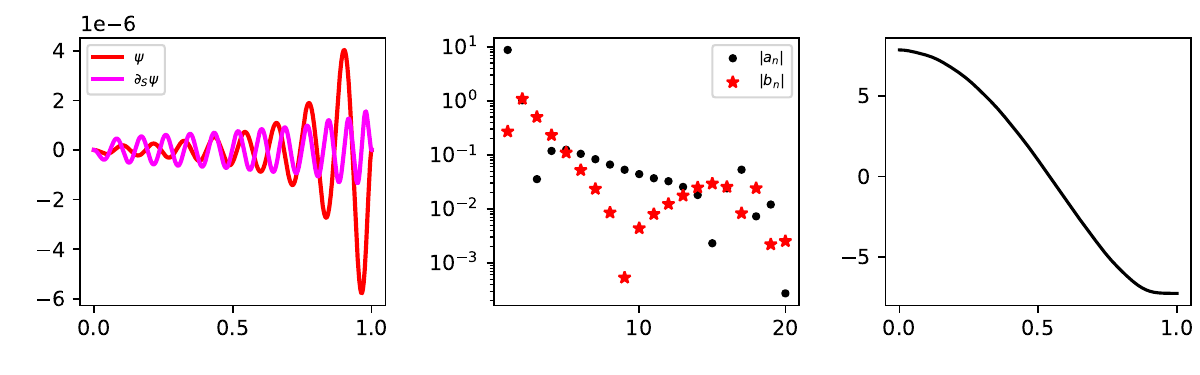}
    \put(19,1){$r$}
    \put(86.5,1){$r$}
    \put(53,1){$n$}

    \put(0., 4){\rotatebox{90}{$\psi(r,S=0),\,\partial_S\psi(r,S=0)$}}
    %\put(34, 14){\rotatebox{90}{$\partial_S\psi(r)$}}
    \put(32.7, 13){\rotatebox{90}{$\vert a_n\vert,$ $\vert b_n\vert$}}
    \put(67.3,13.5){\rotatebox{90}{$\Delta p_1^\text{BL}(r)$}}
    
    \put(12.5,29){Least squares error}
    \put(40,29){Convergence of coefficients}
    \put(75.9,29){BL pressure correction}
        
    \end{overpic}
    \caption{Left panel: least squares error for \eqref{eq:BL_U_boundary_condition} and \eqref{eq:BL_W_bar_boundary_condition}, with the series truncated at $N=20$. Center panel: convergence of the coefficients $a_n$ and $b_n$. Right panel: plot of $\Delta p_1^\text{BL}(r)$ as given in \eqref{eq:boundary_layer_pressure_jump} for $f_1(t)=g_1(\beta)=1$.}
    \label{fig:boundary_layer_fit}
\end{figure}

The above decomposition and the linearity of the Stokes equations implies that  it suffices to compute the coefficients $A_n(t;\beta)$ once say for $f_1(t)=1$ and $f_3(\beta,\nu_s)=1$, obtaining $\bar{A}_n$. The coefficients for other values of $f_1(t)$ and $f_3(\beta,\nu_s)$ are simply $A_n(t)=f_1(t)f_3(\beta,\nu_s)\bar{A}_n$, where
\begin{align}\label{eq:A_bar_n_def}
    \sum_{n=1}^\infty \bar{A}_n \frac{\psi_n'(r)}{r}=f_2(r)=1-4r^2+3r^4.
\end{align}
For convenience, we can integrate \eqref{eq:A_bar_n_def} once in $r$ to obtain
\begin{align}\label{eq:BL_W_bar_boundary_condition}
    %\psi(r,S=0^\pm, t)=\mathrm{Re}\left\{\sum_{n=1}^\infty A_n(t)\psi_n(r)\right\}=\textcolor{red}{-}\int_r^1 \left(\frac{\partial\eta_0}{\partial t}-w_0(r,s=0,t)\right)r\upd r.\\
    \sum_{n=1}^\infty \bar{A}_n\psi_n(r)=-\int_r^1rf_2(r)\upd r=+\frac{1}{2}r^2(1-r^2)^2.
\end{align}
To avoid the convergence issues associated with biorthogonality relationships as discussed in \cite{spence_class_1983}, we find $\bar{A}_n$ by evaluating \eqref{eq:BL_U_boundary_condition} and \eqref{eq:BL_W_bar_boundary_condition} at a discrete set of points and performing a linear regression for $\bar{A}_n\in\mathbb{C}$. Details for the least squares fit and the convergence of $\bar{A}_n$ are given in the left and middle panels of figure~\ref{fig:boundary_layer_fit} for a representative fit where the series are truncated at $N=20$. We remark that the coefficients are the same for either side of the cupula, and for any value of the thickness $\beta$.

\subsection{Boundary layer pressure contribution}
Once they have been obtained, we can compute the axial velocity at either side of the cupula as
\begin{align}
    \bar W_0^\pm(r,S,t)=\frac{1}{r}\frac{\partial \psi^\pm}{\partial r}=\frac{1}{r}\mathrm{Re}\left\{\sum_{n=1}^\infty A_n(t)\psi_n'(r) e^{\mp \mu_n S}\right\}.
\end{align}
Substitution into \eqref{eq:axial_momentum_BL_W_bar} yields \citep{davis_stokes_1990}
\begin{align}\label{eq:BL_pressures}
\begin{split}
    &-\left.\frac{\partial p}{\partial s}\right\vert_{s=0}+\frac{\partial P_1^\pm}{\partial S}=\frac{1}{r}\frac{\partial}{\partial r}\left(L_{-1}\psi^\pm\right)=\mathrm{Re}\left\{\sum_{n=1}^\infty A_n(t)e^{\mp\mu_n S}\Gamma_n(r)\right\},\\
    %&=\frac{1}{r}\mathrm{Re}\left\{\sum_{n=1}^\infty A_n(t)e^{\mp\mu_n S}\left(\psi_n'''(r)-\frac{\psi_n''(r)}{r}+\frac{\psi_n'(r)}{r^2}+\mu_n^2\psi_n'(r)\right)\right\}\\
    &\Gamma_n(r)\equiv  \frac{1}{r}\left(\psi_n'''(r)-\frac{\psi_n''(r)}{r}+\frac{\psi_n'(r)}{r^2}+\mu_n^2\psi_n'(r)\right).
    \end{split}
\end{align}
Computing $\Gamma_n(r)$ and simplifying using the recurrence relation for Bessel functions, $J_2(z)=2/zJ_1(z)-J_0(z)$ we find 
\begin{align}
    \Gamma_n(r)=-\mu_nJ_0(\mu_n r).%\quad \frac{-2\lambda_n^2 J_0(\lambda_n r)}{J_1(\mu_n)}.
\end{align}
Following \cite{damiano_singular_1996} we define $\Delta P_1(r,\vert S\vert, t)=P_1(r,S^-,t)-P_1(r,S^+,t)$, so that integrating \eqref{eq:BL_pressures} and subtracting
\begin{align}
    \Delta P_1(r,\vert S\vert,t)-\Delta P_1(r,\vert0\vert,t)=-2 \mathrm{Re}\left\{\sum_{n=1}^\infty \frac{A_n(t)}{\mu_n}(1-e^{-\mu_n \vert S\vert})\Gamma_n(r)\right\}.
\end{align}
The variable of interest here is $\Delta P_1(r,\vert0\vert,t)$, as this is the $\mathcal{O}(\epsilon)$ pressure jump across the cupula and will thus induce a similarly sized correction to its deformation. To isolate it, we take the limit $\vert S\vert \rightarrow \infty$, yielding
\begin{align}
    \Delta P_1(r,\vert \infty\vert,t)-\Delta P_1(r,\vert0\vert,t)=-2 \mathrm{Re}\left\{\sum_{n=1}^\infty \frac{A_n(t)}{\mu_n}\Gamma_n(r)\right\}.
\end{align}
We identify $\Delta P_1(r,\vert \infty \vert,t)$ as the pressure jump from the outer flow, namely
\begin{align}
    \Delta P_1(r,\vert \infty \vert,t)\equiv\Delta p_1^{\text{outer}}(r,t)=\int_{\epsilon\beta/2}^{2\pi-\epsilon\beta/2}\frac{\partial p_1}{\partial s}\upd s,
\end{align}
which is in fact a function of $t$ only. Therefore, the first order pressure jump across the cupula is
\begin{align}
    \Delta p_1(r,t)=\Delta P_1(r,\vert0\vert,t)=\Delta p_1^{\text{outer}}(t)+2 \mathrm{Re}\left\{\sum_{n=1}^\infty \frac{A_n(t)}{\mu_n}\Gamma_n(r)\right\}\\=\Delta p_1^{\text{outer}}(t)+2f_1(t)f_3(\beta,\nu_s) \mathrm{Re}\left\{\sum_{n=1}^\infty \frac{\bar{A}_n}{\mu_n}\Gamma_n(r)\right\}
    =\Delta p_1^{\text{outer}}(t)+\Delta p_1^{\text{BL}}(r,t),\end{align}
    with the boundary layer contribution to the pressure jump,
    \begin{align}\label{eq:boundary_layer_pressure_jump}
    \Delta p_1^{\text{BL}}(r,t)=-\frac{\pi }{I_4}\left(\dot{\Omega}(t)+\frac{\Delta p_0}{2\pi}\right)f_3(\beta,\nu_s)\mathrm{Re}\left\{\sum_{n=1}^\infty \bar{A}_nJ_0(\mu_nr)\right\}.
\end{align}
Therefore, we find an $r$ dependent correction to the cupular pressure jump, and we give a plot of this $r$-dependence in figure~\ref{fig:boundary_layer_fit}(c), where we can see $\Delta p_1^{\text{BL}}$ is positive (compressive) in the centre of the cupula, and negative (tensile) in regions close to the canal walls. The total pressure jump across the cupula up $\mathcal{O}(\epsilon^2)$ is thus
\begin{align}
    \Delta p =\Delta p_0(t) +\epsilon \Delta p_1^{\text{outer}}(t)-\epsilon\frac{\pi }{I_4}\left(\dot{\Omega}(t)+\frac{\Delta p_0}{2\pi}\right)f_3(\beta,\nu_s)\mathrm{Re}\left\{\sum_{n=1}^\infty \bar{A}_nJ_0(\mu_nr)\right\},
\end{align}
where the leading order pressure jump satisfies \eqref{eq:pressure_jump_leading_order_with_thickness}, repeated here for convenience:
\begin{align}
    \frac{\alpha_0}{\kappa}\frac{\upd \Delta p_0}{\upd t}&=-\frac{1}{16 \tilde{I}_4}(\dot{\Omega}(t)(2\pi-\epsilon\beta)+\Delta p_0).
\end{align}
It may be seen that $f_3(\beta,\nu_s)\propto 1/\beta^2$ as $\beta\rightarrow \infty$, so that the correction due to the presence of the boundary layer decreases in magnitude as the thickness is increased. This is because the deformation profile $\eta_0(r,t)$ of the cupula is a quadratic in $r$ for large $\beta$, and due to the kinematic condition it must be the same quadratic as the leading order outer flow $w_0(r,t)$, so that there is no adjustment to the flow close to the cupula. Conversely, the correction will be greatest for a thin cupula ($\beta\ll1$), when the leading order deformation profile is a quartic in $r$, inconsistent with the outer flow.

\section{Light cupula hypothesis/buoyant cupula}\label{sec:light_cupula}
The body force due to gravity is $\boldsymbol{f}=\Delta \rho\boldsymbol{g}$, where $\Delta \rho=\rho_s-\rho$, so that the vectors $\boldsymbol{f}$ and $\boldsymbol{g}$ have the same sense when the cupula is heavier than the surrounding endolymph (i.e. $\Delta \rho>0$). For a stationary human ($\hat{\Omega}(\hat{t})\equiv0$), the leading order solid problem is 
\begin{align}\label{eq:buoyancy_solid}
    0=\Delta \rho\boldsymbol{g}+\boldsymbol{\nabla}\cdot\boldsymbol{\hat{\tau}}.
\end{align}
Therefore, the deflection will scale as $\boldsymbol{\hat{u}}_s\sim a^2\Delta \rho g \cos\Phi/E$, with $\Phi$ the angle between the normal vector to the cupula's flat faces and gravity (i.e. $g\cos\Phi=\boldsymbol{g}\cdot\mathbf{e}_z$). When we compare this to the rotation induced deflection (which is highest when $\kappa\ll1$) of characteristic size $a^2 R\Omega_0/\nu$, we find that the ``equivalent rotation rate'' for a density change of size $\Delta \rho $ is
\begin{align}
    \Omega_0\sim\frac{\Delta\rho}{\rho}\frac{\mu g}{E R}\cos\Phi.
\end{align}
For conservative estimate $\Delta\rho\cos\Phi/\rho$ of $1\%$ we find $\Omega_0\approx 0.3$ rad$\cdot$s$^{-1}$ for characteristic values of $\mu=10^{-3}$ kg$\cdot$m$^{-1}\cdot$s$^{-1}$, $g=9.81$ m$\cdot$s$^{-1}$, $R=1.6\cdot10^{-3}$ m and $E=20$ Pa. 

The full solution to \eqref{eq:buoyancy_solid} may be found using the techniques from \S\ref{sec:solid_mechanics}. In particular, the deformation will have a radially symmetric component in the vertical direction of the same form as $\eta_1^\text{e}(r,t)$, and a component $\sim\cos\theta$ which does not contribute to the flux computation.

%
%\bibliographystyle{jfm}
%\bibliography{references2}

\end{document}